\documentclass[aps,preprint,showpacs,preprintnumbers,amsmath,amssymb,nofootinbib,showkeys]{revtex4-1}

\RequirePackage[T1]{fontenc}

\usepackage{epsfig,graphicx,amsmath,amssymb}
\RequirePackage{mathptmx}      
\RequirePackage{flushend}
\RequirePackage{color}
\usepackage{changes}
\usepackage{multirow}
\RequirePackage{hyperref}
\hypersetup{
	linktocpage,
    colorlinks,
    citecolor=blue,
    filecolor=black,
    linkcolor=blue,
    urlcolor=blue,
}

\usepackage{slashed}
\usepackage{tikz}
\usetikzlibrary{trees}
\usetikzlibrary{decorations.pathmorphing}
\usetikzlibrary{decorations.markings}
\usepackage{tikz,ifthen}
\tikzset{beamerprimary/.style={structure.fg, thick}}
\tikzset{beamersecondary/.style={structure.bg, thick}}
\tikzset{boson/.style={draw=structure.fg,decorate, decoration={snake}},
    gauge/.style={decorate, decoration={snake} },
    fermion/.style={postaction={decorate},
        decoration={markings,mark=at position .55 with {\arrow{>}}}},
    fermionloop/.style={postaction={decorate},
        decoration={markings,mark=at position .25 with {\arrow{<}}}}, 
    gluon/.style={decorate, 
        decoration={coil,amplitude=4pt, segment length=5pt}},
    scalar/.style={dashed},
    scalarloop/.style={dashed={decorate},
        decoration={markings,mark=at position .25 with {\arrow{<}}}},
    resonance/.style={double,double distance=1.5pt}	
}
\usetikzlibrary{arrows,shapes,positioning}
\usetikzlibrary{decorations.markings}
\tikzstyle arrowstyle=[scale=1]
\tikzstyle directed=[postaction={decorate,decoration={markings,
    mark=at position .65 with {\arrow[arrowstyle]{stealth}}}}]
\tikzstyle reverse directed=[postaction={decorate,decoration={markings,
    mark=at position .65 with {\arrowreversed[arrowstyle]{stealth};}}}]

\begin{document}

\title{Radiative corrections to the 
\texorpdfstring{$\boldsymbol{\tau^-\to (P_1P_2)^-\nu_\tau}$}{Lg} (\texorpdfstring{$\boldsymbol{P_{1,2}=\pi, K}$}{Lg}) decays}

\author{Rafel Escribano$^{1,2}$}\email{rescriba@ifae.es}
\author{Jes\'us Alejandro Miranda$^{2}$}\email{jmiranda@ifae.es}
\author{Pablo Roig$^{3}$}\email{proig@fis.cinvestav.mx}

\affiliation{$^{1}$Grup de F\'isica Te\`orica, Departament de F\'isica, 
Universitat Aut\`onoma de Barcelona, 08193 Bellaterra (Barcelona), Spain.\\
$^{2}$Institut de F\'isica d’Altes Energies (IFAE) and 
The Barcelona Institute of Science and Technology, 
Campus UAB, 08193 Bellaterra (Barcelona), Spain.\\
$^{3}$Departamento de F\'isica, Centro de Investigaci\'on y de
Estudios Avanzados del Instituto Polit\'ecnico Nacional,
Apdo. Postal 14-740, 07000 Ciudad de M\'exico, M\'exico.}


\begin{abstract}
The radiative corrections to the $\tau^-\to (P_1P_2)^-\nu_\tau$ ($P_{1,2}=\pi, K$) decays
are calculated for the first time including the structure-dependent real photon corrections, 
which are obtained using Resonance Chiral Theory.
Our results, whose uncertainty is dominated by the model-dependence of the resummation 
of the radiative corrections and the missing virtual structure-dependent contributions, 
allow for precise tests of CKM unitarity, lepton flavour universality and non-standard interactions.
\end{abstract}
\maketitle
 
\section*{Introduction}
Semileptonic tau decays are well-known to be a clean laboratory for studying QCD 
hadronic matrix elements at energies below $\sim1.8$ GeV \cite{Davier:2005xq,Pich:2013lsa}, 
where the light-flavoured resonances play a key role. 
All non-perturbative information of the one-meson tau decays is encoded in the corresponding 
$P$ decay constants, that are best determined in lattice QCD
\cite{FlavourLatticeAveragingGroupFLAG:2021npn}.
Two-meson tau decays are specified in terms of two form factors, 
whose knowledge has improved over the years thanks to the use of dispersion relations
\cite{Pich:2001pj,GomezDumm:2013sib,Gonzalez-Solis:2019iod,Jamin:2001zq,Moussallam:2007qc,
Boito:2008fq,Boito:2010me,Antonelli:2013usa,Bernard:2013jxa,Escribano:2013bca,
Escribano:2014joa,Descotes-Genon:2014tla,Escribano:2016ntp} 
and nourished with high quality measurements
\cite{OPAL:2000fde,ALEPH:2005qgp,BaBar:2009lyd,OPAL:1998rrm,CLEO:1999dln,
Belle:2008xpe,BaBar:2018qry,Belle:2019bpr}. 
A similar good understanding 
has not yet been achieved in three-meson tau decays  
\cite{GomezDumm:2003ku,Moussallam:2007qc,Dumm:2009va,Dumm:2009kj,GomezDumm:2012dpx,
Shekhovtsova:2012ra,
Nugent:2013hxa,Sanz-Cillero:2017fvr,JPAC:2018zwp,Arteaga:2022xxy} 
or higher-multiplicity modes, preventing for the moment their use in searches for new physics.

On the contrary, one- and two-meson tau decays have enabled significant and 
promising new physics tests in recent years 
\cite{Garces:2017jpz,Cirigliano:2017tqn,Miranda:2018cpf,Cirigliano:2018dyk,Rendon:2019awg,
Chen:2019vbr,
Gonzalez-Solis:2019lze,Gonzalez-Solis:2020jlh,Chen:2020uxi,Arroyo-Urena:2021dfe,
Arroyo-Urena:2021nil,Chen:2021udz,Cirigliano:2021yto}. 
At the precision attained, radiative corrections for these decay modes become necessary, 
which motivated their improved evaluation for the $\tau^-\to P^-\nu_\tau$ cases  
\cite{Guo:2010dv,Guevara:2013wwa,Guevara:2021tpy,Arroyo-Urena:2021dfe,Arroyo-Urena:2021nil}.
For the di-pion tau decays, the need for these corrections first stemmed from their use in the 
dispersive integral rendering the leading order hadronic vacuum polarization contribution 
to the muon $g-2$ \cite{Cirigliano:2001er,Cirigliano:2002pv,Flores-Baez:2006yiq}, 
which was again the target of our recent analysis \cite{Miranda:2020wdg} 
(see also Refs.~\cite{GutierrezSantiago:2020bhy,Chen:2022nxm,Masjuan:2023qsp}). 
Ref.~\cite{Antonelli:2013usa} put forward that, assuming lepton universality, 
semileptonic kaon decay measurements could be used to predict the corresponding 
(crossing-symmetric) tau decays, yielding a $V_{us}$ determination closer to unitarity 
than with the tau decay branching ratios. 
In that work, the model-independent radiative corrections were taken into account and 
the structure-dependent ones were estimated (see also Ref.~\cite{Flores-Baez:2013eba}), 
resulting in a relative large (conservative) uncertainty. 
Including these model-dependent effects is one of our main motivations: 
here we focus on those with a real photon and defer the virtual photon ones to a later dedicated study. 
Instead of relying on lepton universality and checking CKM unitarity \cite{Antonelli:2013usa}, 
one can in principle test the latter, comparing the crossed channels, 
or directly bind new physics non-standard interactions from $\tau^-\to (K\pi)^-\nu_\tau$ decays 
\cite{Rendon:2019awg}. 
For completeness, we also include the radiative corrections to the di-kaon tau decays and 
recall our reference results for the di-pion mode \cite{Miranda:2020wdg}.
As noted in Ref.~\cite{Cirigliano:2021yto}, see Fig.~1 for instance, 
bounds on non-standard interactions from hadronic tau decays are competitive and complementary 
to those coming from LHC searches and electroweak precision observables. 
As a relevant example, the precise comparison of $\tau\to\pi^-\pi^0\nu_\tau(\gamma)$ 
with $e^+e^-\to\pi^+\pi^-(\gamma)$ data, 
which requires the radiative corrections computed in this work (see also Ref.~\cite{Miranda:2020wdg}), 
are able to reduce the allowed new physics area (in the relevant Wilson coefficients plane) 
by a factor $\sim 3$ \cite{Cirigliano:2021yto}.
Real radiation was computed for the $\tau^-\to\eta^{(\prime)}\pi^-\nu_\tau$ decay channels in 
Ref.~\cite{Guevara:2016trs}, showing that it can contend with the non-photon decays, 
as $G$-parity and electromagnetic suppressions compete. 
Finally, we also estimate the corresponding results for the $K^-\eta^{(\prime)}$ channels.

The structure of the paper is as follows. 
In section \ref{sec:PP}, we recall the model-independent description of the 
$\tau^-\to P^-_1 P^0_2\nu_\tau \gamma$ decays and give the leading real-photon model-dependent corrections 
for the $K\pi$, $K\bar{K}$ and $\pi\pi$ cases, 
where only the latter are known (see e.g. Refs.~\cite{Cirigliano:2002pv,Miranda:2020wdg}). 
Branching ratios and spectra for the radiative decays are analysed in 
section \ref{sec:RadTauDec}, 
and the corresponding radiative correction factors are computed in section \ref{sec:RadCors}.
We show the consequences of including them in section \ref{sec:Fits}, 
where we bind new physics couplings in an effective approach. 
Finally, we conclude in section \ref{sec:Concl}. 
Appendices cover $K_{\ell3}$ decays (\ref{AppKl3}), 
structure-independent virtual corrections to di-meson tau decays (\ref{Appx:RC}), 
the non-radiative decays (\ref{AppC}),
and the kinematics of these three- and four-body processes (\ref{App04:Kine}).

\section{The \texorpdfstring{$\boldsymbol{\tau^-\to P_1^-P_2^0\nu_\tau\gamma}$}{Lg} decays}
\label{sec:PP}
The most general structure for the decays 
$\tau(P)\to P_1^-(p_-)P_2^0(p_0)\nu_\tau(q^\prime)\gamma(k)$
is given by
\begin{equation}\label{eq:tau_decay}
\begin{split}
\mathcal{M}&=\frac{e G_F V^{*}_{uD}}{\sqrt{2}}\epsilon^{*}_\mu
\bigg[\frac{H_\nu(p_{-},p_{0})}{k^2-2k\cdot P}
\bar{u}(q')\gamma^\nu (1-\gamma^5)(m_\tau +\slashed{P}-\slashed{k})\gamma^\mu u(P)\\[1ex]
&\quad +\,(V^{\mu\nu}-A^{\mu\nu})\bar{u}(q')\gamma_\nu (1-\gamma^5) u(P)\bigg]\ ,
\end{split}
\end{equation}
with $V_{uD}$ ($D=d,s$) the corresponding CKM matrix element
and where the hadronic matrix element can be written as
\begin{equation}
H^\nu(p_{-},p_{0})=C_V F_{+}(t)Q^\nu+C_S\frac{\Delta_{{-}{0}}}{t}q^\nu F_{0}(t)\ ,
\end{equation}
with $t=q^2$, $Q^\nu=(p_{-}-p_{0})^\nu-\frac{\Delta_{{-}{0}}}{t}q^\nu$, 
$q^\nu=(p_{-}+p_{0})^\nu$, and $\Delta_{ij}=m_i^2-m_j^2$. 
One recovers the usual definition of $H_{K\pi}^\nu$ \cite{Rendon:2019awg} 
by replacing $p_-\to p_K$, $p_0\to p_\pi$
and $\Delta_{-0}\to\Delta_{K\pi}$ for $K^{-}\pi^0$, 
and $p_-\to p_\pi$, $p_0\to p_K$, $C_{V,S}\to -C_{V,S}$ 
and $\Delta_{-0}\to -\Delta_{K\pi}$ for $\bar{K}^{0}\pi^{-}$ 
(we comment on the identifications for the $P_1=P_2$ channels below
). 
In all cases, gauge invariance 
implies $k_\mu V^{\mu\nu}=H^\nu(p_{-},p_{0})$ and $k_\mu A^{\mu\nu}=0$.

\begin{figure}
\begin{center}
\begin{tikzpicture}
	\draw[scalar] (0,0) -- (1.5,1.5) node [right] {$\bar{K}^0$};
	\draw[gauge] (0,0) -- (2.12,0) node[right]{$\gamma$};
	\draw[scalar] (0,0) -- (1.5,-1.5) node [right] {$\pi^-$};
	\draw[fill,white] (0,0) circle [radius=0.12]; 
	\node at (0,0) {\small$\otimes$};
\end{tikzpicture}
\begin{tikzpicture}
	\draw[scalar] (0,0) -- (1.5,1.5) node [right] {$K^-$};
	\draw[gauge] (0,0) -- (2.12,0) node[right]{$\gamma$};
	\draw[scalar] (0,0) -- (1.5,-1.5) node [right] {$\pi^0$};
	\draw[fill,white] (0,0) circle [radius=0.12]; 
	\node at (0,0) {\small$\otimes$};
\end{tikzpicture}
\end{center}
\caption{Feynman diagrams contributing to the term proportional to the metric tensor $g^{\mu\nu}$
in Eq.~(\ref{eq3:VSI}).}
\label{An:fig4}
\end{figure}
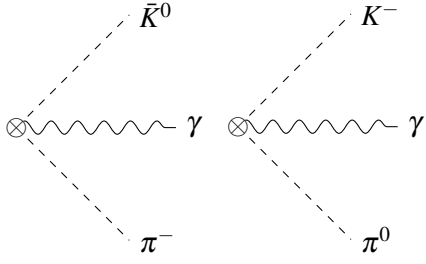

The structure-independent term is given by
\begin{equation}
\label{eq3:VSI}
\begin{split}
V^{\mu\nu}_{\rm SI}&=\frac{ H^\nu(p_-+k,p_0)(2p_-+k)^\mu}{2k\cdot p_-+k^2}
-C_V\frac{F_{+}(t^\prime)-F_{+}(t)}{k\cdot q}Q^\nu q^\mu\\[1ex]
&+\left\{-C_V F_{+}(t^\prime)-\frac{\Delta_{-0}}{t^\prime}
\left[C_S F_{0}(t^\prime)-C_V F_{+}(t^\prime)\right]\right\}g^{\mu\nu}\\[1ex]
&+\frac{\Delta_{-0}}{tt^\prime}
\left\{2\left[C_S F_{0}(t^\prime)-C_V F_{+}(t^\prime)\right]
-\frac{C_S t^\prime}{k\cdot q}\left[F_{0}(t^\prime)-F_{0}(t)\right]\right\} q^\mu q^\nu\ ,
\end{split}
\end{equation}
where $C_V^{K^-\pi^0}=C_S^{K^-\pi^0}=1/\sqrt{2}$ for the $K^-\pi^0$ channel, 
and $C_V^{\bar{K}^0\pi^-}=C_S^{\bar{K}^0\pi^-}=-1$ for the $\bar{K}^0\pi^-$ one,
with $t^\prime=(P-q')^2$. 
The main difference between these two decay modes comes from the overall sign difference 
---except for the first term in the second line of Eq.~(\ref{eq3:VSI})---,
that we absorbed in our definition of $C_{V,S}^{\bar{K}^0\pi^-}$, 
and through the order of the arguments of $H^\nu$ in the above equation.  
At leading order (LO) in Chiral Perturbation Theory (ChPT), 
contributions proportional to $g^{\mu\nu}$ in $V^{\mu\nu}_{\rm SI}$ 
stem from the diagrams in Fig.~\ref{An:fig4}.

For the other tau decay modes, $V_{\rm SI}^{\mu\nu}$ is also obtained from Eq.~(\ref{eq3:VSI}).
In particular, we are also interested in the $\tau^-\to K^- K^0\nu_\tau\gamma$ decays, 
where $C_V^{K^- K^0}=C_S^{K^-K^0}=-1$~\footnote{We 
discuss briefly the $\pi^-\pi^0$ case at the end of Sec.~\ref{VFF:KK};
see Ref.~\cite{Miranda:2020wdg} for further details.}.

The structure-dependent part is given by
\begin{equation}
\label{eq1:VSD}
\begin{split}
V^{\mu\nu}_{\rm SD}&=v_1(k\cdot p_{-} g^{\mu\nu}-k^\nu p_{-}^\mu)+v_2(k\cdot p_{0} g^{\mu\nu}-k^\nu p_{0}^\mu)\\[1ex]
&+v_3(k\cdot p_{0} p_{-}^\mu -k\cdot p_{-} p_{0}^\mu)p_{-}^\nu+v_4(k\cdot p_{0} p_{-}^\mu -k\cdot p_{-} p_{0}^\mu)(p_{-}+p_{0}+k)^\nu\ ,
\end{split}
\end{equation}
and
\begin{equation}
\label{eq1:ASD1}
\begin{split}
A_{\mu\nu}&=i a_1 \epsilon_{\mu\nu\rho\sigma}\left(p_{0}-p_{-}\right)^\rho k^\sigma +i a_2\left(P-q'\right)_\nu\epsilon_{\mu\rho\sigma\tau}k^\rho p_{-}^\sigma p_{0}^\tau\\[1ex]
&+i a_3\epsilon_{\mu\nu\rho\sigma}k^\rho(P-q')^\sigma +i a_4 (p_{0}+k)_\nu\epsilon_{\mu\lambda\rho\sigma}k^\lambda p_{-}^\rho p_{0}^\sigma\ ,
\end{split}
\end{equation}
where $p_{-}$ and $p_{0}$ refer to the momentum of the charged and neutral meson, respectively. 

From Eq.~(\ref{eq3:VSI}), 
it is easy to show that the Low's theorem \cite{Low:1958sn} is manifestly satisfied,

\begin{equation}\label{VSIexp}
\begin{split}
V^{\mu\nu}&=\frac{p_{-}^\mu}{k\cdot p_{-}}H^\nu(p_{-},p_{0})
+\left\lbrace C_V F_{+}(t)+\frac{\Delta_{-0}}{t}
\left[C_S F_{0}(t)-C_V F_{+}(t)\right]\right\rbrace 
\left(\frac{p_{-}^\mu k^\nu}{k\cdot p_{-}}-g^{\mu\nu}\right)\\[1ex]
&-\frac{2\Delta_{-0}}{t^2}\left[C_S F_{0}(t)-C_V F_{+}(t)\right]
\left(\frac{k\cdot p_{0}}{k\cdot p_{-}}p_{-}^\mu-p_{0}^{\mu}\right)(p_{-}+p_{0})^\nu\\[1ex]
&+2\left(\frac{k\cdot p_{0}}{k\cdot p_{-}}p_{-}^\mu-p_{0}^{\mu}\right)
\left[C_V\frac{d F_{+}(t)}{dt}Q^\nu+C_S\frac{\Delta_{{-}{0}}}{t}q^\nu
\frac{d F_{0}(t)}{dt}\right]+\mathcal{O}(k)\ ,
\end{split}
\end{equation}

and the amplitude reads 
\begin{equation}
\label{LowAmp}
\begin{split}
\mathcal{M}&=
\frac{e G_{F} V_{uD}\sqrt{S_{\rm EW}}}{\sqrt{2}}\epsilon^{*}_{\mu}(k) H_\nu(p_{-},p_{0}) 
\bar{u}(q)\gamma^\nu (1-\gamma^5) u(P)\\[1ex]
&\quad\times\left(\frac{p_{-}^\mu}{k\cdot p_{-}+\frac{1}{2}M_\gamma^2}
-\frac{P^\mu}{k\cdot P-\frac{1}{2}M_\gamma^2}\right)+\mathcal{O}(k^0)\ ,
\end{split}
\end{equation}
where $S_{\rm EW}$ encodes the short-distance electroweak corrections 
\cite{Sirlin:1974ni,Sirlin:1977sv,Sirlin:1981ie,Marciano:1985pd,Marciano:1988vm,Marciano:1993sh,
Braaten:1990ef,Erler:2002mv}.

In the Low's limit, one gets
\begin{equation}
\begin{split}
\overline{\left\vert\mathcal{M}\right\vert^2}&= 
2 e^2 G_F^2\left\vert V_{uD}\right\vert^2 S_{\rm EW}
\left\{C_{S}^2\left\vert F_{0}(t)\right\vert^2 D_{0}^{P^-P^0}(t,u)\right.+C_{S}C_{V}\mathrm{Re}\left[F_{+}(t)F_{0}^{*}(t)\right] D_{+0}^{P^-P^0}(t,u)\\[1ex]
&\left.+\,C_{V}^2\left\vert F_{+}(t)\right\vert^2 D_{+}^{P^-P^0}(t,u)\right\}
\sum_{\gamma\,\rm pols.}\left\vert\frac{p_{-}\cdot\epsilon}{p_{-}\cdot k}
-\frac{P\cdot \epsilon}{P\cdot k}\right\vert^2+\mathcal{O}(k^0)\ ,
\end{split}
\end{equation}
where 
\begin{equation}
\label{SqA:eqDp}
\begin{split}
D_{+}^{P^-P^0}(t,u)&=
\frac{m_\tau^2}{2}(m_\tau^2-t)+2m_{0}^2m_{-}^2-2u(m_\tau^2-t+m_{0}^2+m_{-}^2)+2u^2\\[1ex]
&+\frac{\Delta_{-0}}{t}m_\tau^2(2u+t-m_\tau^2-2m_{0}^2)+\frac{\Delta_{-0}^2}{t^2}\frac{m_\tau^2}{2}(m_\tau^2-t)\ ,
\end{split}
\end{equation}
\begin{equation}
\label{SqA:eqD0}
D_{0}^{P^-P^0}(t,u)=\frac{\Delta_{-0}^2 m_\tau^4}{2t^2}\left(1-\frac{t}{m_\tau^2}\right)\ ,
\end{equation}
\begin{equation}
\label{SqA:eqDp0}
D_{+0}^{P^-P^0}(t,u)=
\frac{\Delta_{0-}m_\tau^2}{t}
\left[2u+t-m_\tau^2-2m_{0}^2+\frac{\Delta_{-0}}{t}(m_\tau^2-t)\right]\ ,
\end{equation}
with $u=(P-p_-)^2$.
In this way, besides the Low theorem, the Burnet-Kroll theorem \cite{Burnett:1967km} 
is also explicitly manifest.

Thus, after integration over neutrino and photon 4-momenta, 
the differential decay width in this approximation reads
\begin{equation}
\label{eq:RPhoton}
\begin{split}
    \left.\frac{d\Gamma^{(0)}}{dt\,du}\right\vert_{PP\gamma}&=
    \frac{G_F^2\vert V_{uD}\vert^2 S_{\rm EW}}{128\pi^3m_\tau^3}
    \left\{C_S^2\left\vert F_0(t)\right\vert^2D_{0}^{P^{-}P^{0}}(t,u)\right.+C_V C_S\mathrm{Re}\left[F_{+}^{*}(t)F_{0}(t)\right] D_{+0}^{P^{-}P^{0}}(t,u)\\[1ex]
    &\quad\left.+\,C_V^2\left\vert F_{+}(t)\right\vert^2D_{+}^{P^{-}P^{0}}(t,u)\right\}
    g_{\rm rad}(t,u,M_\gamma)\ ,
\end{split}
\end{equation}
where (see Refs.~\cite{Cirigliano:2001er,Cirigliano:2002pv})
\begin{equation}
    g_{\rm rad}(t,u,M_\gamma)=g_{\rm brems}(t,u,M_\gamma)+g_{\rm rest}(t,u)\ ,
\end{equation}
with
 \begin{subequations}
 \label{eq:JndK}
 \begin{align}
     g_{\rm brems}(t,u,M_\gamma)&=
     \frac{\alpha}{\pi}\left(J_{11}(t,u,M_\gamma)+J_{20}(t,u,M_\gamma)+J_{02}(t,u,M_\gamma)\right)\ ,\\[1ex]
     g_{\rm rest}(t,u)&=\frac{\alpha}{\pi}\left(K_{11}(t,u)+K_{20}(t,u)+K_{02}(t,u)\right)\ .
 \end{align}
 \end{subequations}
The relevant expressions for $J_{ij}(t,u,M_\gamma)$ and $K_{ij}(t,u)$,
which correspond to an integration over $\mathcal{D}^{\rm III}$ and $\mathcal{D}^{\rm IV/III}$,
respectively, can be found in Refs.~\cite{Cirigliano:2002pv,Miranda:2020wdg,Antonelli:2013usa} 
and in App.~\ref{App04:Kine}\footnote{The
function $K_{11}(t,u)$ turns out to be numerically negligible and is not quoted anywhere;
see e.g. Refs~\cite{Cirigliano:2001er, Miranda:2018cpf}.}.

Integrating upon the $u$ variable in Eq.~(\ref{eq:RPhoton}), one gets
\begin{equation}
\label{eq:widthRP}
\begin{split}
\left.\frac{d\Gamma}{dt}\right\vert_{\rm III}&=
\frac{G_F^2 S_{\rm EW}\vert V_{uD}\vert^2 m_\tau^3}{384\pi^3 t}
\left\{\frac{1}{2t^2}\left(1-\frac{t}{m_\tau^2}\right)^2\lambda^{1/2}(t,m_{-}^2,m_0^2)\right.\\[1ex]
&\times\left[C_V^2\left\vert F_{+}(t)\right\vert^2\left(1+\frac{2t}{m_\tau^2}\right)
\lambda(t,m_{-}^2,m_0^2)\overline{\delta}_{+}(t)
+3C_S^2\Delta_{-0}^2\left\vert F_{0}(t)\right\vert^2\overline{\delta}_{0}(t)\right]\left.+\,C_S C_V\frac{4}{\sqrt{t}}\overline{\delta}_{+0}(t)\right\}\ ,
\end{split}
\end{equation} 
with
\begin{subequations}\label{eq:delta_bar}\begin{align}
\overline{\delta}_0(t)=&\ \frac{\int_{u^{-}(t)}^{u^{+}(t)}D_{0}^{P^{-}P^{0}}(t,u)\,
g_{\rm brems}(t,u,M_\gamma)du}{\int_{u^{-}(t)}^{u^{+}(t)}D_{0}^{P^{-}P^{0}}(t,u)\,du}\ ,\\[1ex]
\overline{\delta}_+(t)=&\ \frac{\int_{u^{-}(t)}^{u^{+}(t)}D_{+}^{P^{-}P^{0}}(t,u)\,
g_{\rm brems}(t,u,M_\gamma)du}{\int_{u^{-}(t)}^{u^{+}(t)}D_{+}^{P^{-}P^{0}}(t,u)\,du}\ ,\\[1ex]
\overline{\delta}_{+0}(t)=&\ 
\frac{3t\sqrt{t}}{4m_\tau^6}\int_{u^{-}(t)}^{u^{+}(t)}D_{+0}^{P^{-}P^{0}}(t,u)\,
g_{\rm brems}(t,u,M_\gamma)\mathrm{Re}\left[F_{+}^{*}(t) F_{0}(t)\right]du\ .
\end{align}
\end{subequations}

The remaining contribution, $\left.d\Gamma/dt\right\vert_{\rm IV/III}$, 
which corresponds to the integration over $\mathcal{D}_{\rm IV/III}$ with $g_{\rm rest}(t,u)$ 
instead of $g_{\rm brems}(t,u,M_\gamma)$,
is almost negligible and only becomes relevant near threshold. 
In Ref.~\cite{Cirigliano:2008wn}, 
the subleading contributions in the Low's approximation were studied, 
showing that they are not negligible and need to be taken into account to get a reliable 
estimation.
\section{Structure-dependent contributions}
The evaluation of the structure-dependent tensors, $V^{\mu\nu}_{SD}$ and $A^{\mu\nu}$ in eqs.~(\ref{eq1:VSD}) and (\ref{eq1:ASD1}), requires non-perturbative methods or lattice QCD (which has only been explored for the $\pi\pi$ case, see \cite{Bruno:2018ono}). The tau lepton mass value probes the hadronization of QCD currents in its semileptonic decays beyond the regime of validity of Chiral Perturbation Theory~\cite{Weinberg:1978kz,Gasser:1983yg,Gasser:1984gg,Bijnens:1999sh,Bijnens:2001bb} ($\chi PT$), which is the low-energy effective field theory of QCD describing $P$ meson physics. To extend the applicability of the chiral Lagrangians to the energy region where meson states resonate, a successful strategy has been to include the corresponding fields as explicit degrees of freedom into the action, using approximate flavor symmetry, without additional assumptions affecting the possible resonance dynamics \cite{Ecker:1988te,Ecker:1989yg}. This procedure was later christened Resonance Chiral Theory ($R\chi T$, see e.g. ref.~\cite{Portoles:2010yt}) and yields as a result the saturation of the $\chi PT$ low-energy constants upon resonance integration.

Explicitly, the construction of the relevant Lagrangian pieces including resonances uses the chiral tensors \cite{Bijnens:2001bb} (we only quote those relevant in our computation)
\begin{eqnarray}\label{eq.chiraltensors}
&u_\mu=i\left[u^\dagger(\partial_\mu-ir_\mu)u-u(\partial_\mu-i\ell_\mu)u^\dagger\right],\nonumber\\[1ex]
&f_{\pm}^{\mu\nu}=uF_L^{\mu\nu}u^\dagger\pm u^\dagger F_R^{\mu\nu}u,\\[1ex]
&\chi_{\pm} = u^\dagger\chi u^\dagger \pm u \chi u\,,\nonumber
\end{eqnarray}
where the pseudo-Goldstone bosons are included in $u$ via ($\lambda_ a$ are the Gell-Mann matrices, so that $\phi^3=\pi^0$ in the isospin symmetry limit)
\begin{equation}
u=\mathrm{exp}\left(\frac{i\Phi}{\sqrt{2}f}\right),\qquad \Phi=\sum_{a=0}^8\frac{\lambda_a}{\sqrt{2}}\phi^a\,,
\end{equation}
and left- and right-handed sources $\ell_\mu$ and $r_\mu$ enter through
\begin{equation}
F_X^{\mu\nu}=\partial^\mu x^\nu-\partial^\nu x^\mu-i[x^\mu,x^\nu]=eQF^{\mu\nu}+\dots,\quad x=\ell,\,r;\quad Q=\mathrm{diag}\left(\frac{2}{3},-\frac{1}{3},-\frac{1}{3}\right),
\end{equation}
with $x_\mu=eQA_\mu+\dots$, being $A_\mu$ the photon field and $F^{\mu\nu}=\partial^\mu A^\nu-\partial^\nu A^\mu$ the corresponding field-strength tensor.
Spin-zero sources ($s$ and $p$) appear in $\chi_{\pm}$ through $\chi=2B(s+ip)$, where we recall that the two leading chiral low-energy constants ($f$ and $B$) determine the light-quark condensate in the chiral limit, $-Bf^2=<0|\bar{q}q|0>$, with $f\sim90$ MeV the pion decay constant. The numerical value of $B$ is not needed, as it only enters the pseudoscalar meson squared masses, which are proportional to it. Indeed, in the isospin symmetry limit, diag$(2Bs)=(m_\pi^2,\,m_\pi^2,\,2m_K^2-m_\pi^2)$, where $s$ accounts for the (diagonal) light-quark mass matrix, which ensure chiral symmetry breaking as in QCD.

The $R\chi T$ Lagrangian includes the lowest order $\chi PT$ Lagrangian in both parity sectors, which is
\begin{equation}
\mathcal{L}_{\mathrm{non-res}}^{\mathrm{even}}=\frac{f^2}{4}<u^\mu u_\mu+\chi_+>\,,\qquad \mathcal{L}_{\mathrm{non-res}}^{\mathrm{odd}}=\mathcal{L}_{\mathrm{WZW}}\,,\;
\end{equation}
where WZW stands for the chiral anomaly contribution worked out by Wess-Zumino and Witten \cite{Wess:1971yu, Witten:1983tw}.

In addition, $\mathcal{L}_{R\chi T}$ has pieces including resonance fields and chiral tensors. These are usually incorporated taking into account the order (within the chiral counting) of their contributions -upon resonance exchange- to the $\chi PT$ couplings \cite{Ecker:1988te, Ecker:1989yg,Cirigliano:2006hb, Kampf:2011ty}, as well as their behaviour \cite{Pich:2002xy} in the limit of a large number of colors \cite{tHooft:1973alw,tHooft:1974pnl}. QCD asymptotic behaviour forbids (linear combinations of) operators with increasing number of derivatives. In this way $R\chi T$ bridges between the low-energy behaviour of $\chi PT$ and the high-energy constraints of perturbative QCD, keeping predictivity for a set of related observables, to a given precision, without unnecessary assumptions (like, for instance vector meson dominance \cite{Sakurai:1960ju} or any additional symmetry of gauge type related to them \cite{Bando:1987br}). Within this framework, the need for non-resonant contributions was explored e. g. in ref.~\cite{Roig:2014uja} for the $\omega-\pi^0$ transition form factor, where it was found that they could play a r\^ole above 2 GeV only. Taking this into account and the fact that we are limited kinematically by the tau mass, we neglect non-resonant pieces in the following.

Apart from contributions that are suppressed by approximate flavor symmetries (more on this below), the next-to-leading order $\chi PT$ couplings are saturated by spin-one resonance exchange \cite{Ecker:1988te, Ecker:1989yg} coming from the following operators
\begin{equation}
\mathcal{L}_V=\frac{F_V}{2\sqrt{2}}\left\langle V_{\mu\nu}f_+^{\mu\nu}\right\rangle +i \frac{G_V}{2\sqrt{2}}\left\langle V_{\mu\nu}\left[u^\mu,u^\nu\right]\right\rangle\,,\qquad \mathcal{L}_A=\frac{F_A}{2\sqrt{2}}\left\langle A_{\mu\nu}f_-^{\mu\nu}\right\rangle\,,
\end{equation}
where $V_{\mu\nu}=\sum_{a=0}^8\frac{\lambda^a}{\sqrt{2}}V_{\mu\nu}^a$, and analogously for the axial-vector resonances. For convenience, spin-one resonances are worked out in the antisymmetric tensor field formalism \cite{Ecker:1988te, Ecker:1989yg}. For completeness we note that the kinetic terms come from the Lagrangian (which also includes interactions, hidden in the covariant derivative, that are however not needed in what follows, so $\nabla^\alpha\sim\partial^\alpha$)
\begin{equation}
\mathcal{L}_{\mathrm{Res}}^{\mathrm{Kin}}=-\frac{1}{2}\left\langle \nabla_\lambda V^{\lambda\nu}\nabla^\rho V_{\rho\nu} \right\rangle+\frac{M_V^2}{4}\left\langle V_{\mu\nu}V^{\mu\nu} \right\rangle,
\end{equation}
with obvious replacements for axial-vector resonances, $A_{\mu\nu}$.

$U(3)$ symmetry in the (axial-)vector resonance masses is broken by operators of the form $\left\langle V_{\mu\nu}V^{\mu\nu}\chi_+\right\rangle$, $V\leftrightarrow A$ \cite{Cirigliano:2003yq,Guo:2009hi,Roig:2019reh}, which can accommodate the measured spectra (in this way we will simply replace the different spin-one resonance masses by the PDG values~\cite{ParticleDataGroup:2022pth} in the following). Analogously, breaking of this flavor symmetry also affects $F_{V,A}$. However, in the vector case, the coupling giving this shift ~\footnote{It is $\lambda_6^V$ in the notation of ref.~\cite{Cirigliano:2006hb} and $\lambda_V$ in ref.~\cite{Guevara:2018rhj}, with $\lambda_6^V=\frac{\lambda_V}{\sqrt{2}}$.}\cite{Cirigliano:2006hb} is constrained by short-distance QCD to vanish within the scheme considered in ref.~\cite{Guevara:2018rhj} (see also ref.~\cite{Roig:2014uja}, where these contributions were neglected \textit{a priori}). Based on this, flavor symmetry on $F_{V,A}$ is assumed in the remainder of the paper.

All resonance contributions to the $v_i$ form factors (those to the $a_i$ are differed by one chiral order and thus neglected) depend on ratios of resonance couplings over the meson decay constant $\left(\frac{F_VG_V}{f^2},\frac{F_V^2}{f^2},\frac{F_A^2}{f^2}\right)$, which are constrained by asymptotic QCD. Therefore the well-known $SU(3)$ symmetry breaking which causes $f_K\sim1.2f_\pi$ \cite{ParticleDataGroup:2022pth} cannot be accounted for, within the considered simplified scheme, without conflicting with short-distance QCD requirements. Thus, we are using $f\sim90$ MeV for the $v_i$ also in the strangeness-changing channels~\footnote{The associated error is, however, much smaller than the uncertainty that we will find in our results, see section \ref{sec:RadCors}.}. The dispersive constructions giving the $F_{+,0}$ form factors account for flavor symmetry breaking in the $K\pi$ channels, as required by the precision of the corresponding measurements, which fed the phaseshifts entering the dispersive integrals.

\subsubsection{Vector contributions}
\label{VFF:KK}
Including those Lagrangian terms that, upon resonance integration, 
contribute to the ChPT $\mathcal{O}(p^4)$ low-energy constants (LECs), 
we have found the following contributions to the vector form factors $v_i$ in Eq.~(\ref{eq1:VSD}),
which are depicted in Fig.~\ref{An:fig3}:

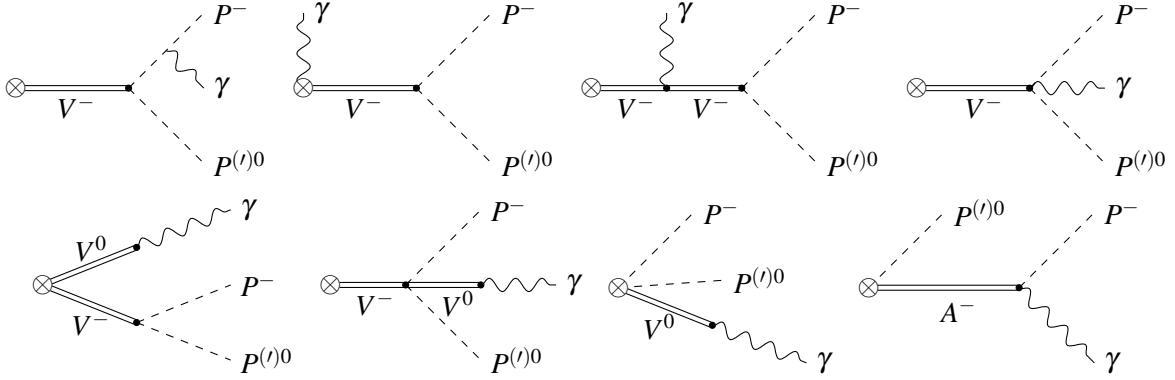
\begin{figure}
\begin{center}
\begin{tikzpicture}
	\draw[resonance] (-1.5,0)--(0,0);
	\draw[scalar] (0,0) -- (1,1) node[right] {\small$P^-$};
	\draw[scalar] (0,0) -- (1,-1) node[right] {\small$P^{(\prime)0}$};
	\draw[fill] (0,0) circle [radius=0.04];
	\draw[fill,white] (-1.5,0) circle [radius=0.12];
	\draw[gauge] (0.5,0.5) -- (1,0) node[right] {\small$\gamma$};
	\node at (-1.5,0) {\small$\otimes$};
	\node at (-0.69,-0.25) {\small$V^{-}$};
\end{tikzpicture}
\begin{tikzpicture}
	\draw[resonance] (-1.5,0)--(0,0);
	\draw[gauge] (-1.5,0) -- (-1.5,1) node[right] {\small$\gamma$};
	\draw[scalar] (0,0) -- (1,1) node[right] {\small$P^-$};
	\draw[scalar] (0,0) -- (1,-1) node[right] {\small$P^{(\prime)0}$};
	\draw[fill] (0,0) circle [radius=0.04];
	\draw[fill,white] (-1.5,0) circle [radius=0.12];
	\node at (-1.5,0) {\small$\otimes$};
	\node at (-0.69,-0.25) {\small$V^{-}$};
\end{tikzpicture}
\begin{tikzpicture}
	\draw[resonance] (3,0)--(5,0);
	\draw[fill,white] (3,0) circle [radius=0.12];
	\node at (3,0) {\small$\otimes$};
	\draw[gauge] (4,0) -- (4,1) node[right] {\small$\gamma$};
	\draw[fill] (4,0) circle [radius=0.04];
	\draw[scalar] (5,0) -- (6,1) node[right] {\small$P^-$};
	\draw[scalar] (5,0) -- (6,-1) node[right] {\small$P^{(\prime)0}$};
	\draw[fill] (5,0) circle [radius=0.04];
	\node at (3.6,-0.25) {\small$V^{-}$};
	\node at (4.6,-0.25) {\small$V^{-}$};
\end{tikzpicture}
\begin{tikzpicture}
	\draw[resonance] (7.5,0)--(9,0);
	\draw[fill,white] (7.5,0) circle [radius=0.12];
	\node at (7.5,0) {\small$\otimes$};
	\draw[gauge] (9,0) -- (10,0) node[right] {\small$\gamma$};
	\draw[scalar] (9,0) -- (10,1) node[right] {\small$P^-$};
	\draw[scalar] (9,0) -- (10,-1) node[right] {\small$P^{(\prime)0}$};
	\draw[fill] (9,0) circle [radius=0.04];
	\node at (8.4,-0.25) {\small$V^{-}$};
\end{tikzpicture}	
\begin{tikzpicture}
	\draw[resonance] (-1.5,-3)--(-0.25,-2.5);
	\draw[resonance] (-1.5,-3)--(-0.25,-3.5);
	\draw[fill,white] (-1.5,-3) circle [radius=0.12];
	\node at (-1.5,-3) {\small$\otimes$};
	\draw[gauge] (-0.25,-2.5)--(1,-2) node[right] {\small$\gamma$};
	\draw[fill] (-0.25,-2.5,0) circle [radius=0.04];
	\draw[scalar] (-0.25,-3.5)--(1,-4) node[right] {\small$P^{(\prime)0}$};
	\draw[scalar] (-0.25,-3.5)--(1,-3) node[right] {\small$P^-$};
	\draw[fill] (-0.25,-3.5) circle [radius=0.04];
	\node at (-0.85,-3.52) {\small$V^{-}$};
	\node at (-0.85,-2.52) {\small$V^0$};
\end{tikzpicture}
\begin{tikzpicture}
	\draw[resonance] (3,-3)--(5,-3);
	\draw[fill,white] (3,-3) circle [radius=0.12];
	\node at (3,-3) {\small$\otimes$};
	\draw[gauge] (5,-3) -- (6,-3) node[right] {\small$\gamma$};
	\draw[fill] (5,-3) circle [radius=0.04];
	\draw[scalar] (4,-3) -- (5,-2) node[right] {\small$P^-$};
	\draw[scalar] (4,-3) -- (5,-4) node[right] {\small$P^{(\prime)0}$};
	\draw[fill] (4,-3) circle [radius=0.04];
	\node at (3.6,-3.25) {\small$V^{-}$};
	\node at (4.7,-3.25) {\small$V^0$};
\end{tikzpicture}
\begin{tikzpicture}
	\draw[scalar] (7.5,-3)--(8.5,-2) node[right] {\small$P^-$};
	\draw[scalar] (7.5,-3)--(8.9,-2.9) node[right] {\small$P^{(\prime)0}$};
	\draw[resonance] (7.5,-3)--(8.75,-3.5);
	\draw[fill,white] (7.5,-3) circle [radius=0.12];
	\node at (7.5,-3)  {\small$\otimes$};
	\draw[gauge] (8.75,-3.5)--(10,-4) node[right] {\small$\gamma$};
	\draw[fill] (8.75,-3.5) circle [radius=0.04];
	\node at (8.05,-3.52) {\small$V^0$};
\end{tikzpicture}
\begin{tikzpicture}
	\draw[resonance] (3,-6)--(5,-6);
	\node at (4.2,-6.3) {\small$A^-$};
	\draw[scalar] (3,-6) -- (4,-5) node[right] {\small$P^{(\prime)0}$};
	\draw[fill,white] (3,-6) circle [radius=0.12];
	\node at (3,-6) {\small$\otimes$};
	\draw[scalar] (5,-6) -- (6,-5) node[right] {\small$P^-$};
	\draw[gauge] (5,-6) -- (6,-7) node[right] {\small$\gamma$};
	\draw[fill] (5,-6) circle [radius=0.04];
\end{tikzpicture}
\end{center}
\caption{Vector and axial-vector meson exchange diagrams contributing to the 
$\tau^-\to P_1^- P_2^0\nu_\tau\gamma$ decays at $\mathcal{O}\left(p^4\right)$. 
$V^0$ stands for the $\rho^0$, $\omega$ and $\phi$ resonances, 
$V^-=K^{*-}$ for the $K\pi$ modes and $V^-=\rho^{-}$ for the $K^-K^0$ one, 
and $A^{-}=K_1^{-}$ in $K^-K^0$ and $K^-\pi^0$, and $A^{-}=a_1^{-}$ in $\pi^-\bar{K}^0$.}
\label{An:fig3}
\end{figure}

\begin{subequations}
\allowdisplaybreaks
\begin{align}
v_1=&\ \frac{F_V G_V}{\sqrt{2}f^2 M_\rho^2}\left\{
\left(1+\frac{1}{3}\frac{M_\rho^2}{M_\omega^2}+\frac{2}{3}\frac{M_\rho^2}{M_\phi^2}\right)
\left[1+\frac{1}{2}\left(t-\Delta_{K\pi}\right)D_{K^{*}}^{-1}(t)\right]\right.\nonumber\\[1ex]
&+2M_\rho^2D_{K^{*}}^{-1}(t^\prime)
+M_\rho^2\left(t-\Delta_{K\pi}\right)D_{K^{*}}^{-1}(t)D_{K^{*}}^{-1}(t^\prime)\left.\right\}
\nonumber\\[1ex]
&+\frac{F_V^2}{2\sqrt{2}f^2M_\rho^2}\left[
-\frac{1}{2}\left(1+\frac{1}{3}\frac{M_\rho^2}{M_\omega^2}
+\frac{2}{3}\frac{M_\rho^2}{M_\phi^2}\right)
\left(1-t^\prime\,D_{K^{*}}^{-1}(t^\prime)\right)-M_\rho^2D_{K^{*}}^{-1}(t^\prime)\right]
\nonumber\\[1ex]
&+\frac{F_A^2}{\sqrt{2}f^2 M_{K_1}^2}\left(M_{K_1}^2
-\frac{1}{2}\Sigma_{K\pi}+\frac{1}{2}t\right)D_{K_1}^{-1}[(p_K+k)^2]\ ,\\[1ex]
v_2=&\ \frac{F_V G_V}{\sqrt{2}f^2 M_\rho^2}\left(t+\Delta_{K\pi}\right)\left[
-\frac{1}{2}\left(1+\frac{1}{3}\frac{M_\rho^2}{M_\omega^2}
+\frac{2}{3}\frac{M_\rho^2}{M_\phi^2}\right)D_{K^{*}}^{-1}(t)
-M_\rho^2D_{K^{*}}^{-1}(t)D_{K^{*}}^{-1}(t^\prime)\right]\nonumber\\[1ex]
&+\frac{F_V^2}{2\sqrt{2}f^2M_\rho^2}\left[ 
-\frac{1}{2}\left(1+\frac{1}{3}\frac{M_\rho^2}{M_\omega^2}
+\frac{2}{3}\frac{M_\rho^2}{M_\phi^2}\right)\left(1+t^\prime\,D_{K^{*}}^{-1}(t^\prime)\right)
-M_\rho^2D_{K^{*}}^{-1}(t^\prime)\right]\nonumber\\[1ex]
&+\frac{F_A^2}{\sqrt{2}f^2 M_{K_1}^2}\left(M_{K_1}^2-m_K^2-k\cdot p_K\right)
D_{K_1}^{-1}[(p_K+k)^2]\ ,\\[1ex]
v_3=&\ \frac{F_A^2}{\sqrt{2}f^2 M_{K_1}^2} D_{K_1}^{-1}[(p_K+k)^2]\ ,\\[1ex]
v_4=&\ -\frac{2F_V G_V}{\sqrt{2}f^2}D_{K^{*}}^{-1}(t)D_{K^{*}}^{-1}(t^\prime)
+\frac{F_V^2}{2\sqrt{2}f^2M_\rho^2}\left(1+\frac{1}{3}\frac{M_\rho^2}{M_\omega^2}
+\frac{2}{3}\frac{M_\rho^2}{M_\phi^2}\right)D_{K^{*}}^{-1}(t^\prime)\ ,
\end{align}
\end{subequations}
for $K^-\pi^0$,
\begin{subequations}
\allowdisplaybreaks
\begin{align}
v_1=&-\frac{F_V G_V}{f^2 M_\rho^2}\left[2+2M_\rho^2D_{K^{*}}^{-1}(t^\prime)+\frac{1}{2}
\left(1+\frac{1}{3}\frac{M_\rho^2}{M_\omega^2}+\frac{2}{3}\frac{M_\rho^2}{M_\phi^2}\right)
\left(t+\Delta_{K\pi}\right)D_{K^{*}}^{-1}(t)\right.\nonumber\\[1ex]
&+\left(t+\Delta_{K\pi}\right)M_\rho^2D_{K^{*}}^{-1}(t)D_{K^{*}}^{-1}(t^\prime)\left.\right]
\nonumber\\[1ex]
&-\frac{F_V^2}{2f^2M_\rho^2}\left[-M_\rho^2D_{K^{*}}^{-1}(t^\prime)+\frac{1}{2}
\left(1+\frac{1}{3}\frac{M_\rho^2}{M_\omega^2}+\frac{2}{3}\frac{M_\rho^2}{M_\phi^2}\right)
t^\prime\,D_{K^{*}}^{-1}(t^\prime)
+\frac{1}{2}\left(-3+\frac{1}{3}\frac{M_\rho^2}{M_\omega^2}
+\frac{2}{3}\frac{M_\rho^2}{M_\phi^2}\right)\right]
\nonumber\\[1ex]
&-\frac{F_A^2}{f^2 M_{a_1}^2}\left(M_{a_1}^2
-\frac{1}{2}\Sigma_{K\pi}+\frac{1}{2}t\right)D_{a_1}^{-1}[(p_\pi+k)^2]\ ,\\[1ex]
v_2=&-\frac{F_V G_V}{f^2 M_\rho^2}\left\{\left(t-\Delta_{K\pi}\right)
\left[-M_\rho^2D_{K^{*}}^{-1}(t) D_{K^{*}}^{-1}(t^\prime)-\frac{1}{2}
\left(1+\frac{1}{3}\frac{M_\rho^2}{M_\omega^2}+\frac{2}{3}\frac{M_\rho^2}{M_\phi^2}\right)
D_{K^{*}}^{-1}(t)\right]\right.\nonumber\\[1ex]
&+1-\frac{1}{3}\frac{M_\rho^2}{M_\omega^2}-\frac{2}{3}\frac{M_\rho^2}{M_\phi^2}\left.\right\}
\nonumber\\[1ex]
&-\frac{F_V^2}{2f^2M_\rho^2}\left[-M_\rho^2D_{K^{*}}^{-1}(t^\prime)-\frac{1}{2}
\left(1+\frac{1}{3}\frac{M_\rho^2}{M_\omega^2}+\frac{2}{3}\frac{M_\rho^2}{M_\phi^2}\right)
t^\prime\,D_{K^{*}}^{-1}(t^\prime)
+\frac{1}{2}\left(-3+\frac{1}{3}\frac{M_\rho^2}{M_\omega^2}
+\frac{2}{3}\frac{M_\rho^2}{M_\phi^2}\right)\right]
\nonumber\\[1ex]
&-\frac{F_A^2}{f^2 M_{a_1}^2}
\left(M_{a_1}^2-m_\pi^2 k\cdot p_\pi\right)D_{a_1}^{-1}[(p_\pi+k)^2]\ ,\\[1ex]
v_3&=-\frac{F_A^2}{f^2 M_{a_1}^2} D_{a_1}^{-1}[(p_\pi+k)^2]\ ,\\[1ex]
v_4&=\frac{2F_V G_V}{f^2}D_{K^{*}}^{-1}(t)D_{K^{*}}^{-1}(t^\prime)
-\frac{F_V^2}{2 f^2M_\rho^2}\left(1+\frac{1}{3}\frac{M_\rho^2}{M_\omega^2}
+\frac{2}{3}\frac{M_\rho^2}{M_\phi^2}\right)D_{K^{*}}^{-1}(t^\prime)\ ,
\end{align}
\end{subequations}
for $\bar{K}^0\pi^-$, and
\begin{subequations}
\allowdisplaybreaks
\begin{align}
v_1=&-\frac{F_V G_V}{f^2 M_\rho^2}\left[
1+\frac{1}{3}\frac{M_\rho^2}{M_\omega^2}+\frac{2}{3}\frac{M_\rho^2}{M_\phi^2}
+\left(t-\Delta_{K^{-}K^{0}}\right)D_{\rho}^{-1}(t)+2M_\rho^2D_{\rho}^{-1}(t^\prime)\right.
\nonumber\\[1ex]
&+M_\rho^2\left(t-\Delta_{K^{-}K^{0}}\right)D_{\rho}^{-1}(t)D_{\rho}^{-1}(t^\prime)
\left.\right]
\nonumber\\[1ex]
&-\frac{F_V^2}{2f^2M_\rho^2}\left[
-\frac{1}{3}\frac{M_\rho^2}{M_\omega^2}-\frac{2}{3}\frac{M_\rho^2}{M_\phi^2}
+t^\prime\,D_{\rho}^{-1}(t^\prime)-M_\rho^2D_{\rho}^{-1}(t^\prime)\right]\nonumber\\[1ex]
&-\frac{F_A^2}{f^2 M_{K_1}^2}\left(M_{K_1}^2-\frac{1}{2}\Sigma_{K^{-}K^{0}}
+\frac{1}{2}t\right)D_{K_1}^{-1}[(p_{-}+k)^2]\ ,\\
v_2=&-\frac{F_V G_V}{f^2 M_\rho^2}\left[
-1+\frac{1}{3}\frac{M_\rho^2}{M_\omega^2}+\frac{2}{3}\frac{M_\rho^2}{M_\phi^2}
-\left(t+\Delta_{K^{-}K^{0}}\right)D_{\rho}^{-1}(t)-M_\rho^2
\left(t+\Delta_{K^{-}K^{0}}\right)D_{\rho}^{-1}(t)D_{\rho}^{-1}(t^\prime)\right]
\nonumber\\[1ex]
&-\frac{F_V^2}{2f^2M_\rho^2}\left[
-\frac{1}{3}\frac{M_\rho^2}{M_\omega^2}-\frac{2}{3}\frac{M_\rho^2}{M_\phi^2}
-t^\prime\,D_{\rho}^{-1}(t^\prime)-M_\rho^2D_{\rho}^{-1}(t^\prime)\right]\nonumber\\[1ex]
&-\frac{F_A^2}{f^2 M_{K_1}^2}\left(M_{K_1}^2-m_{K^{-}}^2
-k\cdot p_{-}\right)D_{K_1}^{-1}[(p_{-}+k)^2]\ ,\\[1ex]
v_3=&-\frac{F_A^2}{f^2 M_{K_1}^2} D_{K_1}^{-1}[(p_{-}+k)^2]\ ,\\[1ex]
v_4=&\ \frac{2F_V G_V}{f^2}D_{\rho}^{-1}(t)D_{\rho}^{-1}(t^\prime)
-\frac{F_V^2}{f^2M_\rho^2}D_{\rho}^{-1}(t^\prime)\ ,
\end{align}
\end{subequations}
for $K^{-}K^0$, where $\Sigma_{-0}=m_{-}^2+m_{0}^2$ and $D_R(x)=M_R^2-x-iM\Gamma_R(x)$. 
Off-shell resonance widths\footnote{The 
on-shell width corresponds to the imaginary part of the pole position of the resonance.
The imaginary part of the corresponding loop function provides an off-shell width function, 
which extends off the pole \cite{GomezDumm:2000fz}.}
are given in terms of the leading pseudo-Goldstone boson 
cuts~\cite{GomezDumm:2000fz, Dumm:2009va, Dumm:2009kj}.

It is straightforward to show that, except for a Clebsch-Gordan coefficient (CGC) factor, 
one recovers the expressions found in Refs.~\cite{Cirigliano:2002pv,Miranda:2020wdg} 
for the vector form factors of the $\tau^-\to\pi^-\pi^0\nu_\tau\gamma$ decays 
in the isospin-symmetry limit.

All the former resonance contributions are given in terms of three couplings: 
$F_V$, responsible for instance of the coupling of the vector resonance to the vector current; 
$F_A$ for the couplings of the axial resonance; and 
$G_V$ which yields, among others, vertices between the vector resonance and a couple of 
pseudo-Goldstone bosons (see e.g.~Ref.~\cite{Ecker:1988te} for further details).

\subsubsection{Axial contributions}
The Feynman diagrams that contribute to these decays are depicted in 
Figs.~\ref{An:fig03}--\ref{An:fig32KK}. 
At $\mathcal{O}(p^4)$, the axial form factors $a_i$ in Eq.~(\ref{eq1:ASD1}), 
which receive contributions from the Wess-Zumino-Witten functional
\cite{Wess:1971yu,Witten:1983tw}, are given by
\begin{equation}
a_1=\frac{N_c}{12\sqrt{2}\pi^2 f^2}\ ,\quad 
a_2=-\frac{N_c}{6\sqrt{2}\pi^2 f^2(t^\prime-m_K^2)}\ ,\quad 
a_3=-\frac{N_c}{24\sqrt{2}\pi^2f^2}\ ,
\end{equation}
for $K^-\pi^0$, 
\begin{equation}
a_3=-\frac{N_c}{24\pi^2 f^2}\ ,
\end{equation}
for $\bar{K}^0\pi^-$, and 
\begin{equation}
a_3=\frac{N_c}{24\pi^2 f^2}\ ,
\end{equation}
for $K^-K^0$, 
where $N_c=3$ is the number of colours and $f$ is the pion decay constant in the chiral limit, 
$f\sim 90$ MeV.

\begin{figure}
\begin{center}
\begin{tikzpicture}
	\draw[gauge] (9,0) -- (10,0) node[right] {\small$\gamma$};
	\draw[scalar] (9,0) -- (10,1) node[right] {\small$K^-$};
	\draw[scalar] (9,0) -- (10,-1) node[right] {\small$\pi^0$};
	\draw[fill] (9,0) circle [radius=0.04];
	\draw[fill,white] (9,0) circle [radius=0.12];
	\node at (9,0) {\small$\otimes$};
\end{tikzpicture}
\begin{tikzpicture}
	\draw[scalar] (7.5,0)--(9,0);
	\draw[fill,white] (7.5,0) circle [radius=0.12];
	\node at (7.5,0) {\small$\otimes$};
	\draw[gauge] (9,0) -- (10,0) node[right] {\small$\gamma$};
	\draw[scalar] (9,0) -- (10,1) node[right] {\small$K^-$};
	\draw[scalar] (9,0) -- (10,-1) node[right] {\small$\pi^0$};
	\draw[fill] (9,0) circle [radius=0.04];
	\node at (8.4,-0.25) {\small$K^{-}$};
\end{tikzpicture}
\end{center}
\caption{Axial contributions to the $\tau^-\to K^-\pi^0\nu_\tau\gamma$ decays 
at $\mathcal{O}\left(p^4\right)$.}
\label{An:fig03}
\end{figure}

\begin{figure}
\begin{center}
\begin{tikzpicture}
	\draw[gauge] (9,0) -- (10,0) node[right] {\small$\gamma$};
	\draw[scalar] (9,0) -- (10,1) node[right] {\small$\bar{K}^0$};
	\draw[scalar] (9,0) -- (10,-1) node[right] {\small$\pi^-$};
	\draw[fill] (9,0) circle [radius=0.04];
	\draw[fill,white] (9,0) circle [radius=0.12];
	\node at (9,0) {\small$\otimes$};
\end{tikzpicture}
\end{center}
\caption{Axial contributions to the $\tau^-\to \bar{K}^0\pi^-\nu_\tau\gamma$ decays 
at $\mathcal{O}\left(p^4\right)$.}
\label{An:fig32}
\end{figure}

\begin{figure}
\begin{center}
\begin{tikzpicture}
	\draw[gauge] (9,0) -- (10,0) node[right] {\small$\gamma$};
	\draw[scalar] (9,0) -- (10,1) node[right] {\small$K^0$};
	\draw[scalar] (9,0) -- (10,-1) node[right] {\small$K^-$};
	\draw[fill] (9,0) circle [radius=0.04];
	\draw[fill,white] (9,0) circle [radius=0.12];
	\node at (9,0) {\small$\otimes$};
\end{tikzpicture}
\end{center}
\caption{Axial contributions to the $\tau^-\to K^-K^0\nu_\tau\gamma$ decays 
at $\mathcal{O}\left(p^4\right)$.}
\label{An:fig32KK}
\end{figure}
In Fig.~\ref{An:fig32KK}, only one diagram contributes to the 
$\tau^-\to K^-K^0\nu_\tau\gamma$ decays similarly to the 
$\tau^-\to \bar{K}^0\pi^-\nu_\tau\gamma$ decays.
This is because the $K^-\to\bar{K}^0\pi^-\gamma$ (or $\pi^-\to K^-K^0\gamma$) vertex 
is absent in the WZW Lagrangian\footnote{This 
feature was already studied for the $K_{\ell 3}$ decays in Ref.~\cite{Bijnens:1992en}, 
where the non-local kaon pole term is only present in $A^+_{\mu\nu}$ 
for $K^+\to\pi^0\ell^+\nu_\ell\gamma$ decays.}. 
We reproduce the known anomalous contributions \cite{Cirigliano:2002pv,Miranda:2020wdg} 
for the $\tau^-\to\pi^-\pi^0\nu_\tau\gamma$ case. 
We neglect resonance contributions in the anomalous sector, 
which start at $\mathcal{O}(p^6)$ in the chiral power counting~\cite{Kampf:2011ty}.

\section{Radiative hadronic tau decays}
\label{sec:RadTauDec}
The differential rate for the $\tau^-\to P_1^-P_2^0\nu_\tau\gamma$ decays 
in the $\tau$ rest frame is given by
\begin{equation}
\label{Four_body_decays}
    d\Gamma=\frac{(2\pi)^4}{4m_\tau}\sum_{\text{spin}}\overline{\vert\mathcal{M}\vert^2}d\Phi_4\ ,
\end{equation}
where $d\Phi_4$ is the corresponding  4-body phase space, given by
\begin{equation}\begin{split}
    d\Phi_4=&\delta^{(4)}(P-p_{-}-p_{0}-q^\prime-k)\frac{d^3p_{-}}{(2\pi)^3 2E_{-}}\frac{d^3p_{0}}{(2\pi)^3 2E_{0}}
    \frac{d^3 q^\prime}{(2\pi)^3 2E_\nu}\frac{d^3 k}{(2\pi)^3 2E_\gamma}\ ,
\end{split}\end{equation}
and $\overline{\vert\mathcal{M}\vert^2}$ is the unpolarized spin-averaged squared amplitude. 
Inasmuch as this amplitude is not IR finite, we follow the same procedure as in 
Refs.~\cite{Cirigliano:2002pv,Miranda:2020wdg} 
where a photon energy cut, $E_\gamma^{\text{cut}}$, was introduced to study the dynamics of the 
$\tau^-\to\pi^-\pi^0\nu_\tau\gamma$ decays.

In this analysis, we call ``complete bremsstrahlung'' or simply ``SI'' the amplitude with 
$v_{1,2,3,4}=a_{1,2,3,4}=0$. 
For the $\mathcal{O}(p^4)$ contributions, as in Ref.~\cite{Miranda:2020wdg}, 
we distinguish between using the set of short-distance constraints $F_V=\sqrt{2}F$, $G_V=F/\sqrt{2}$ 
\cite{Ecker:1989yg} and $F_A=F$; 
or $F_V=\sqrt{3}F$, $G_V=F/\sqrt{3}$ and $F_A=\sqrt{2}F$ 
\cite{Cirigliano:2004ue,Cirigliano:2006hb,Kampf:2011ty,Roig:2013baa}. 
The former corresponds to the constraints from $2$-point Green functions 
and the second to the values consistent up to 3-point Green functions, 
which include operators that contribute at $\mathcal{O}(p^6)$ 
(that we are not including in this work). 
The difference between both sets of constraints has been employed to estimate roughly 
the model-dependent error of this approach 
\cite{Roig:2019reh,Miranda:2020wdg,Arroyo-Urena:2021nil,Arroyo-Urena:2021dfe,Guevara:2021tpy}. 
In all our subsequent analyses, 
the $\mathcal{O}(p^4)$ results include the SI part and the structure dependent part 
(either with the $F_V=\sqrt{2}F$ or with the $F_V=\sqrt{3}F$ set of constraints).

\begin{figure}
\centering			
\includegraphics[width=0.65\textwidth]{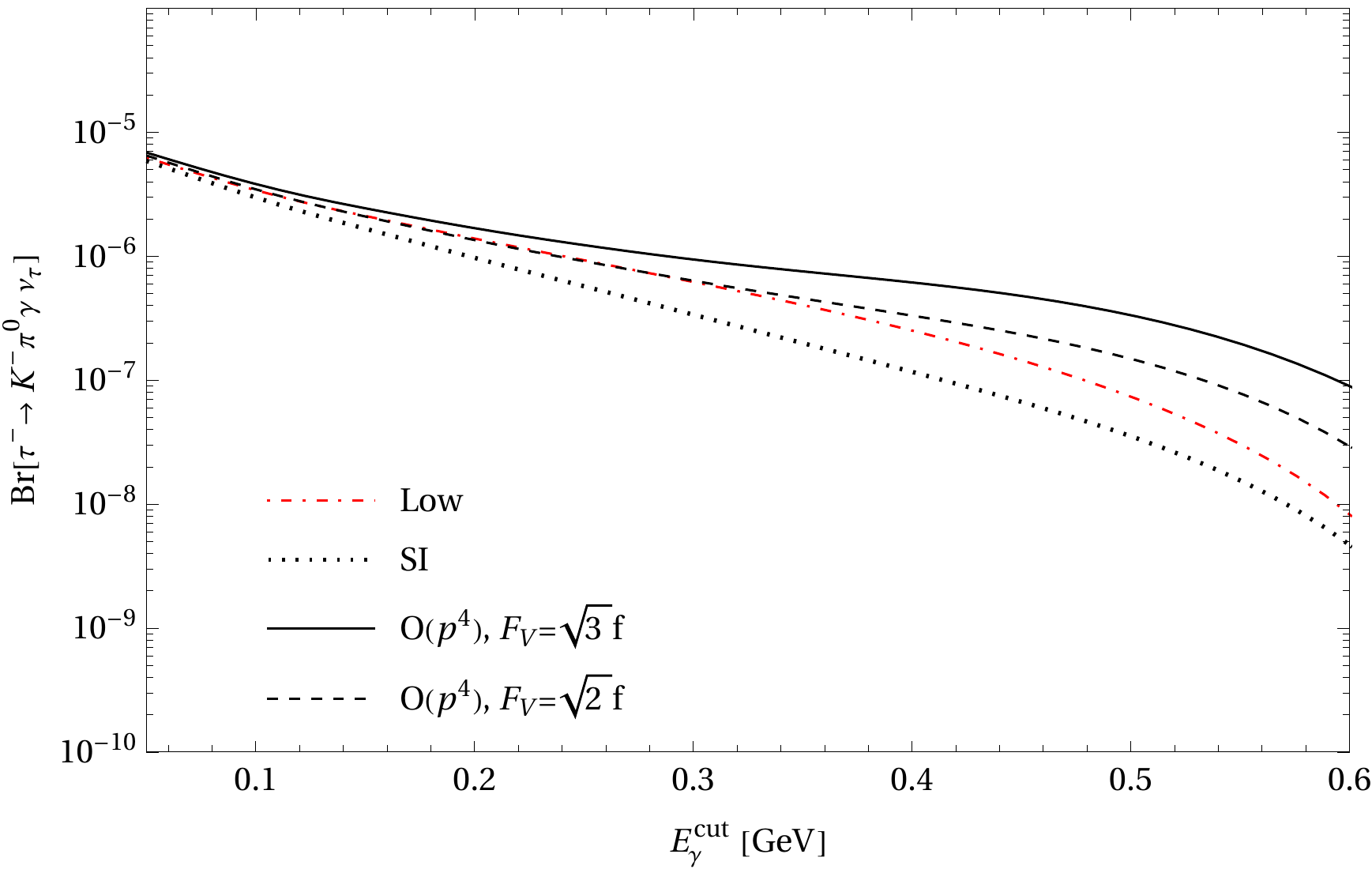}\\[1ex]
\includegraphics[width=0.65\textwidth]{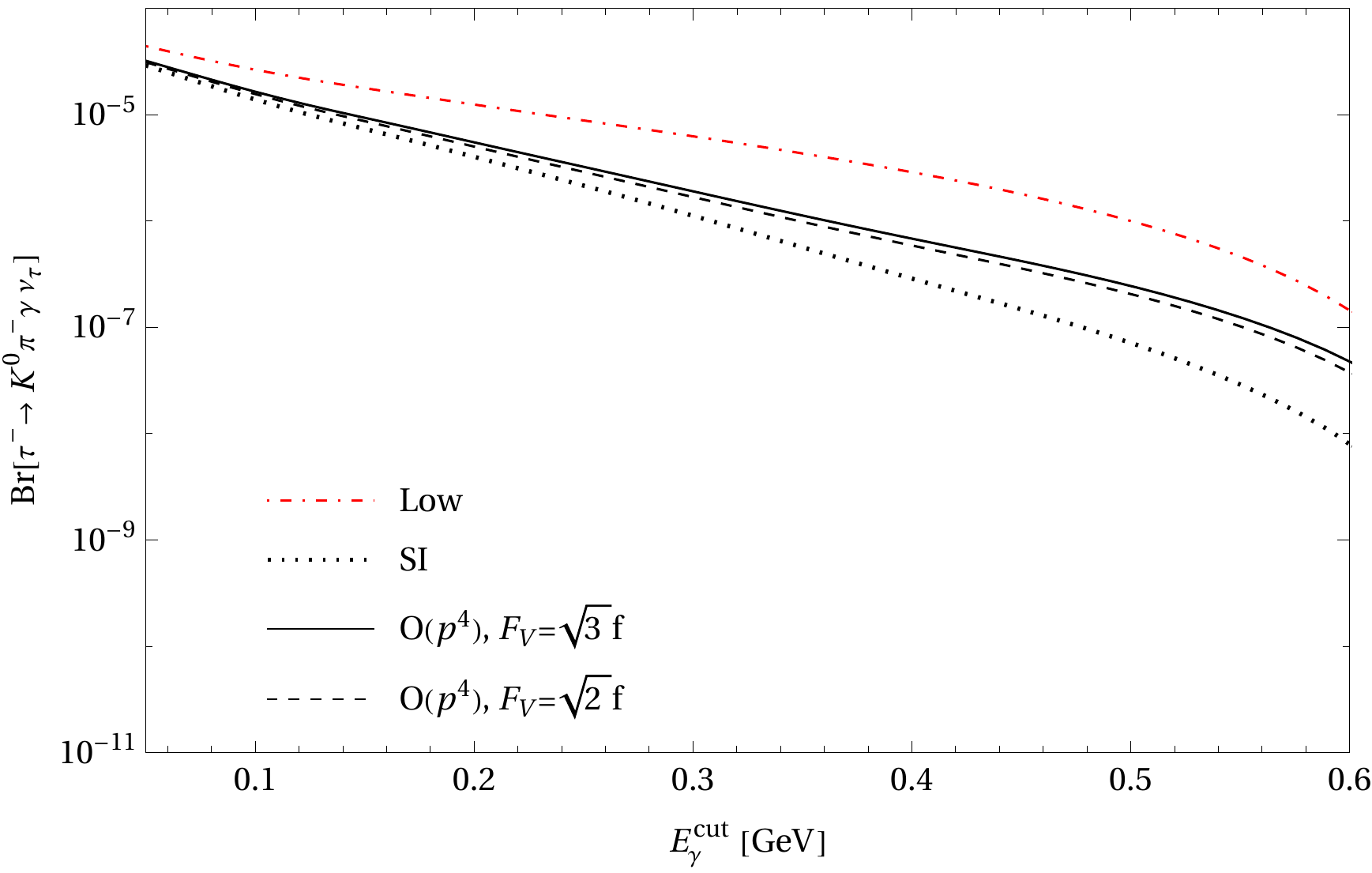}\\[1ex]
\includegraphics[width=0.65\textwidth]{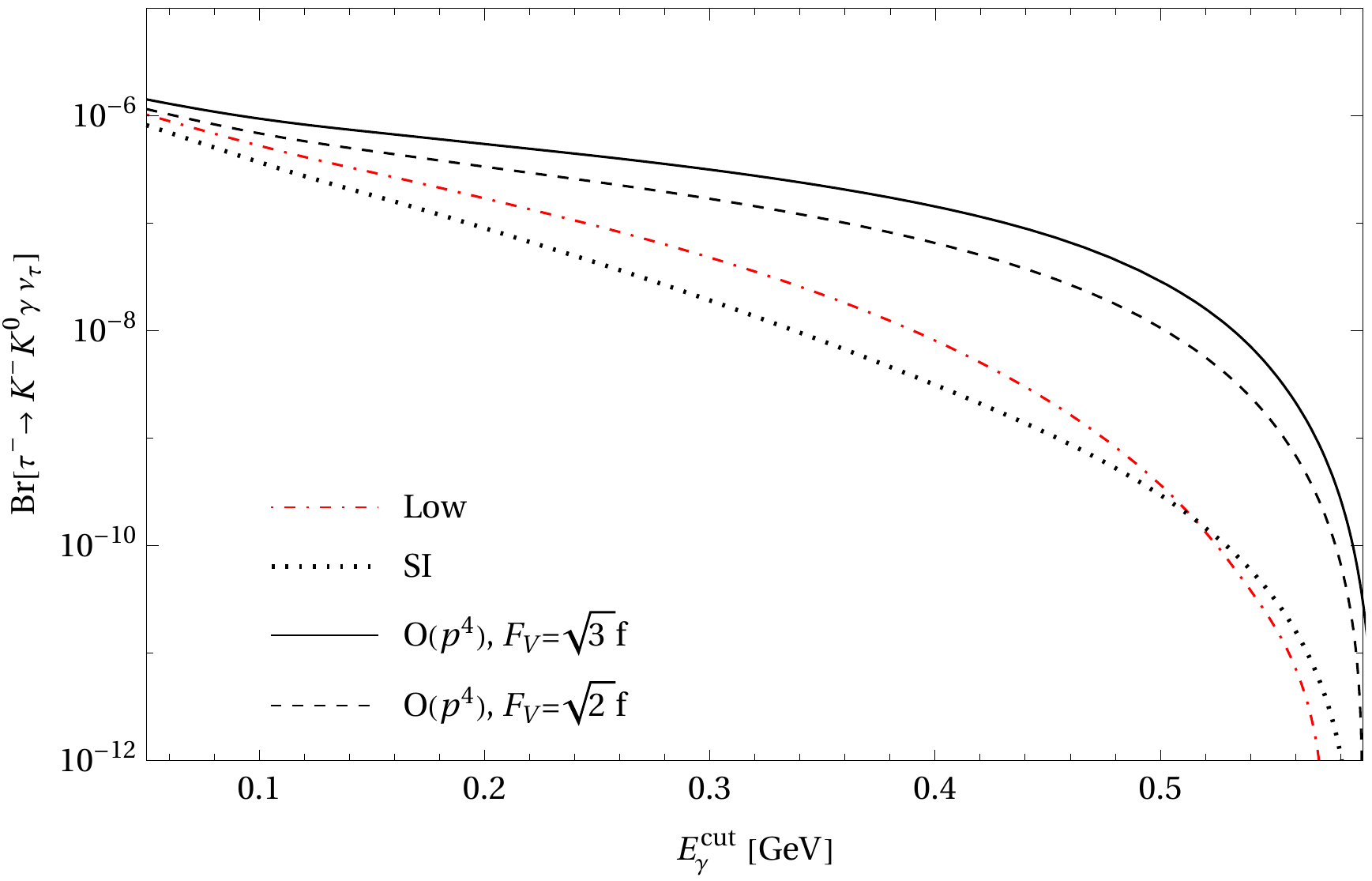}	
\caption{Branching ratio for the $\tau^-\to K^-\pi^0\nu_\tau\gamma$ (top), 
the $\tau^-\to \bar{K}^0\pi^{-}\nu_\tau\gamma$ (centre) and 
the $\tau^-\to K^-K^{0}\nu_\tau\gamma$ (bottom) decays as a function of $E^{\rm cut}_\gamma$.
The dotted line represents the bremsstrahlung contribution, 
the solid line and dashed line represent the $\mathcal{O}\left(p^4\right)$ corrections using 
$F_V=\sqrt{3}F$ and $F_V=\sqrt{2}F$, respectively. The red one corresponds to the Low approximation.}
\label{Appx6:fig6}
\end{figure}

\begin{figure}
\centering			
\includegraphics[width=0.65\textwidth]{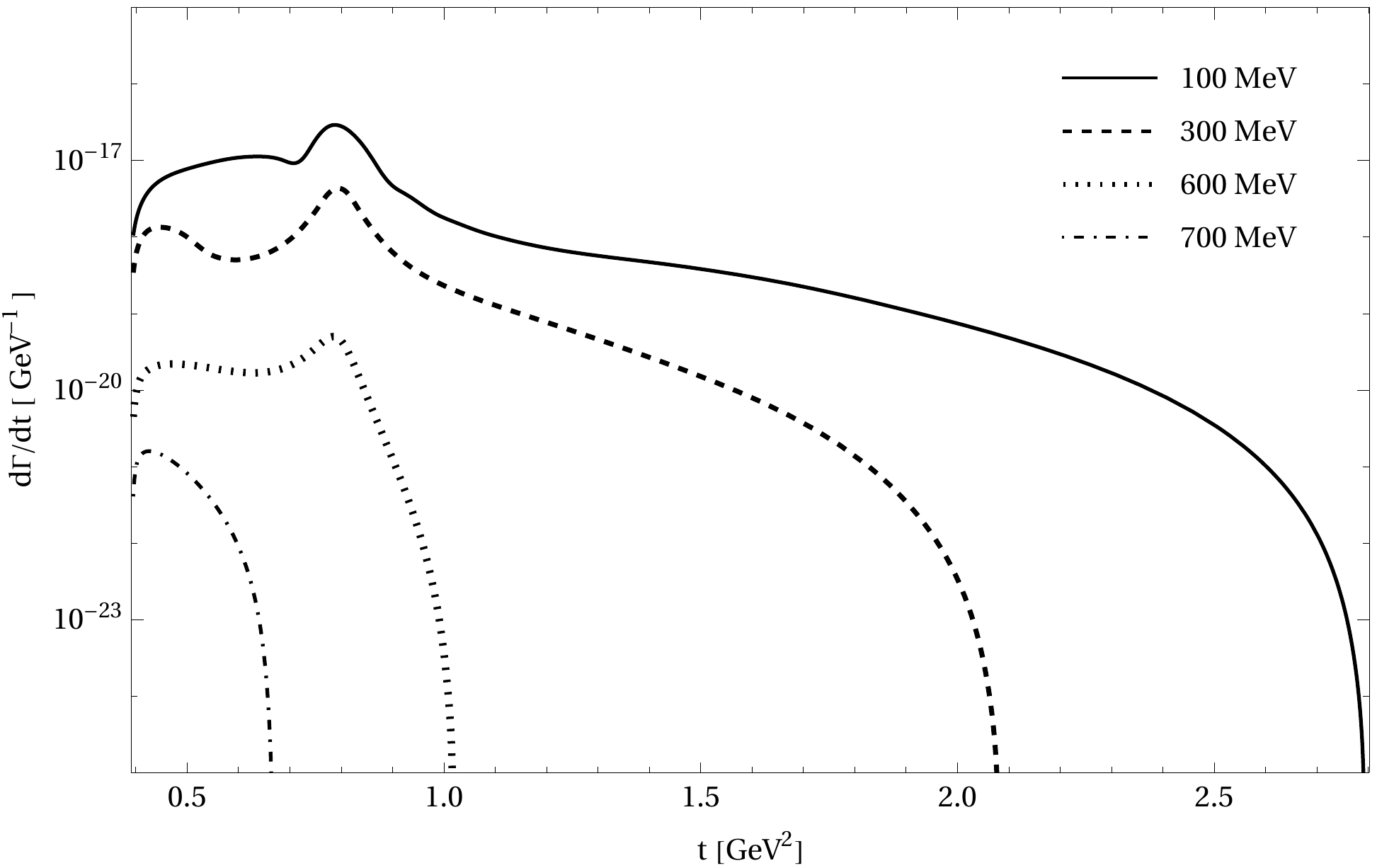}\\[1ex]
\includegraphics[width=0.65\textwidth]{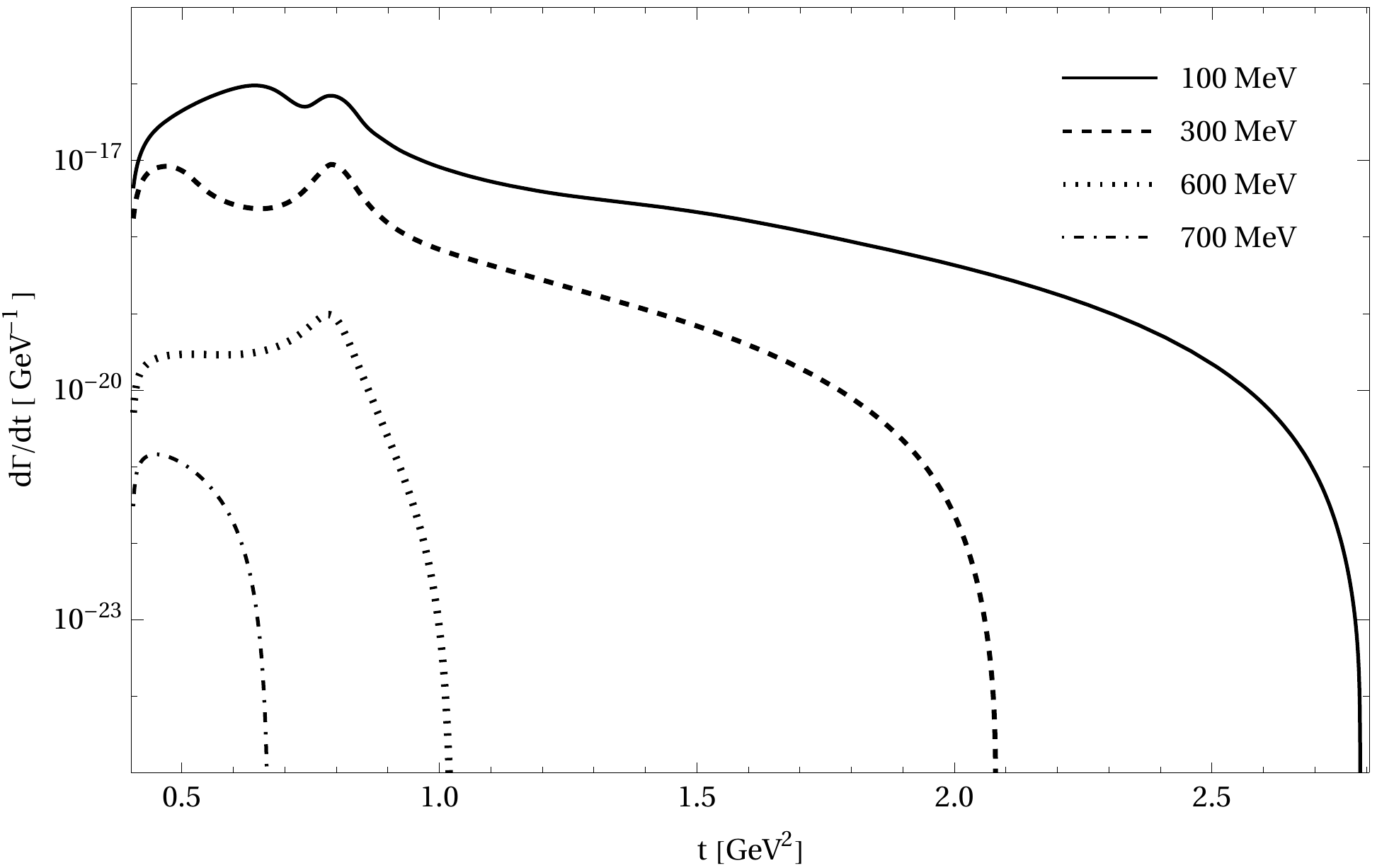}\\[1ex]	
\includegraphics[width=0.65\textwidth]{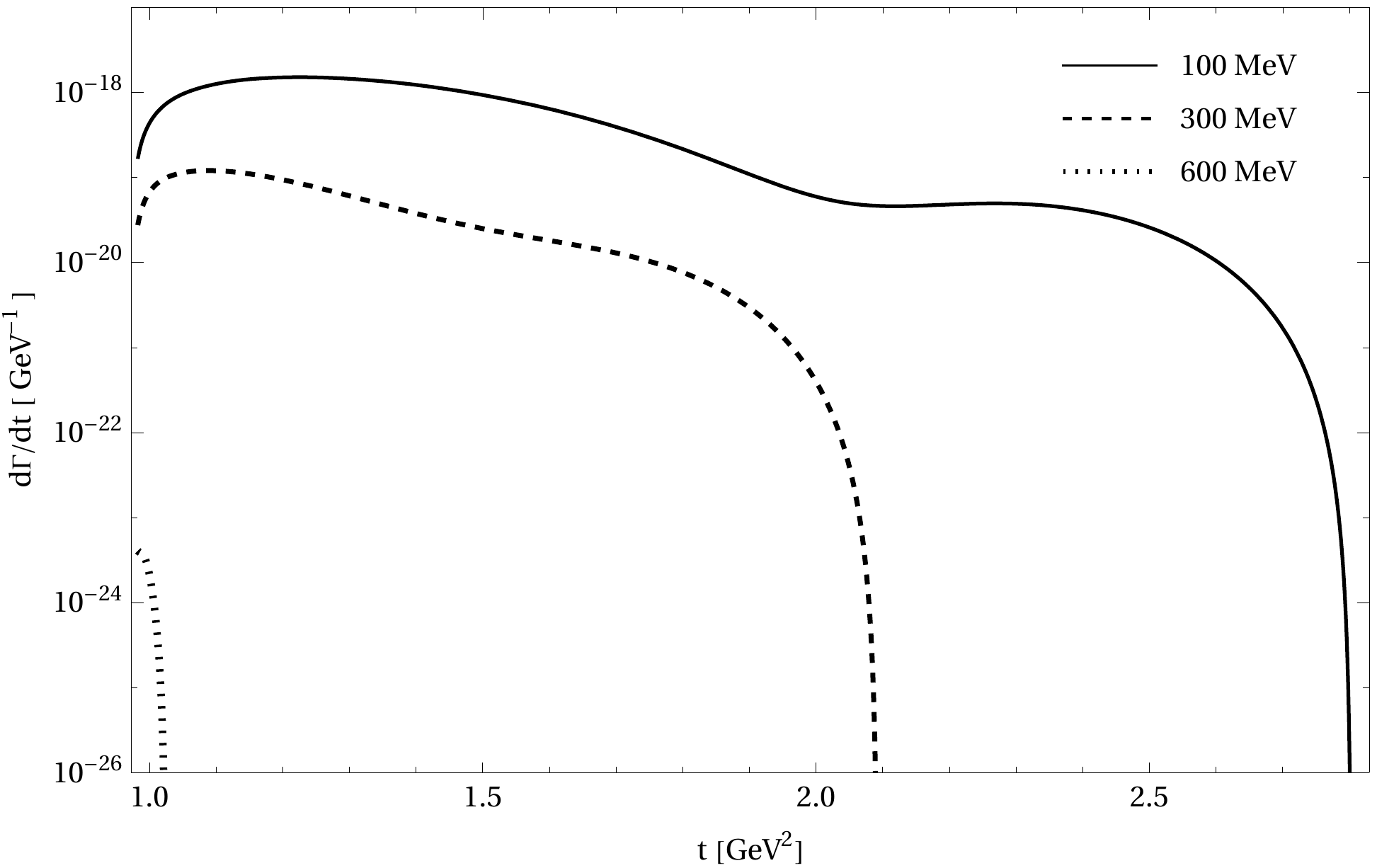}		
\caption{The $K^-\pi^0$ (top), $\bar{K}^0\pi^{-}$ (centre) and $K^-K^{0}$ (bottom) 
SI hadronic invariant mass distributions for several $E_\gamma^{\rm cut}$ values.}
\label{Appx06:fig5}
\end{figure}

\begin{figure}
\centering			
\includegraphics[width=0.6\textwidth]{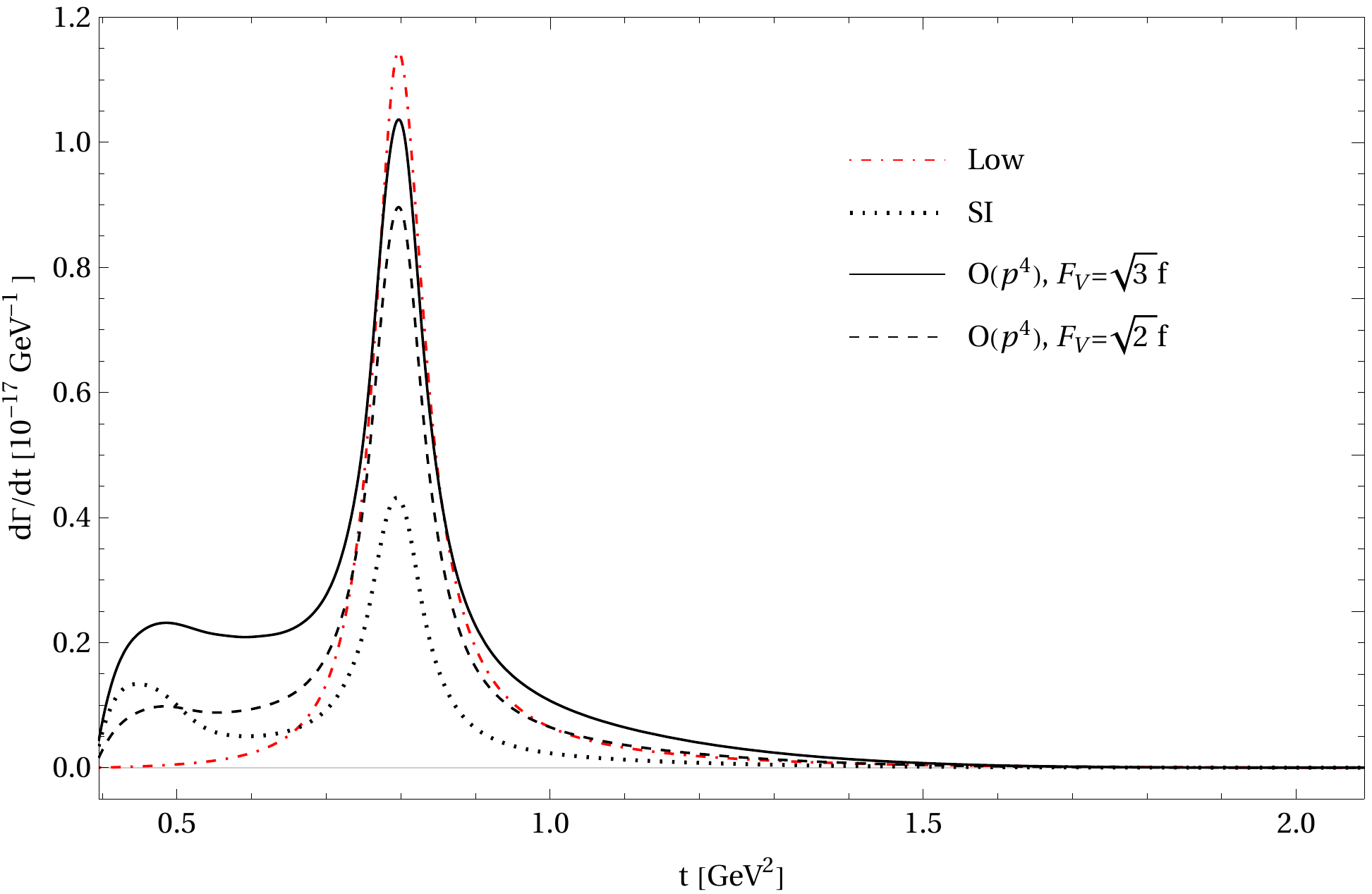}\\[1ex]
\includegraphics[width=0.6\textwidth]{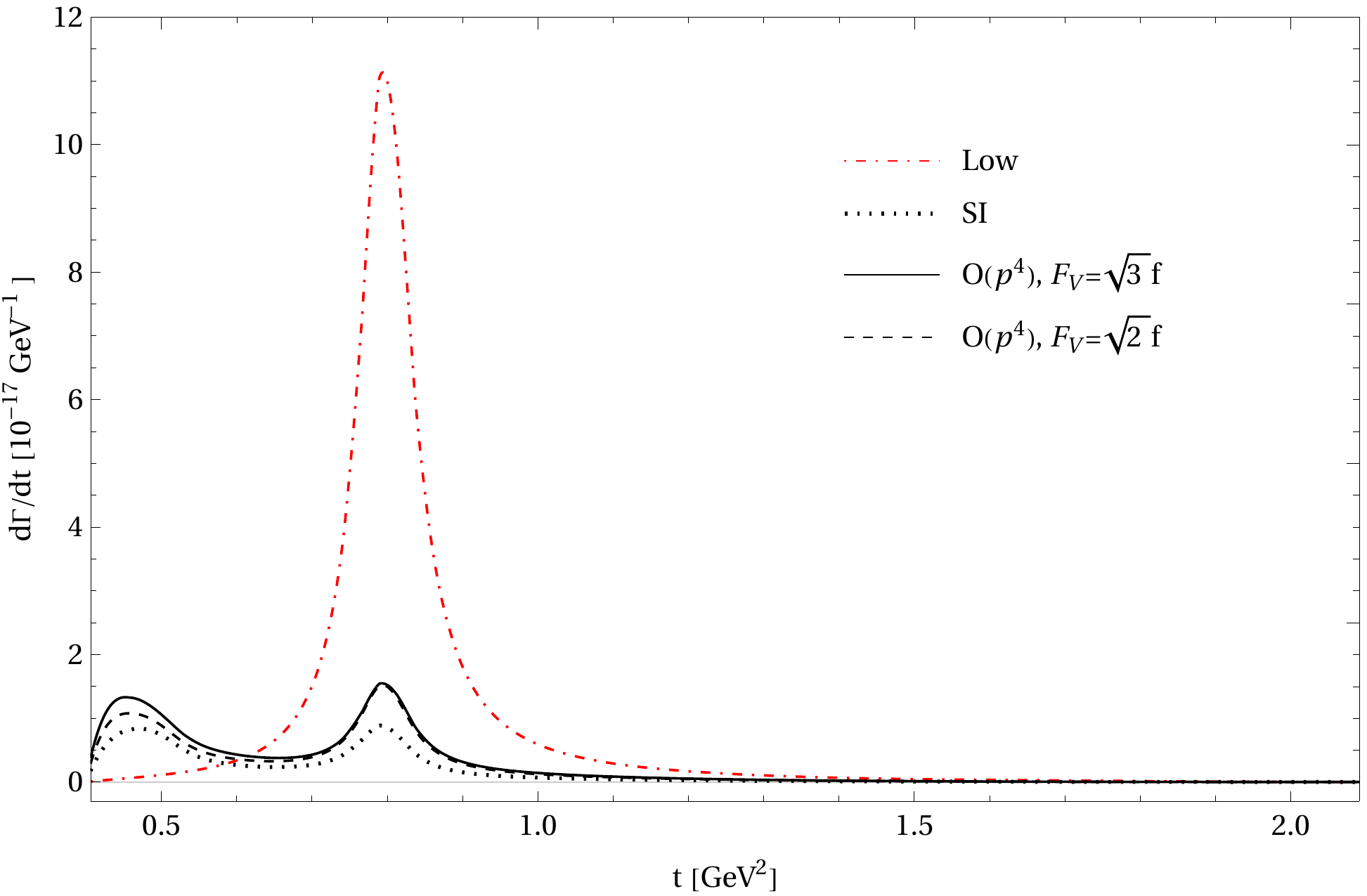}\\[1ex]
\includegraphics[width=0.6\textwidth]{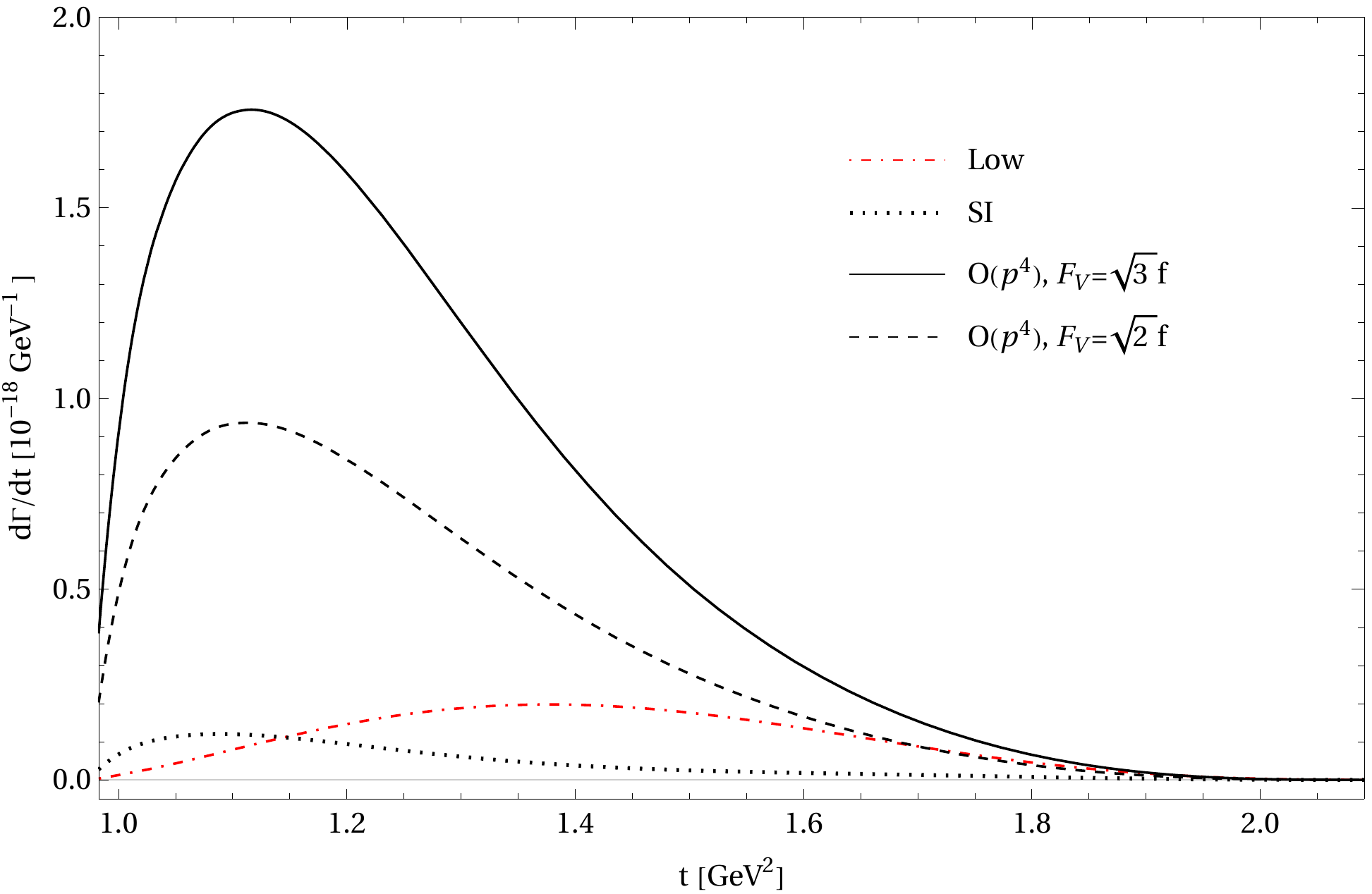}			
\caption{The $K^-\pi^0$ (top), $\bar{K}^0\pi^{-}$ (centre) and $K^-K^{0}$ (bottom) 
hadronic invariant mass distributions for $E_\gamma^{\rm cut}\geq 300\,\mathrm{MeV}$. 
The solid and dashed line represent the $\mathcal{O}\left(p^4\right)$ corrections using 
$F_V=\sqrt{3}F$ and $F_V=\sqrt{2}F$, respectively. 
The dotted line represents the bremsstrahlung contribution (SI). 
The red one corresponds to the Low approximation.}
\label{Appx6:fig5}
\end{figure}

\begin{figure}
\centering			
\includegraphics[width=0.60\textwidth]{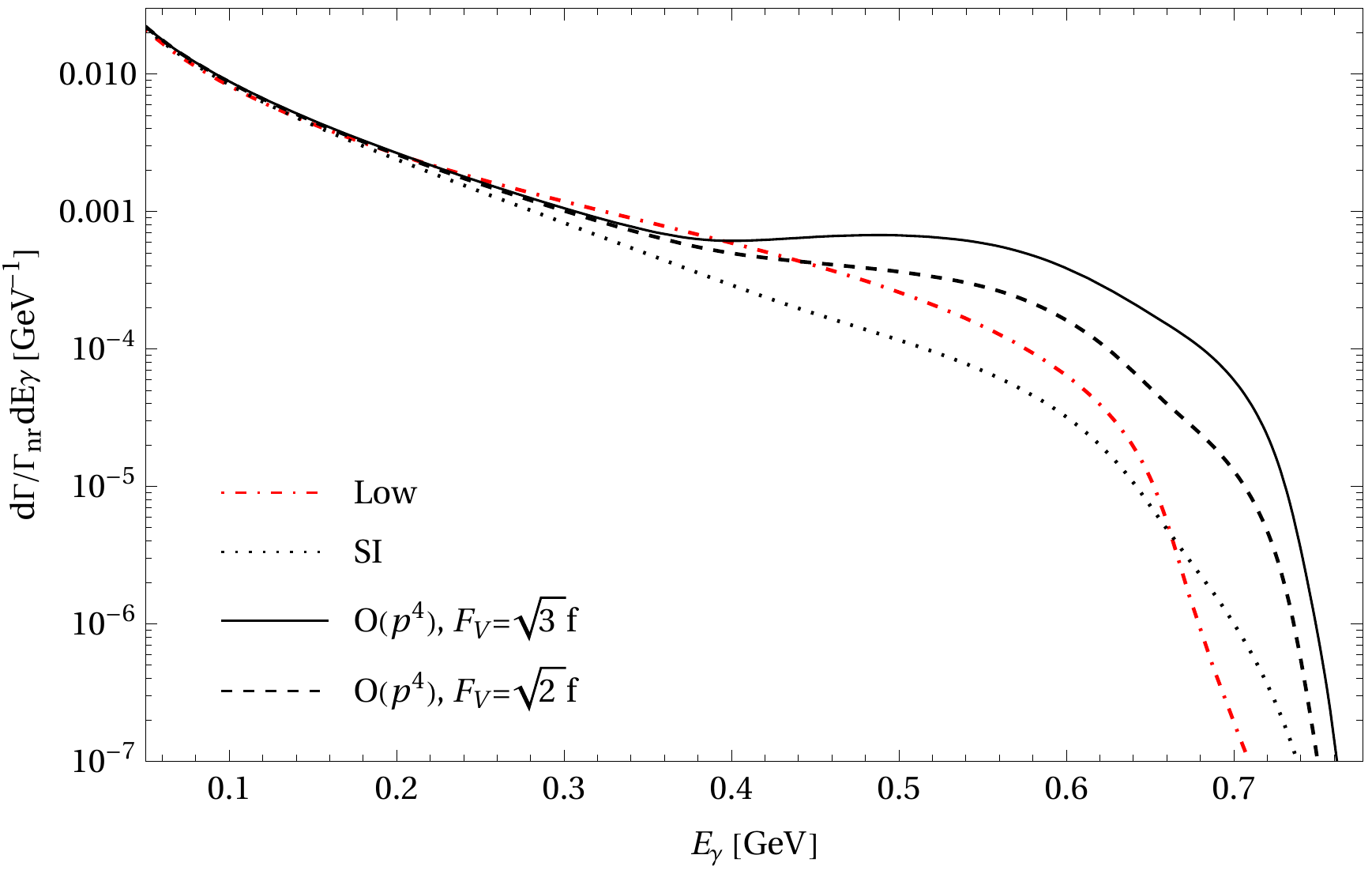}\\[1ex]
\includegraphics[width=0.60\textwidth]{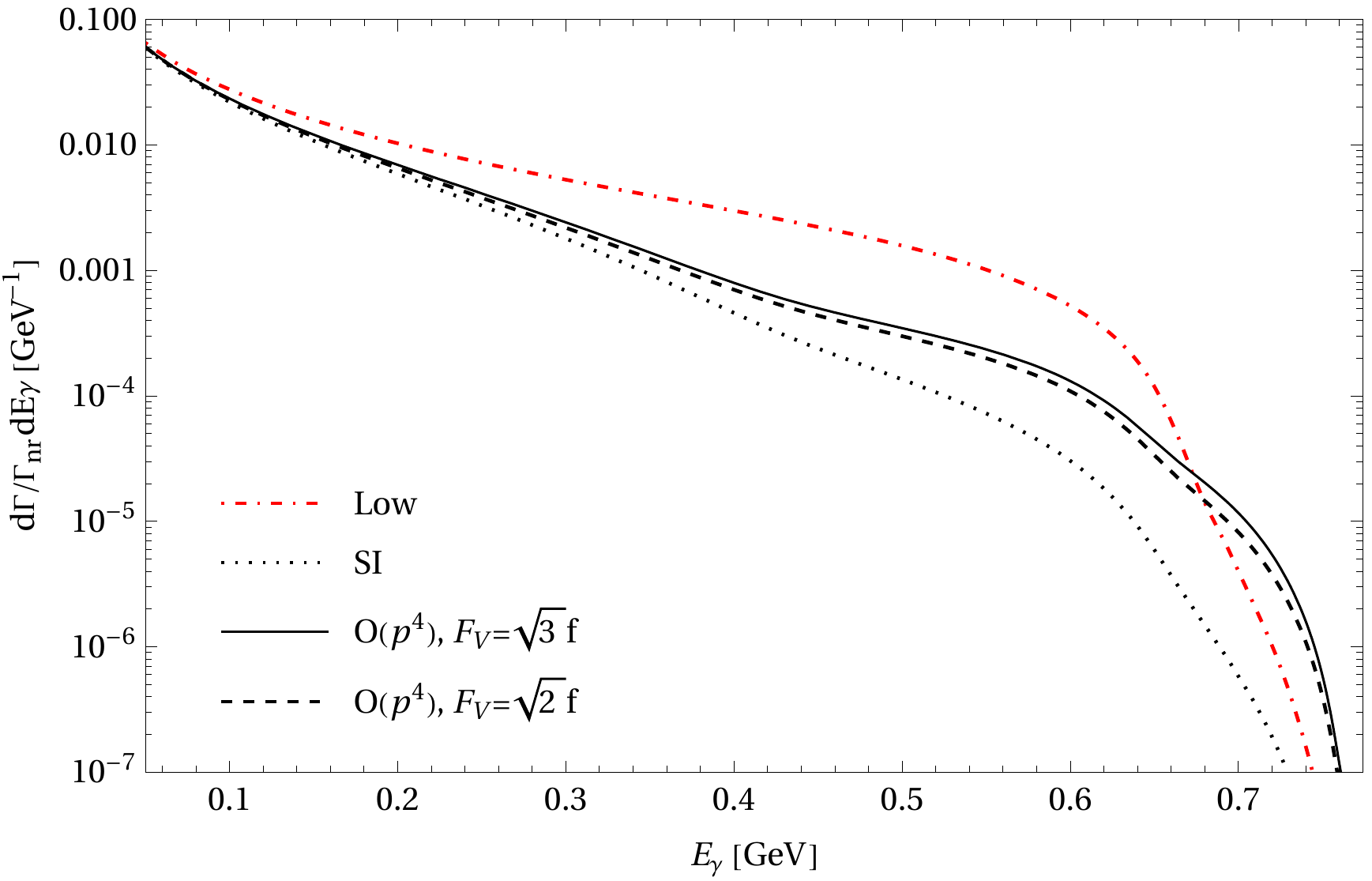}\\[1ex]
\includegraphics[width=0.60\textwidth]{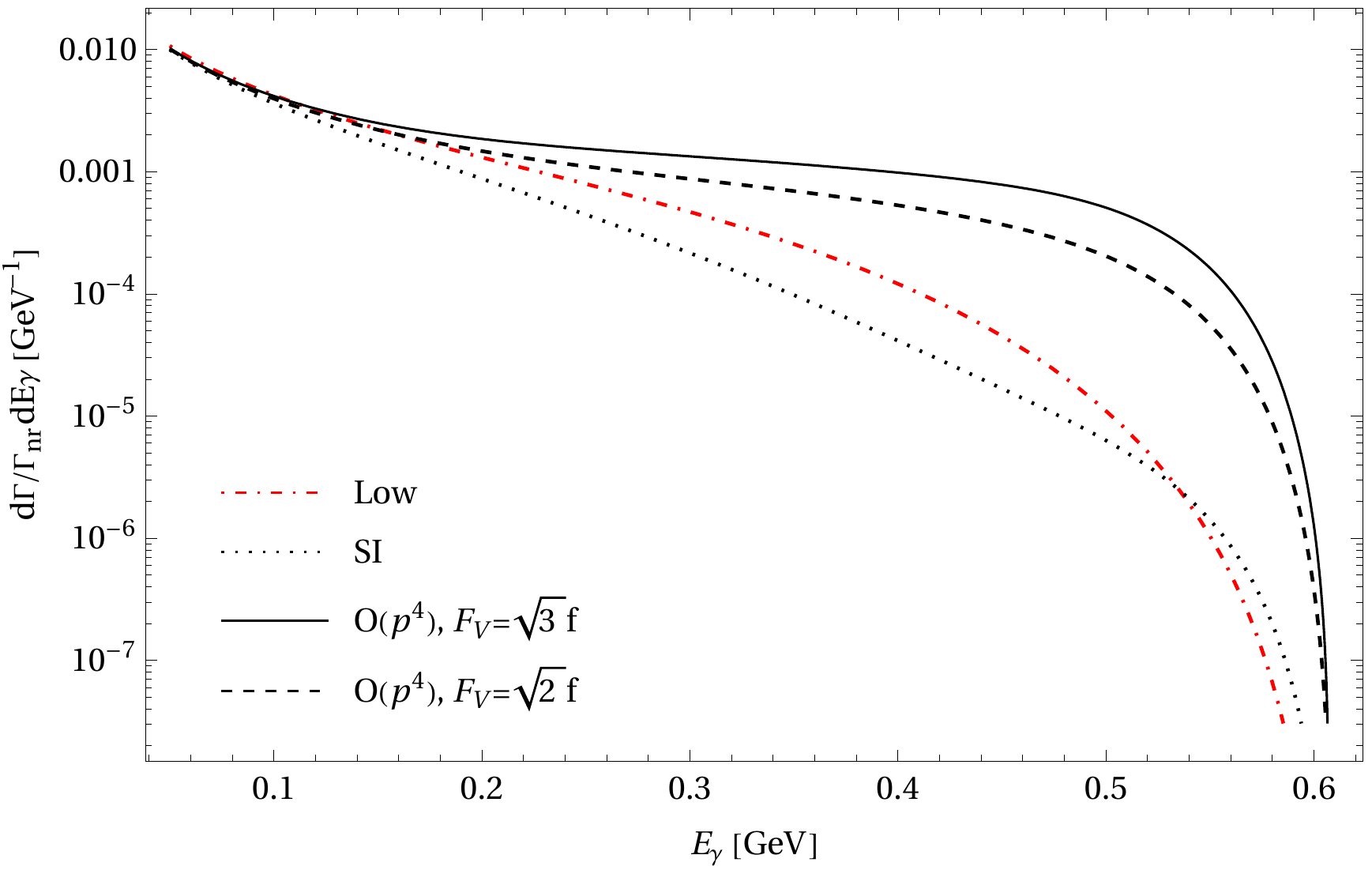}
\caption{Photon energy distribution for the $\tau^-\to K^-\pi^0\nu_\tau\gamma$ (top), 
the $\tau^-\to \bar{K}^0\pi^{-}\nu_\tau\gamma$ (centre) and 
the $\tau^-\to K^-K^{0}\nu_\tau\gamma$ (bottom) decays 
normalized with the non-radiative decay width. 
The dotted line represents the bremsstrahlung contribution
and the red one the Low approximation.
The solid and dashed lines represent the $\mathcal{O}\left(p^4\right)$ corrections using 
$F_V=\sqrt{3}F$ and $F_V=\sqrt{2}F$, respectively.}
\label{Appx6:fig5.5}
\end{figure}

Integrating Eq.~(\ref{Four_body_decays}) using the dispersive vector and scalar form factors
\cite{Gonzalez-Solis:2019iod,Boito:2008fq,Escribano:2014joa,Escribano:2013bca,Jamin:2001zq,Guo:2011pa, 
Guo:2012yt, Guo:2016zep}, 
we get the $P_1^-P_2^0$ invariant mass distribution, 
the photon energy distribution and the branching fraction as a function of $E_\gamma^{\text{cut}}$. 
The outcomes are depicted in Figs.~\ref{Appx6:fig6}, \ref{Appx06:fig5}, \ref{Appx6:fig5} 
and \ref{Appx6:fig5.5}, and summarized in Tables \ref{Appx6:tab1}, \ref{Appx6:tab1.2} and 
\ref{Appx4:tab1K}.

\begin{table}
\begin{center}
\small{\begin{tabular}{|c|c|c|c|c|}
\hline
$E_{\gamma}^{\rm cut}$ & Br(Low) & BR(SI) & 
BR($F_V=\sqrt{2}F$) $\left[\mathcal{O}\left(p^4\right)\right]$ & 
BR($F_V=\sqrt{3}F$) $\left[\mathcal{O}\left(p^4\right)\right]$\\
\hline
\hline
$100\,\mathrm{MeV}$ & $3.4\times 10^{-6}$ & $3.0\times 10^{-6}$ & $3.5\times 10^{-6}$  
& $3.8\times 10^{-6}$\\
$300\,\mathrm{MeV}$ & $6.2\times 10^{-7}$ & $3.4\times 10^{-7}$ & $6.3\times 10^{-7}$  
& $9.4\times 10^{-7}$\\
$500\,\mathrm{MeV}$ & $7.4\times 10^{-8}$ & $3.5\times 10^{-8}$ & $1.5\times 10^{-7}$  
& $3.3\times 10^{-7}$\\
\hline
\end{tabular}}
\caption{Branching ratios Br$(\tau^-\to K^-\pi^0\nu_\tau\gamma)$ for different values of 
$E_\gamma^{\rm cut}$. 
The third column corresponds to the complete bremsstrahlung, 
and the fourth and fifth to the $\mathcal{O}\left(p^4\right)$ contributions.}
\label{Appx6:tab1}
\end{center}
\end{table}

\begin{table}
\begin{center}
\small{\begin{tabular}{|c|c|c|c|c|}
\hline
$E_{\gamma}^{\rm cut}$ & Br(Low) & BR(SI) & 
BR($F_V=\sqrt{2}F$) $\left[\mathcal{O}\left(p^4\right)\right]$ &  
BR($F_V=\sqrt{3}F$) $\left[\mathcal{O}\left(p^4\right)\right]$\\
\hline
\hline
$100\,\mathrm{MeV}$ & $2.6\times 10^{-5}$ & $1.4\times 10^{-5}$ & $1.6\times 10^{-5}$  
& $1.6\times 10^{-5}$\\
$300\,\mathrm{MeV}$ & $6.2\times 10^{-6}$ & $1.1\times 10^{-6}$ & $1.7\times 10^{-6}$  
& $1.9\times 10^{-6}$\\
$500\,\mathrm{MeV}$ & $1.0\times 10^{-6}$ & $7.1\times10^{-8} $ & $2.0\times 10^{-7}$  
& $2.4\times 10^{-7}$\\
\hline
\end{tabular}}
\caption{Branching ratios Br$(\tau^-\to\bar{K}^0\pi^{-}\nu_\tau\gamma)$ 
for different values of $E_\gamma^{\rm cut}$. 
The third column corresponds to the complete bremsstrahlung, 
and the fourth and fifth to the $\mathcal{O}\left(p^4\right)$ contributions.}
\label{Appx6:tab1.2}
\end{center}
\end{table}

\begin{table}
\begin{center}
\small{\begin{tabular}{|c|c|c|c|c|}
\hline
$E_{\gamma}^{\rm cut}$ & BR(Low) & BR(SI) & 
BR($F_V=\sqrt{2}F$) $\left[\mathcal{O}\left(p^4\right)\right]$ &  
BR($F_V=\sqrt{3}F$) $\left[\mathcal{O}\left(p^4\right)\right]$\\
\hline
\hline
$100\,\mathrm{MeV}$ & $5.3\times 10^{-7}$ & $3.7\times 10^{-7}$ & $6.8\times 10^{-7}$  
& $9.4\times 10^{-7}$\\
$300\,\mathrm{MeV}$ & $4.8\times 10^{-8}$ & $1.9\times 10^{-8}$ & $1.7\times 10^{-7}$  
& $3.1\times 10^{-7}$\\
$500\,\mathrm{MeV}$ & $3.7\times 10^{-10}$ & $3.0\times 10^{-10}$ & $1.1\times 10^{-8}$  
& $2.9\times 10^{-8}$\\
\hline
\end{tabular}}
\caption{Branching ratios Br$(\tau^-\to K^- K^0\nu_\tau\gamma)$ 
for different values of $E_\gamma^{\rm cut}$. 
The third column corresponds to the complete bremsstrahlung, 
and the fourth and fifth to the $\mathcal{O}\left(p^4\right)$ contributions.}
\label{Appx4:tab1K}
\end{center}
\end{table}

The branching fractions of the radiative decays as a function of $E_\gamma^\text{cut}$ 
are shown in Fig.~\ref{Appx6:fig6}.
In Tables \ref{Appx6:tab1} and \ref{Appx6:tab1.2}, 
one can see that for $E_\gamma^{\text{cut}}\lesssim 100\,\text{MeV}$ the main contribution at 
$\mathcal{O}(p^4)$ comes from the complete bremsstrahlung (SI) amplitude in agreement with 
the results in Refs.~\cite{Cirigliano:2002pv,Miranda:2020wdg,Chen:2022nxm} 
for the $\tau^-\to\pi^-\pi^0\nu_\tau\gamma$ decays (see also the recent ref.~\cite{Esparza-Arellano:2023dps}). 
It is also seen that the Low's approximation is sufficient to describe the $K^-\pi^0$ decays
up to these energies, 
while this is not the case for the $\bar{K}^0\pi^{-}$ ones.
Contrary to the $\tau^-\to(K\pi)^-\nu_\tau$ transitions, 
where the $K^-\pi^0$ and $\pi^-\bar{K}^0$ decay modes 
differ only by a squared CGC factor in the isospin symmetry limit, 
the radiative decays are more subtle. 
At low photon energies, these two modes are approximately related by 
$\mathrm{Br}(\tau\to\bar{K}^0\pi^-\nu_\tau\gamma)/\mathrm{Br}(\tau^-\to K^-\pi^0\nu_\tau\gamma)
\approx 2(m_K/m_\pi)\sim 7$, which explains their hierarchy.
In both decay channels, the SD contributions seem to be subdominant,
while the $\tau^-\to K^-K^0\nu_\tau\gamma$ decays are more susceptible to these contributions 
(see Table \ref{Appx4:tab1K}).

In Fig.~\ref{Appx06:fig5}, the decay spectrum is depicted with $v_i=a_i=0$ 
for different $E_\gamma^{\text{cut}}$ values. 
For the $\tau^-\to(K\pi)^-\nu_\tau\gamma$ decays,
the first peak is due to bremsstrahlung off the charged meson, 
i.e.~$K^-$ or $\pi^-$, 
and the second one receives contributions from bremsstrahlung off the $\tau$ lepton and 
resonance exchange.
In Fig.~\ref{Appx6:fig5}, we compare the distributions for $E_\gamma^{\text{cut}}=300\,\text{MeV}$ 
using the Low's approximation (red dashed line), the SI amplitude (dotted line), 
and the $\mathcal{O}(p^4)$ amplitude 
with $F_V=\sqrt{2}F$ (dashed line) and $F_V=\sqrt{3}F$ (solid line). 
The most important contribution for the $(K\pi)^-$ decay channels comes from the 
$K^*(892)$ resonance exchange around $s\sim 0.79\,\text{GeV}^2$. 
It is worth noting that for the $\tau^-\to\bar{K}^0\pi^-\nu_\tau\gamma$ decays, 
there is a huge suppression around the $K^*(892)$ peak 
when the full distribution is compared to the Low one.
The reason for that is the following.
While the Low approximation in Eq.~(\ref{LowAmp}) includes only the $\mathcal{O}(k^{-1})$
dominant contributions from the bremsstrahlung off the initial tau lepton and the final charged meson,
the full amplitude in Eq.~(\ref{eq:tau_decay}) contains also the $\mathcal{O}(k^0)$
subdominant contribution from the first line of this equation,
which is common to both $(K\pi)^-$ channels,
plus all the $\mathcal{O}(k^0)$ contributions from Eq.~(\ref{VSIexp}).
Among the latter, the leading numerical contribution comes from the first term in the second line of
Eq.~(\ref{eq3:VSI}),
which has different sign depending on the channel, 
as already mentioned after that equation.
This fact makes that in the case of the $K^-\pi^0$ mode 
these two leading subdominant contributions approximately 
cancel each other and the Low and full distributions are quite similar at the $K^*(892)$ peak.
Conversely, for the $\bar{K}^0\pi^-$ mode, the two contributions combine 
with the overall effect of decreasing considerably the peak.
Finally, we 
just mention that the $K^-K^0$ invariant mass distribution is more sensitive to SD contributions, 
although the $\rho(1450)$ effect is hidden in the spectrum because of the 
corresponding kinematical suppression.

The photon energy distribution is shown in Fig.~\ref{Appx6:fig5.5}. 
The SI amplitude in all these decays governs the distribution for $E_\gamma\lesssim 100\,\text{MeV}$, 
in agreement with the outcomes for the branching ratio. 
However, the SD contributions become relevant for $E_\gamma\gtrsim 250\ \text{MeV}$. 
This feature makes these decays an excellent probe for testing SD effects.
The same analysis for the $\tau^-\to\pi^-\pi^0\nu_\tau\gamma$ decays
can be found for instance in Ref.~\cite{Miranda:2020wdg}.

\section{Radiative Corrections}
\label{sec:RadCors}
The overall differential decay width is given by
\begin{equation}
\label{eq:widthoverall}
\left.\frac{d\Gamma}{dt}\right\vert_{PP(\gamma)}=
\left.\frac{d\Gamma}{dt}\right\vert_{PP}+
\left.\frac{d\Gamma}{dt}\right\vert_{\rm III}+
\left.\frac{d\Gamma}{dt}\right\vert_{\rm IV/III}+
\left.\frac{d\Gamma}{dt}\right\vert_{\rm rest}\ ,
\end{equation}
where the first term is the non-radiative differential width in Eq.~(\ref{eq:AppNR}), 
the second and third terms correspond to the Low approximation integrated according to the kinematics 
in Refs.~\cite{Cirigliano:2002pv,Miranda:2020wdg}, Eq. (\ref{eq:widthRP}), 
and the last term includes the remaining contributions.

To evaluate the first term in the right-hand side of Eq.~(\ref{eq:widthoverall}) 
we use two models for the factorization of the radiative corrections to the form factors. Both of them were pioneered in $K_{\ell3}$ decays and have also been employed in the $\tau^-\to (K\pi)^-\nu_\tau$ processes. 
As we will see, the difference between both results will saturate the uncertainty of our 
real-photon radiative corrections.
To the best of our knowledge, 
the importance of the factorization model for the former corrections
was not previously recognized in the literature. 

The contributions to the form factors due to virtual corrections can be written as 
\begin{equation}
\label{F+0tu}
    F_{+/0}(t,u)=F_{+/0}(t)+\delta F_{+/0}(t,u)\ ,
\end{equation}
where 
$\delta F_0(t,u)=\delta F_{+}(t,u)+\frac{\displaystyle t}{\displaystyle\Delta_{-0}}\delta\bar{f}_{-}(u)$. 
In model $1$, $\delta F_{+}(t,u)$ is given by~\cite{Cirigliano:2002pv}
\begin{equation}
    \frac{\delta F_+(t,u)}{F_+(t)}=\frac{\displaystyle\alpha}{\displaystyle 4\pi}
    \left[2(m_{-}^2+m_{\tau}^2-u)
    \mathcal{C}(u,M_\gamma)+2\log\left(\frac{\displaystyle m_{-}m_\tau}{\displaystyle M_\gamma^2}\right)
    \right]+\delta \bar{f}_{+}(u)\ ,
\end{equation}
while in model 2, it is written as~\cite{Antonelli:2013usa} (we note that in this second case, the correction $\delta \bar{f}_+(u)$ to $\delta F_+(t,u)$ is not modulated by the vector form factor $F_+(t)$)
\begin{equation}
    \frac{\delta F_+(t,u)}{F_+(t)}=\frac{\displaystyle\alpha}{\displaystyle 4\pi}
    \left[2(m_{-}^2+m_{\tau}^2-u)
    \mathcal{C}(u,M_\gamma)+2\log\left(\frac{\displaystyle m_{-}m_\tau}{\displaystyle M_\gamma^2}\right)
    \right]+\frac{\delta \bar{f}_{+}(u)}{F_+(t)}\ ,
\end{equation}
where $\mathcal{C}(u,M_\gamma)$, $\delta \bar{f}_{+}(u)$ and $\delta \bar{f}_{-}(u)$ 
are defined in App.~\ref{Appx:RC}.  
A similar factorization prescription was used in Ref.~\cite{Cirigliano:2001mk}, 
where model 2 was preferred over model 1 for the $K_{\mu3}$ decays since the loop contributions 
to $\bar f_{+/-}(u)$ are different~\footnote{Both 
prescriptions were studied for the $K_{e3}$ decays in Ref.~\cite{Cirigliano:2008wn}, 
their outcomes for $\delta^{K\ell}_{\text{EM}}(\mathcal{D}_3)(\%)$ are shifted from 0.41 to 0.56 for
$K^0_{e3}$ and from -0.564 to -0.410 for $K^\pm_{e3}$ modes 
where the former numbers correspond to model 2.}.
We will see here that model 1 factorization warrants smoother corrections than model 2 
when resonance contributions are included, 
as resonance enhancements will cancel in the long-distance radiative correction factor $G_{\rm EM}(t)$ 
in Eq.~(\ref{eqGEM}), as opposed to model 2. 
This motivates our preference of model 1 over model 2 in our following phenomenological analysis. 

A couple of points are worth to stress in connection to both factorization models and the preferred one. First, a measurement of di-meson or photon energy spectra in the considered decays will be really helpful in reducing the model-dependence of our results (particularly on the factorization prescription). Second, we will present elsewhere the corresponding computations of the virtual photon structure-dependent corrections, which will complete these at the one-loop order. We expect that the model-dependence is reduced in the sum of all radiative corrections, so having this last missing piece available will also be valuable for diminishing the model-dependence (again with particular emphasis on the precise factorization in the considered decays).

\begin{figure}
\centering			
\includegraphics[width=0.45\textwidth]{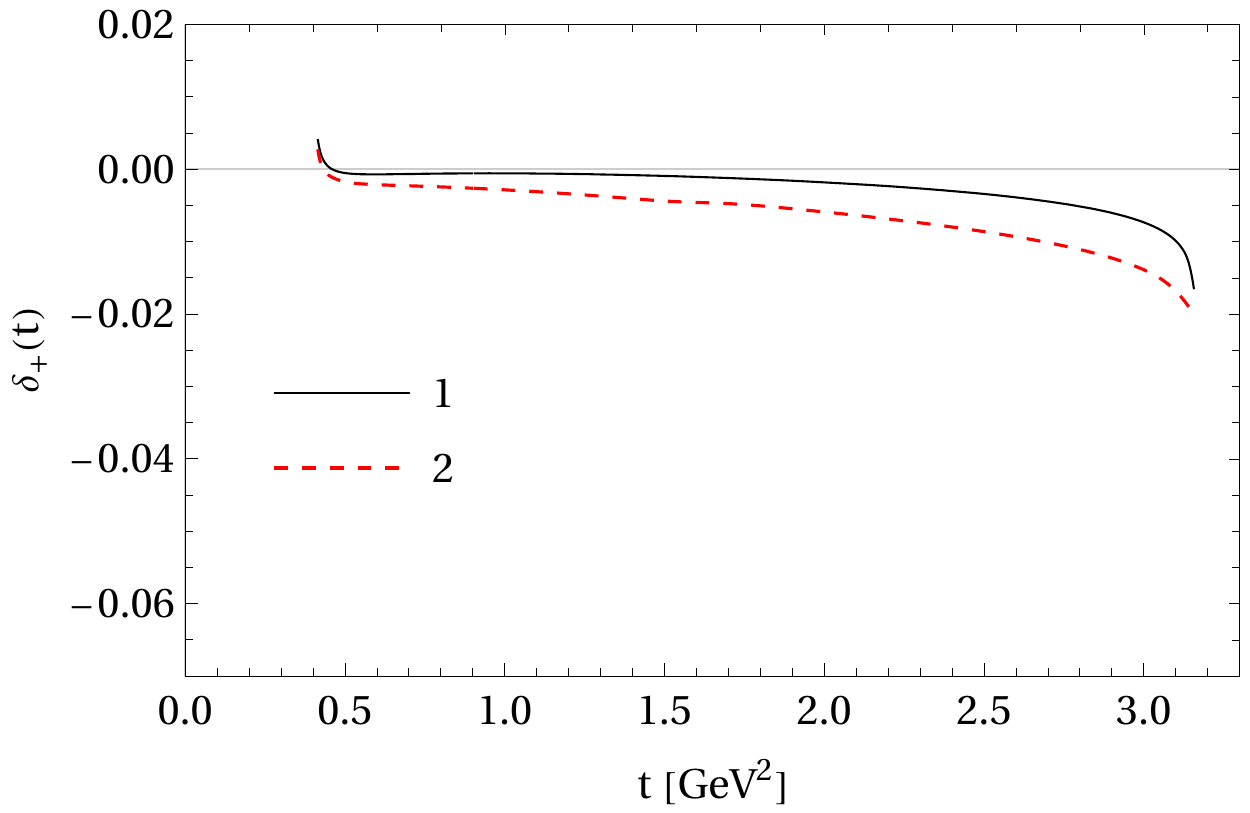}\qquad
\includegraphics[width=0.45\textwidth]{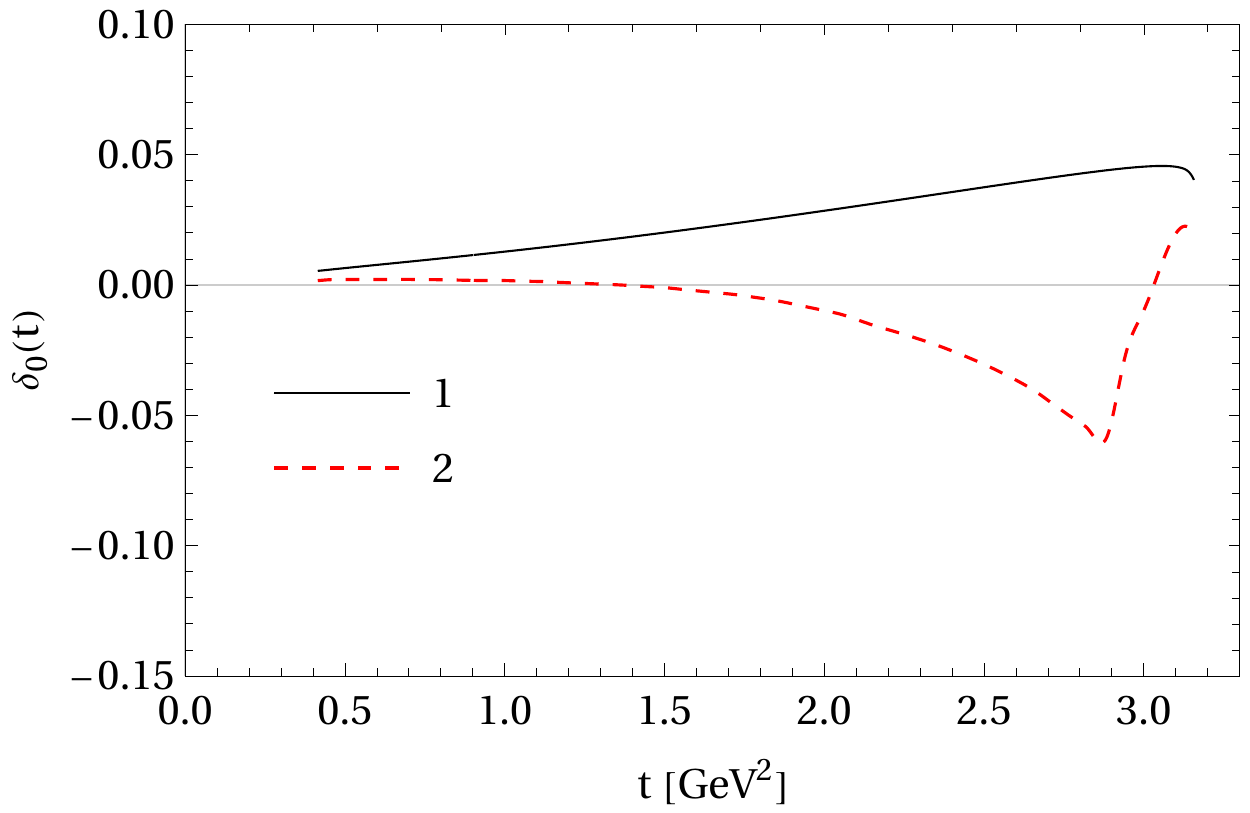}\\[1ex]
\includegraphics[width=0.45\textwidth]{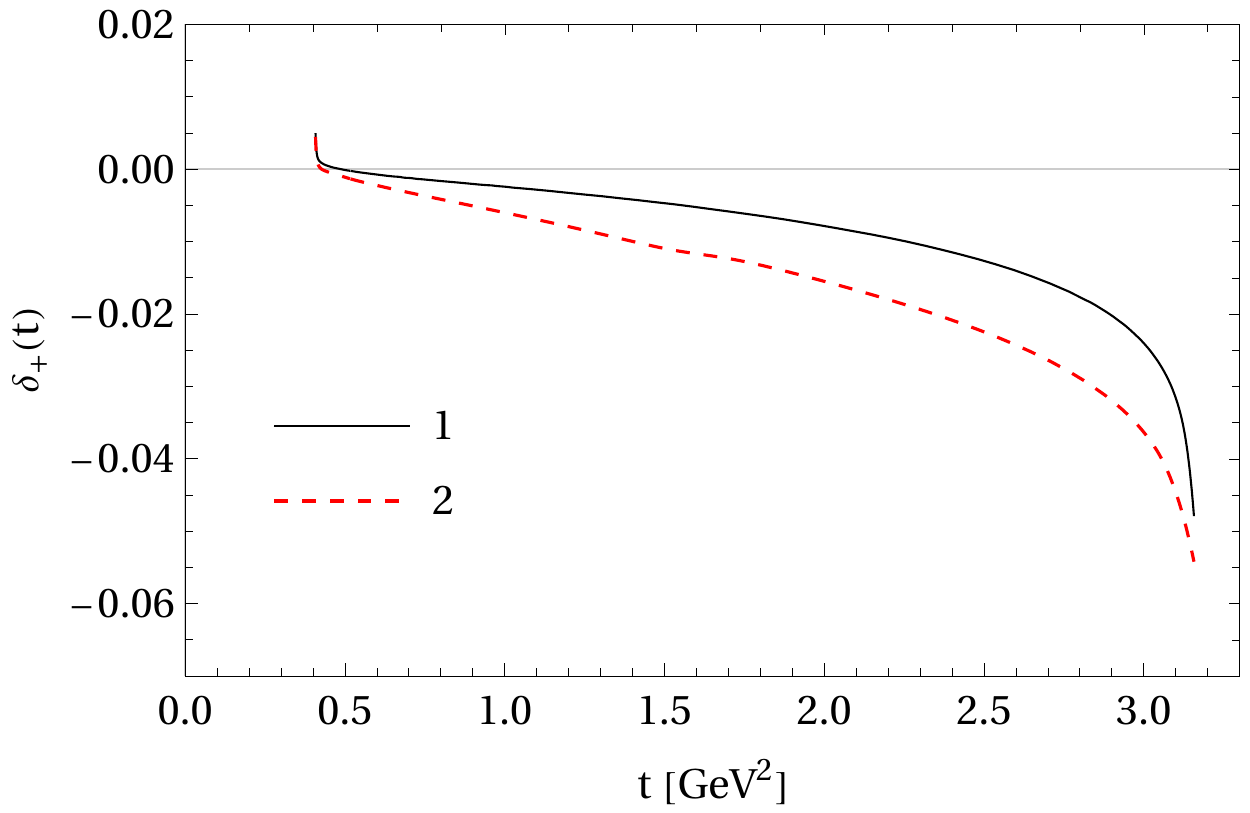}\qquad
\includegraphics[width=0.45\textwidth]{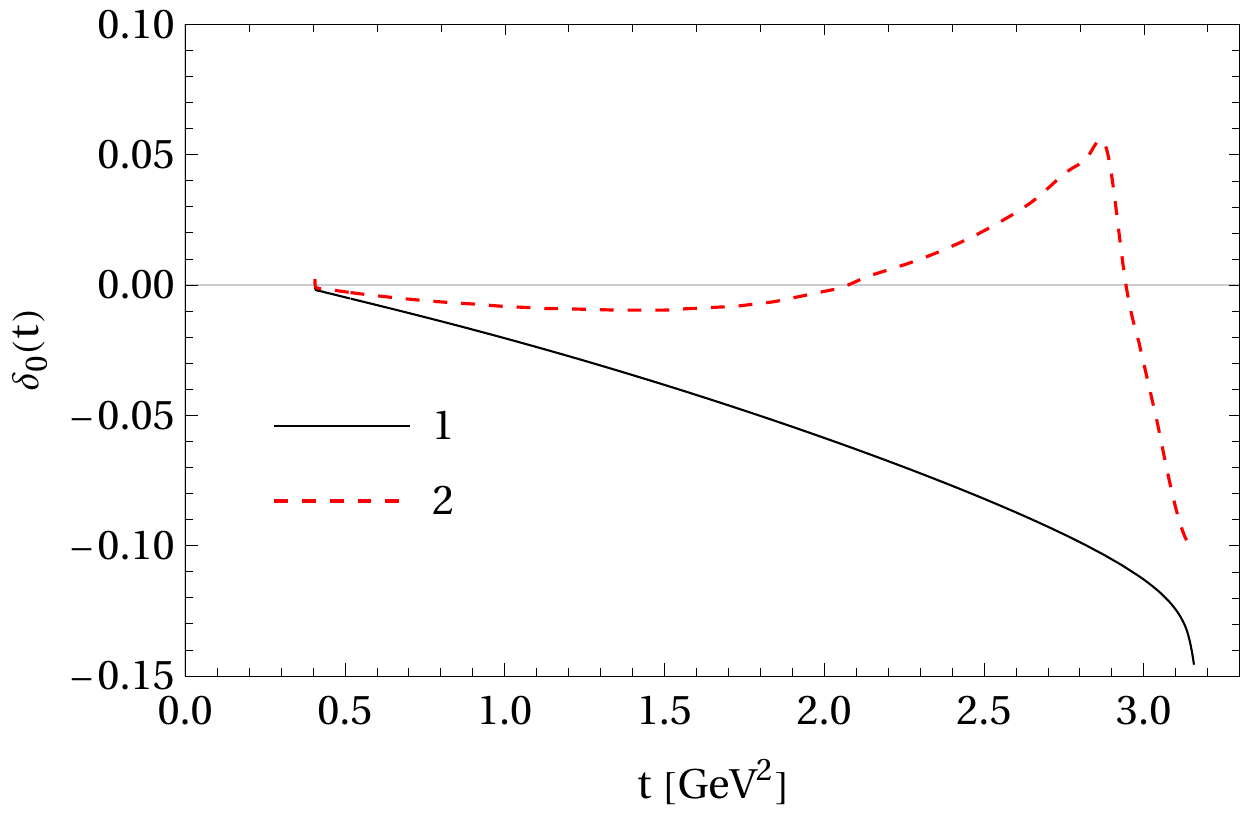}\\[1ex]
\includegraphics[width=0.45\textwidth]{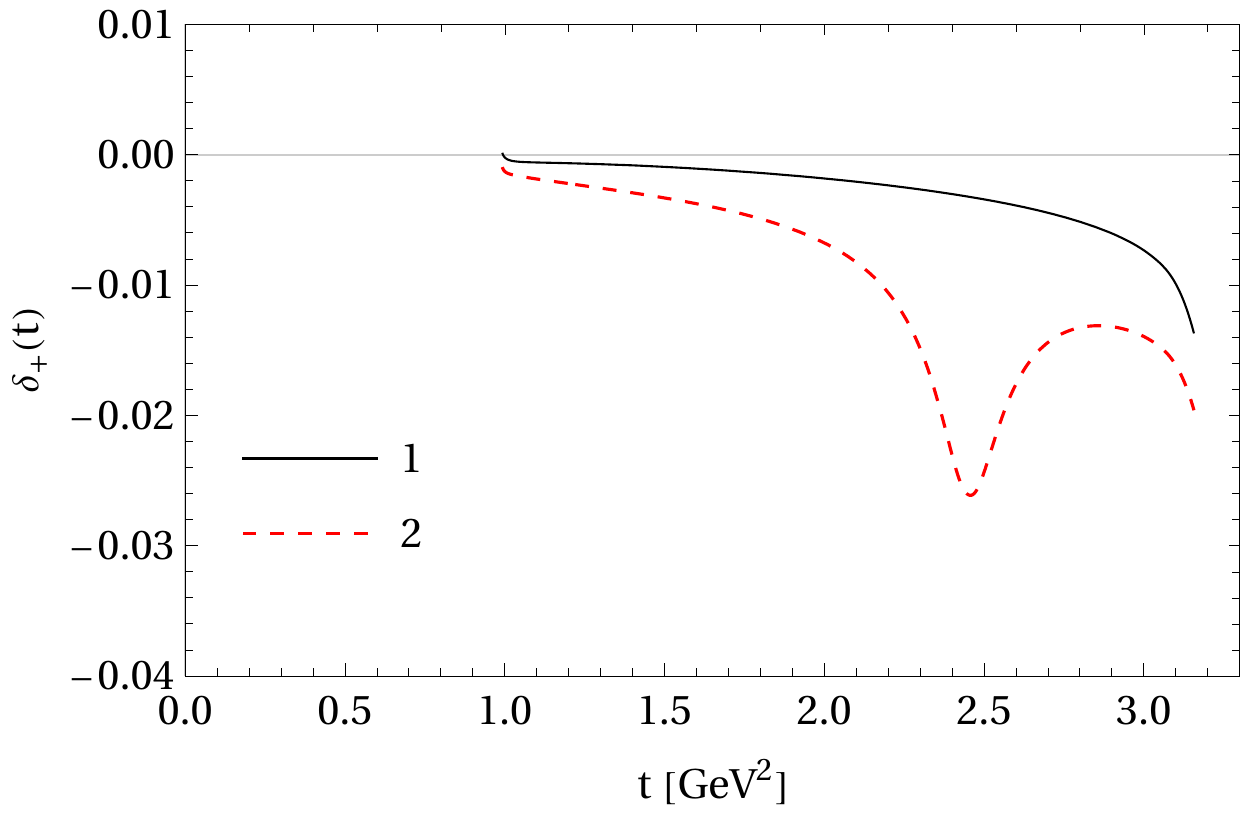}\qquad
\includegraphics[width=0.45\textwidth]{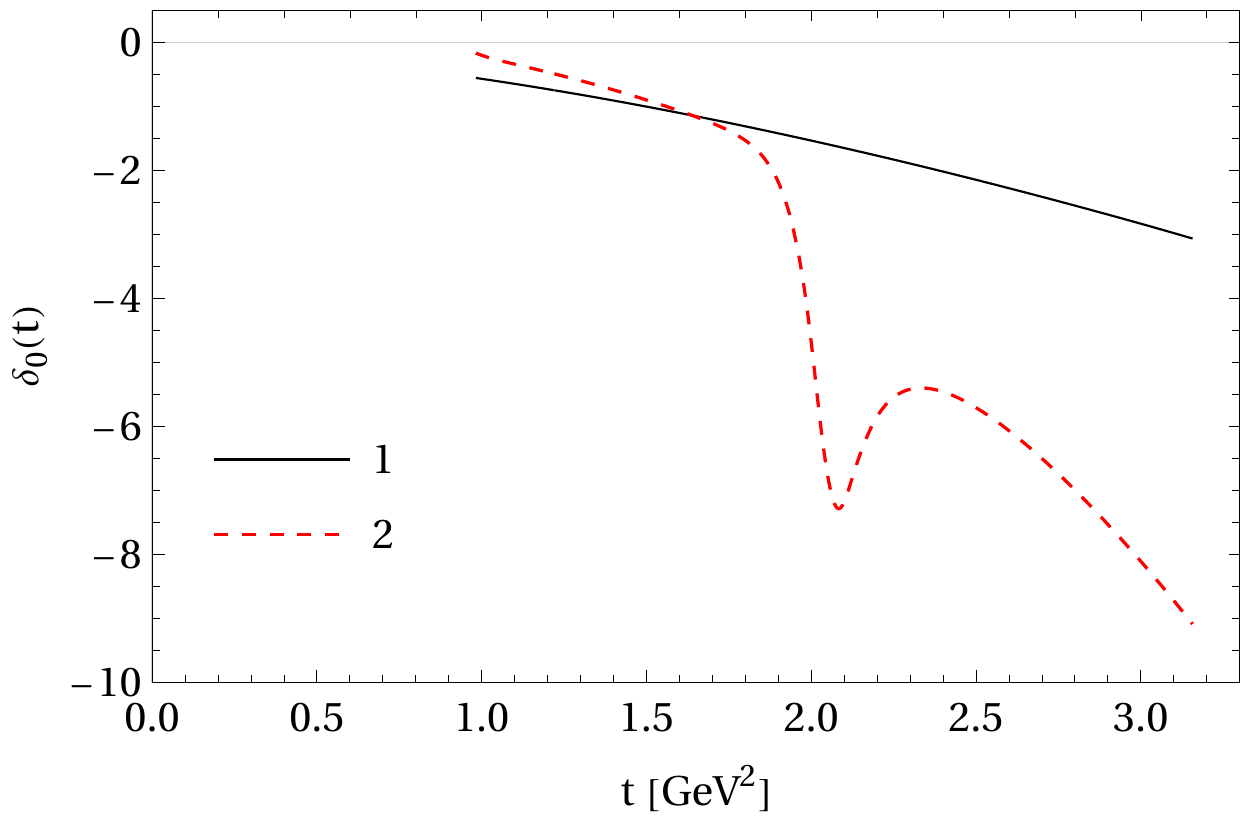}	
\caption{Correction factors $\delta_{+}(t)$ (left) and $\delta_{0}(t)$ (right) 
to the differential decay rates of the $K^{-}\pi^0$, $\bar{K}^{0}\pi^{-}$ and 
$K^{-}K^0$ modes from top to bottom, according to models 1 (solid black) and 2 (dashed red).}
\label{RADCOR:fig1}
\end{figure}

The correction factors $\overline{\delta}_{A}(t)$ for the four-body decays 
and $\widetilde{\delta}_{A}(t)$ for the three-body processes
appearing in Eqs.~(\ref{eq:delta_bar}) and (\ref{eq:delta_tilde}), respectively, 
where $A={+,0,+0}$, are both IR divergent when $M_\gamma\to 0$.
Nevertheless, the overall contribution,
$\delta_{A}(t)=\overline{\delta}_A(t)+\widetilde{\delta}_A(t)$, is finite. 
In Fig.~\ref{RADCOR:fig1}, we show the predictions for $\delta_{A}(t)$ for the 
$K^-\pi^0$, $\bar{K}^0\pi^-$ and $K^-K^0$ decay modes using the form factors in model $1$ and $2$. 
Whilst our results for $\delta_{+}(t)$ in model $2$ agree with those in Ref.~\cite{Antonelli:2013usa} 
in Figure $2$,
the predictions for $\delta_0(t)$ are slightly different as a consequence of the parameterization 
of the scalar form factor~\footnote{This 
effect is mainly responsible for the slight difference between our results for model 2 
in Table \ref{RADCOR:fig1} and those in Ref.~\cite{Antonelli:2013usa}.}.
 
\begin{figure}
\centering			
\includegraphics[width=0.45\textwidth]{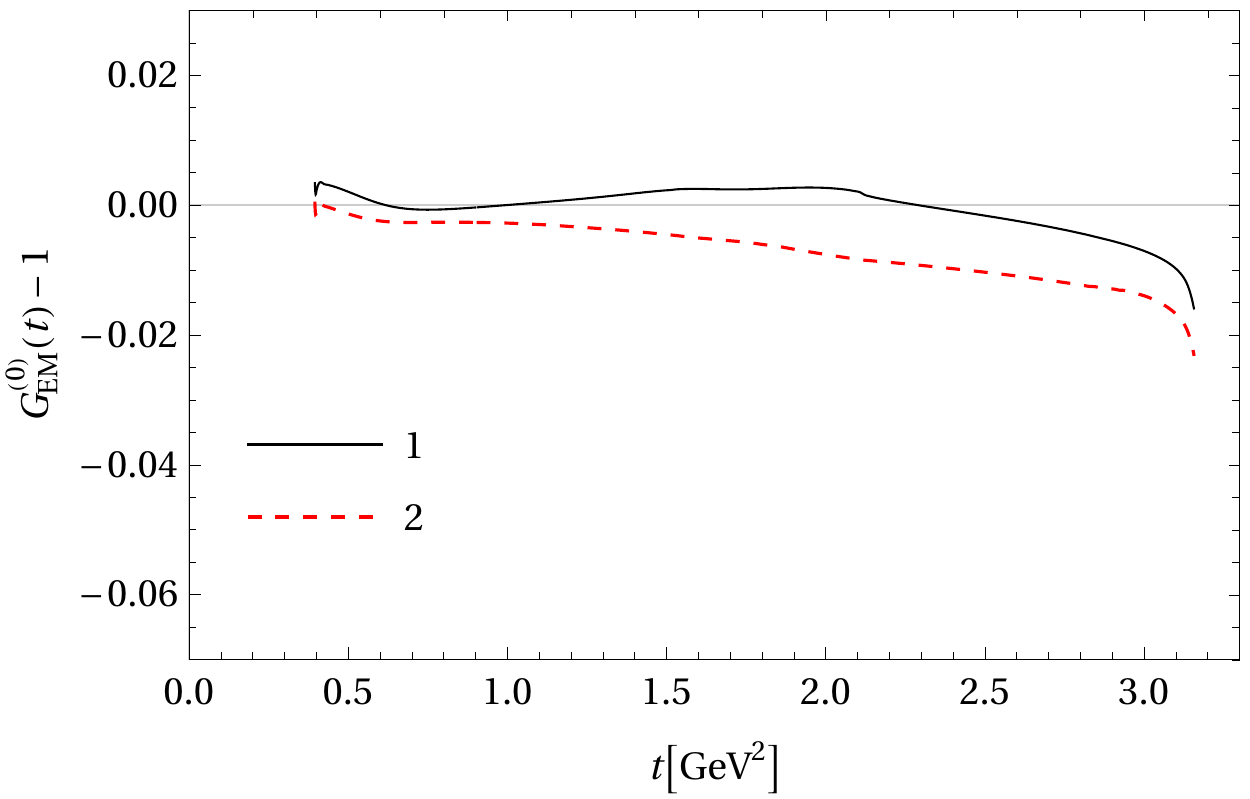}\qquad
\includegraphics[width=0.45\textwidth]{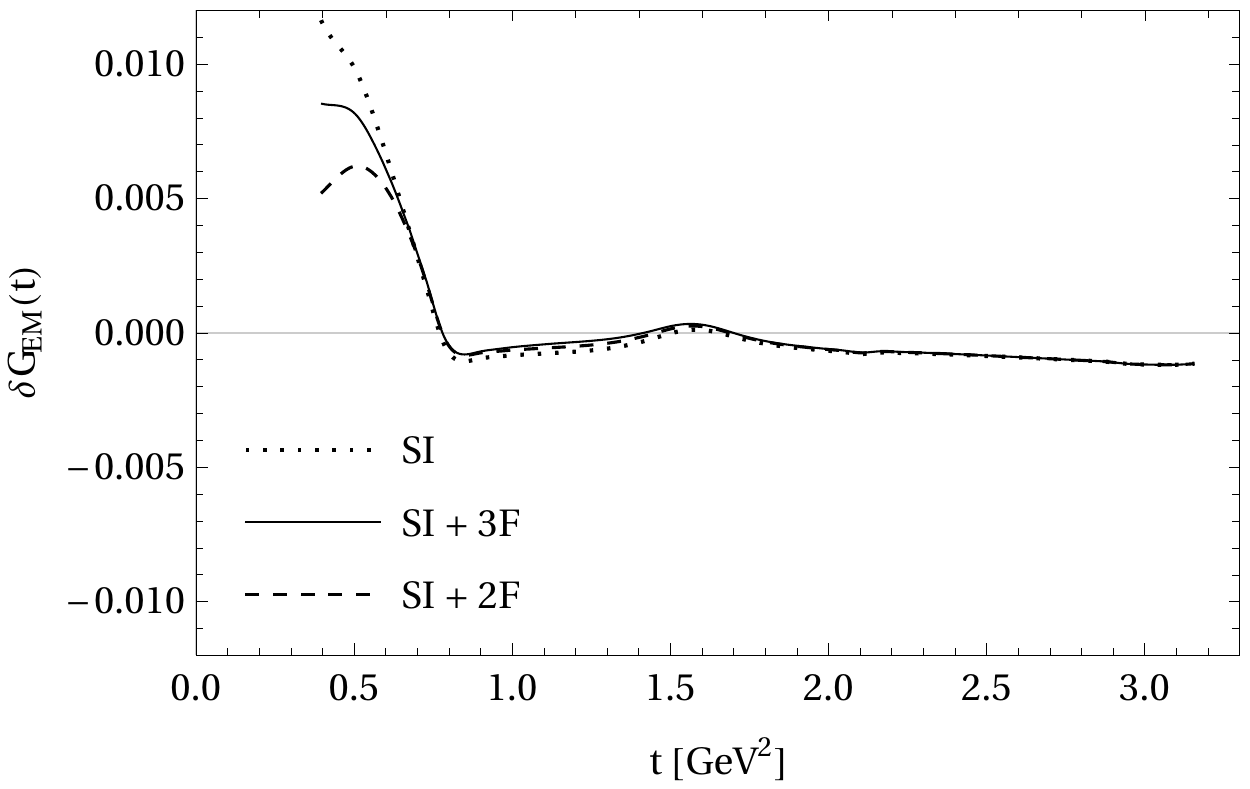}\\[1ex]	
\includegraphics[width=0.45\textwidth]{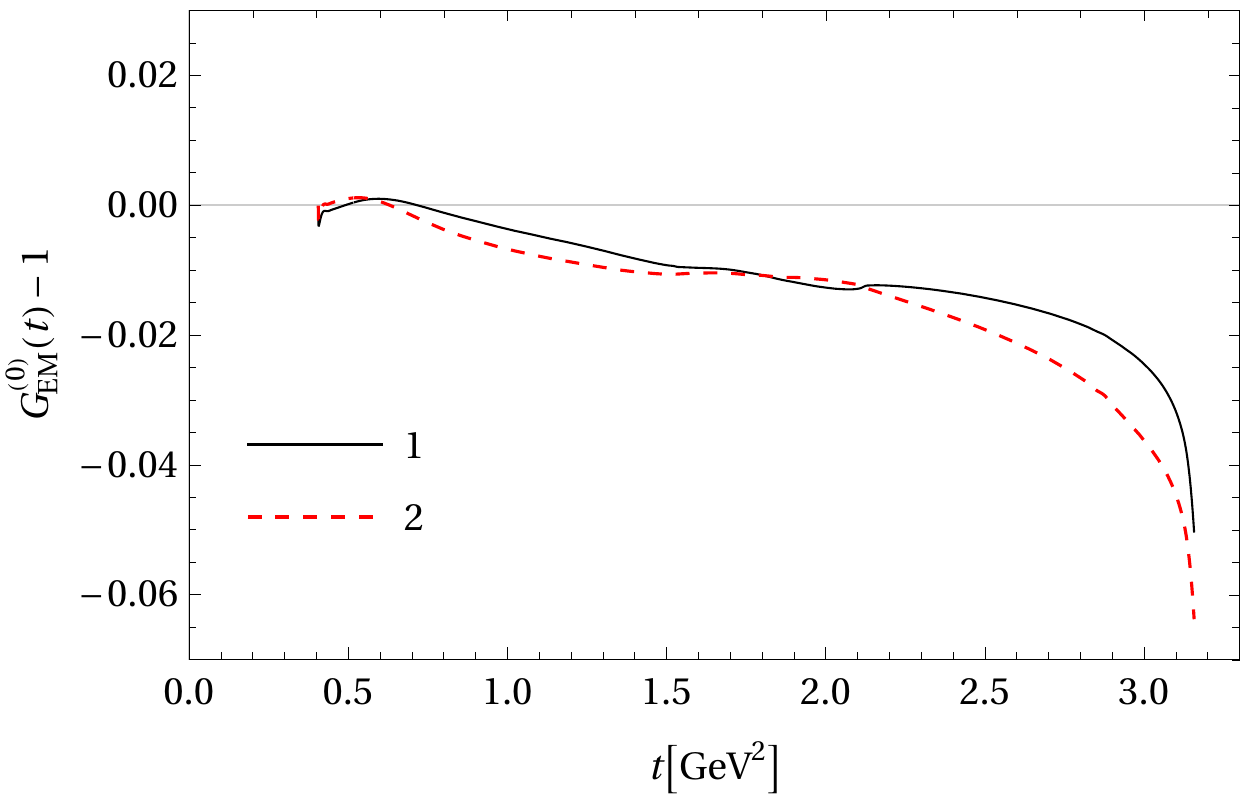}\qquad
\includegraphics[width=0.45\textwidth]{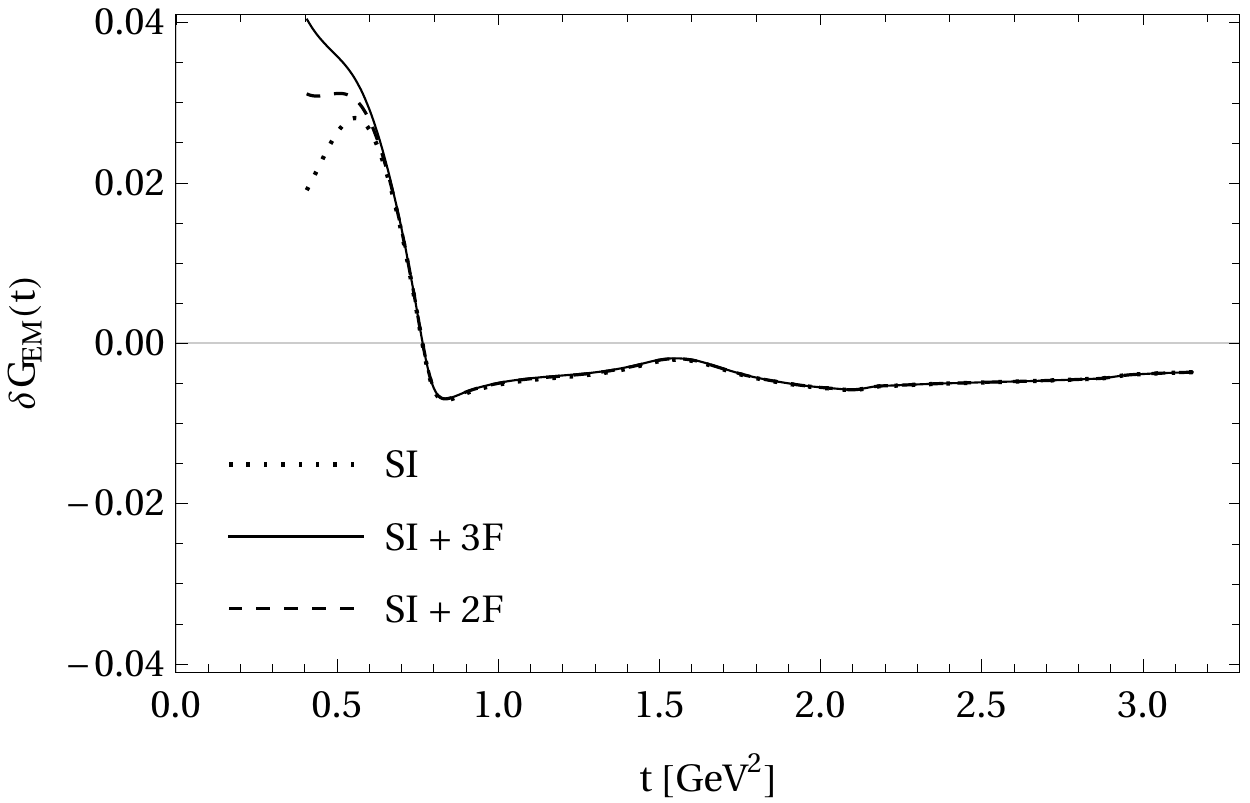}\\[1ex]	
\includegraphics[width=0.45\textwidth]{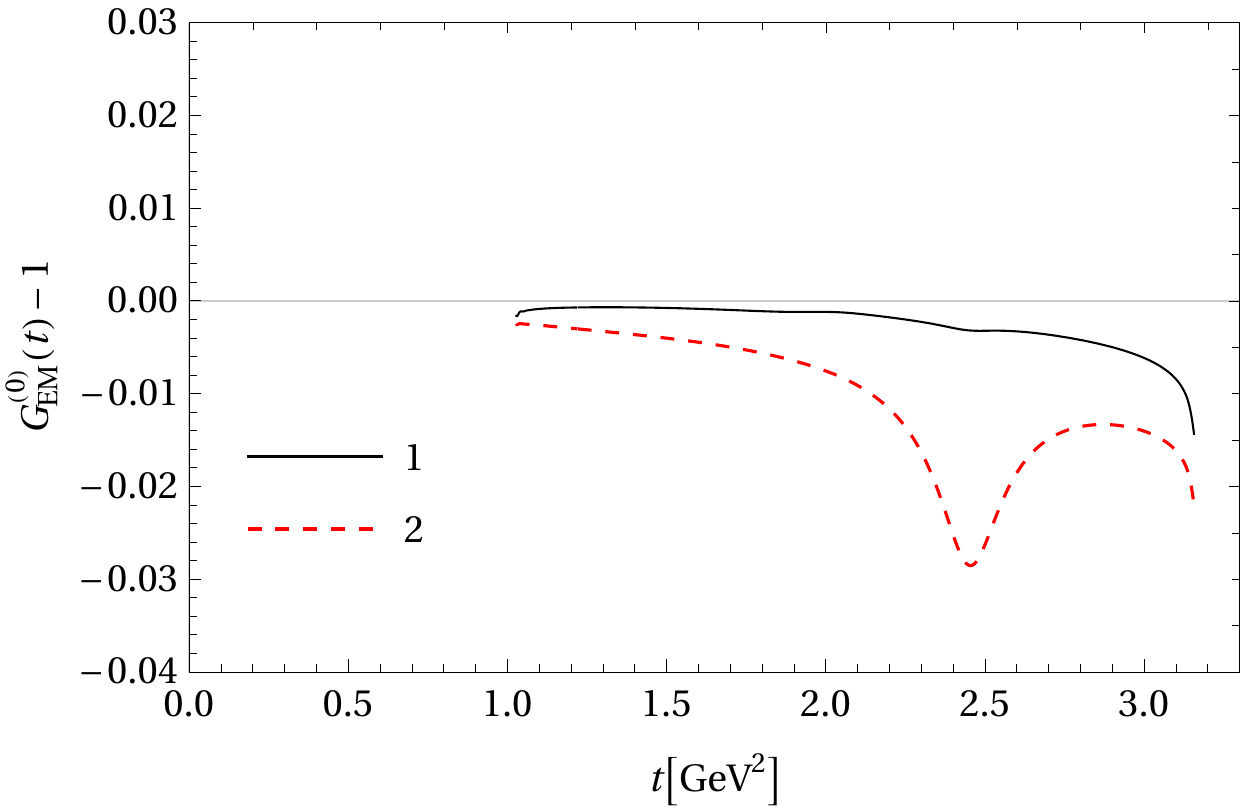}\qquad
\includegraphics[width=0.45\textwidth]{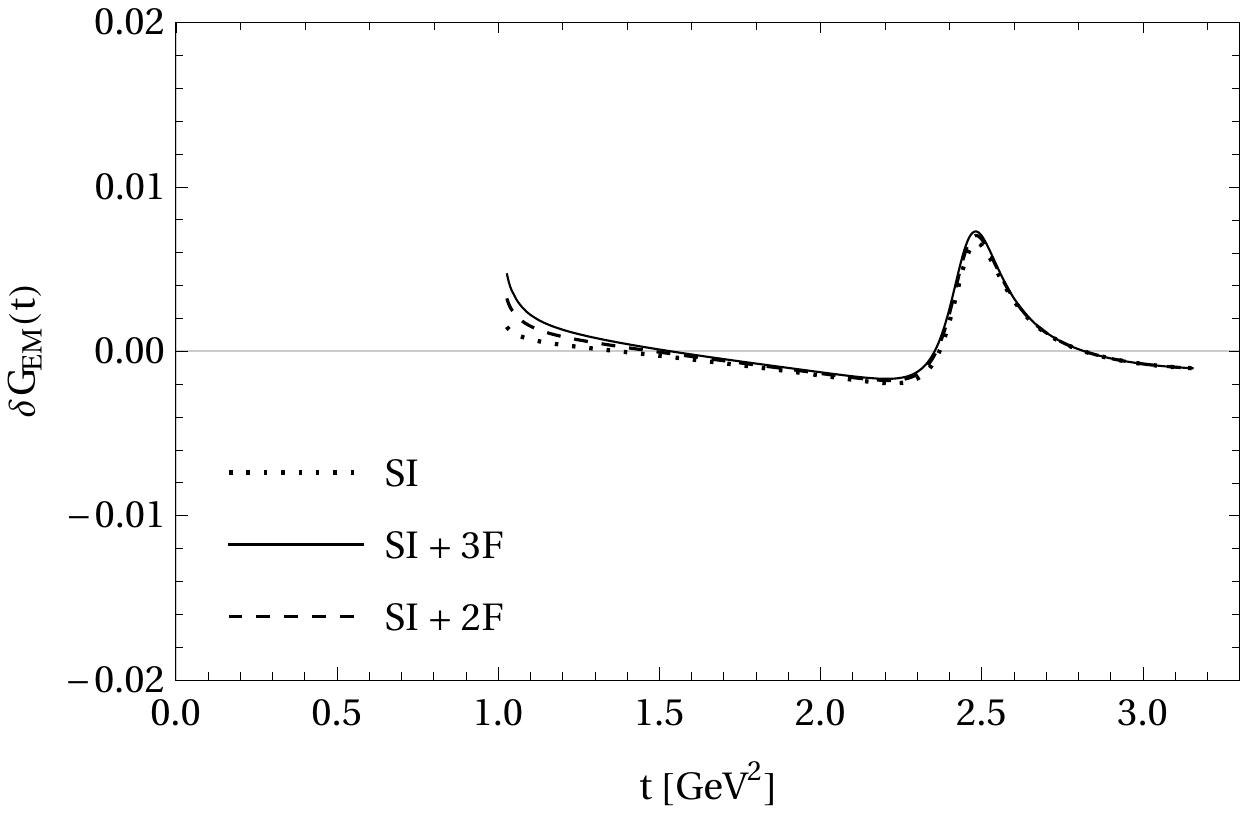}\\[1ex]	
\includegraphics[width=0.45\textwidth]{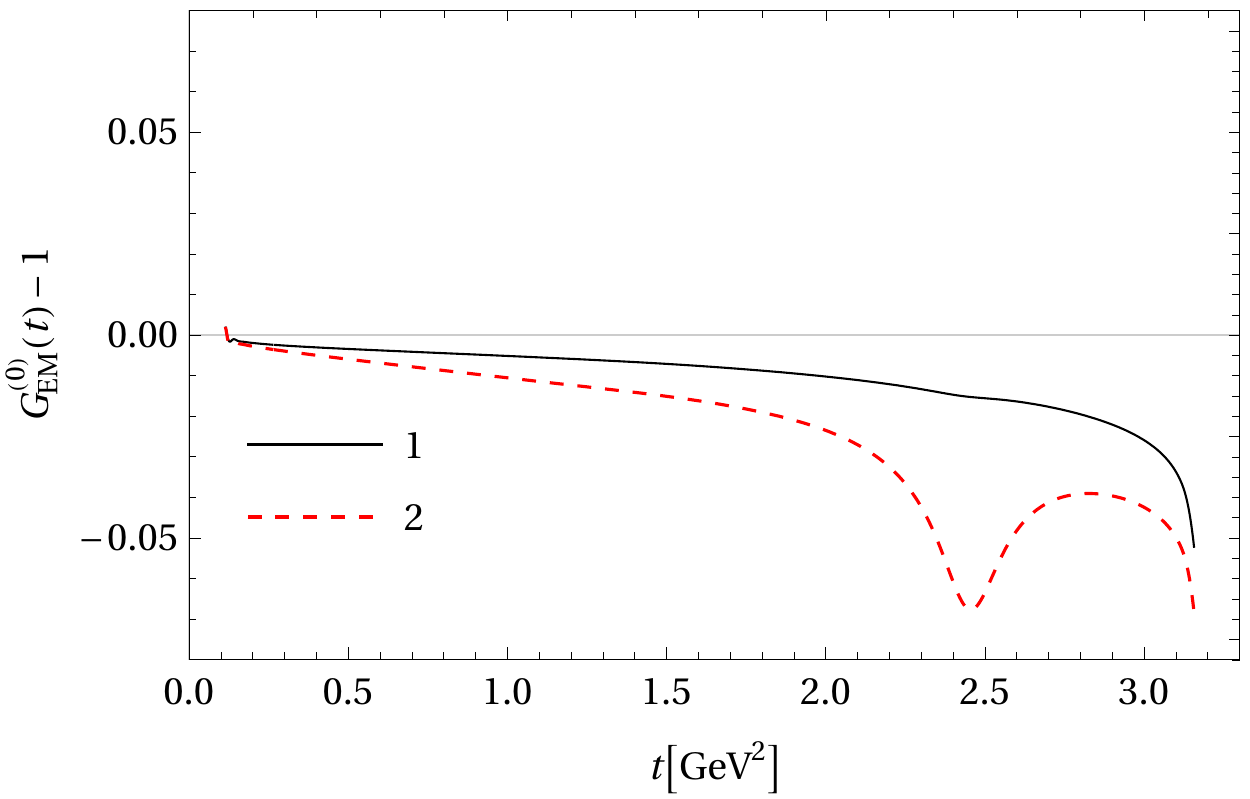}\qquad
\includegraphics[width=0.45\textwidth]{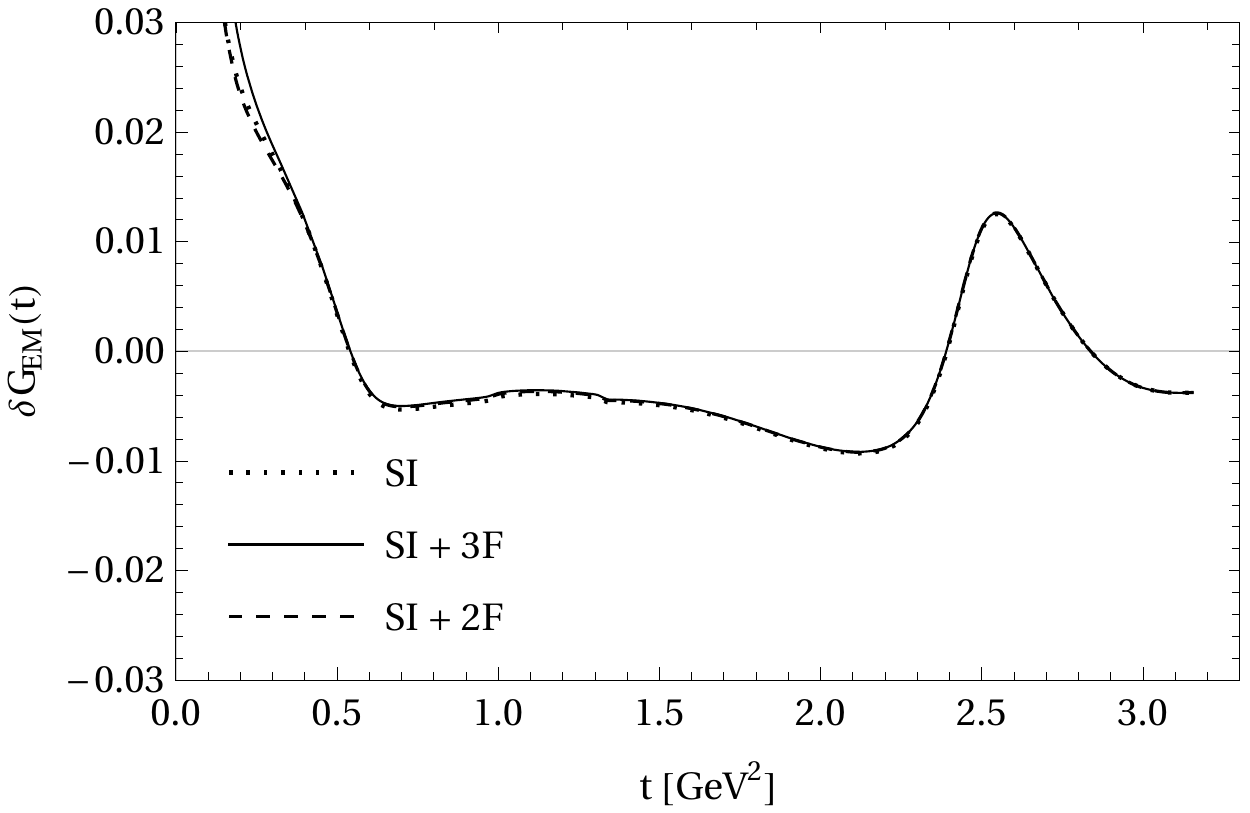}	
\caption{Correction factors $G_{\rm EM}^{(0)}(t)-1$ (left) and $\delta G_{\rm EM}(t)$ (right) 
to the differential decay rates of the $K^{-}\pi^0$, $\bar{K}^{0}\pi^{-}$, $K^{-}K^0$, 
and $\pi^-\pi^0$ modes from top to bottom.}
\label{RADCOR:fig2}
\end{figure}

The differential decay width can be written as
\begin{equation}
\label{eqGEM}
\begin{split}
\left.\frac{d\Gamma}{dt}\right\vert_{PP(\gamma)}&=
\frac{G_{F}^2\left\vert V_{uD}F_{+}(0)\right\vert^2 S_{\rm EW}m_\tau^3}{768\pi^3 t^3}
\left(1-\frac{t}{m_\tau^2}\right)^2\lambda^{1/2}(t,m_{-}^2,m_{0}^2)\\[1ex]
&\times\left[C_V^2\vert\tilde{F}_{+}(t)\vert^2\left(1+\frac{2t}{m_\tau^2}\right)
\lambda(t,m_{-}^2,m_{0}^2)+3 C_S^2\Delta_{-0}^2\vert\tilde{F}_{0}(t)\vert^2\right] G_{\rm EM}(t)\ ,
\end{split}
\end{equation}
where $G_{\rm EM}(t)$ encodes the electromagnetic corrections due to real and virtual photons.
For simplicity, we have split $G_{\rm EM}(t)$ in two parts: 
the leading Low approximation plus non-radiative contributions, $G_{\rm EM}^{(0)}(t)$, 
and the remainder,  $\delta G_{\rm EM}(t)$, which includes the SD contributions to the amplitude.
The predictions for both are shown in Fig.~\ref{RADCOR:fig2}. 

Integrating upon $t$, we get
\begin{equation}
\Gamma_{PP(\gamma)}=
\frac{G_F^2 S_{\rm EW}m_\tau^5}{96\pi^3}\left\vert V_{uD} F_{+}(0)\right\vert^2 I^{\tau}_{PP}
\left(1+\delta_{\rm EM}^{PP}\right)^2\ ,
\end{equation}
where 
\begin{equation}
\begin{split}   
I^\tau_{PP}=
&\frac{1}{8m_\tau^2}\int^{m_\tau^2}_{t_{\rm thr}}\frac{d t}{t^3}\left(1-\frac{t}{m_\tau^2}\right)^2
\lambda^{1/2}(t,m_{-}^2,m_{0}^2)\\[1ex]
&\times\left[C_V^2 \vert \tilde{F}_{+}(t)\vert^2\left(1+\frac{2t}{m_\tau^2}\right)
\lambda(t,m_{-}^2,m_{0}^2)+3 C_S^2 \Delta_{-0}^2\vert \tilde{F}_{0}(t)\vert^2\right]\ .
\end{split}
\end{equation}

The results for $\delta_{\rm EM}^{PP}$ are shown in Table \ref{RADCOR:tab1}, 
where the third and fourth columns correspond to the sum of the first three terms 
in Eq.~(\ref{eq:widthoverall}), and the last three columns to the fourth term in that equation.
The value in model $1$ for the $\bar{K}^0\pi^-$ channel agrees with the result in 
Ref.~\cite{Flores-Baez:2013eba}, 
which is related to our definition by 
$\delta_{\rm EM}^{\bar{K}^0\pi^-}=\delta^{\rm m.i.}_{\rm EM}/2\simeq -0.063\%$. 
Although our outcomes for the $(K\pi)^-$ modes agree within errors with those 
in Refs.~\cite{Flores-Baez:2013eba,Antonelli:2013usa}, 
the value in model $2$ (and also model $1$) for the $K^-\pi^0$ decay channel is larger 
than the $K^0\pi^-$ one, 
which is at odds with Ref.~\cite{Antonelli:2013usa}\footnote{Incidentally, 
our results would agree more closely swapping the numbers for 
$\delta_{\text{EM}}^{K^-\tau}\leftrightarrow\delta_{\text{EM}}^{\bar{K}^0\tau}$
in Ref.~\cite{Antonelli:2013usa}.}.


\begin{table}
\begin{center}
\small{\begin{tabular}{|c|c|cc|ccc|}
\hline
$\delta_{\rm EM}^{PP}$ & Ref.~\cite{Antonelli:2013usa}  &  \multicolumn{2}{c|}{$G_{\rm EM}^{(0)}(t)$} 
& \multicolumn{3}{c|}{$\delta G_{\rm EM}(t)$}\\
&  & Model 1 & Model 2
& SI & SI + 2F & SI + 3F\\
\hline
\hline
$K^-\pi^0$ & $-0.20(20)$ & $-0.019$ & $-0.137$ & $-0.001$ & $+0.006$ 
& $+0.010$\\
$\bar{K}^0\pi^-$ & $-0.15(20)$ & $-0.086$ & $-0.208$ & $-0.098$ & $-0.085$ 
& $-0.080$\\
$K^-K^0$ & $-$ & $-0.046$ & $-0.223$ & $-0.012$ & $+0.003$ & $+0.016$\\
$\pi^-\pi^0$ & $-$ & $-0.196$ & $-0.363$ & $-0.010$ & $-0.002$ & $+0.010$\\
\hline
\end{tabular}}
\caption{Electromagnetic corrections to hadronic $\tau$ decays in $\%$.}
\label{RADCOR:tab1}
\end{center}
\end{table}

The complete radiative corrections (that we always quote in $\%$) 
are obtained adding to the model 1/2 results,
which comprise the (negligible) $\mathcal{D}^{\rm IV/III}$ part, 
the $2F/3F$ contributions, which include the SI part. 
We explained before why we prefer the model 1 over the model 2 results. 
We will take the difference with respect to model 2 as an asymmetric error on the model 1 results. 
For the structure-dependent contributions, we consider the $3F$ results as our central values 
and the difference with respect to $2F$ as a symmetric error for our model-dependence. 
To be on the safe side, we will take twice this error as our corresponding uncertainty. 
Finally, we have to account for the uncertainty associated to the 
missing structure-dependent virtual-photon corrections. 
According to the results in Refs.~\cite{Arroyo-Urena:2021nil,Arroyo-Urena:2021dfe} 
for the one-meson tau decays, this contribution is of the same size as the structure-independent correction.
We will thus estimate its absolute value as the sum of the `Model 1' and `SI+3F' results 
in Table~\ref{RADCOR:tab1}, allowing it to have either sign. 
We will assign an additional $60\%$ uncertainty on it according to the results 
in Ref.~\cite{Arroyo-Urena:2021nil,Arroyo-Urena:2021dfe}. 
Proceeding this way, our main results are
\begin{equation}
\label{eq:MAINRESULT}
\begin{split}
&\delta_{\rm EM}^{K^-\pi^0}=-\left(0.009^{+0.010}_{-0.118}\right)\ ,
\qquad\delta_{\rm EM}^{\bar{K}^0\pi^-}=-\left(0.166^{+0.100}_{-0.157}\right)\ ,\\[1ex]
&\delta_{\rm EM}^{K^-K^0}=-\left(0.030^{+0.032}_{-0.180}\right)\ ,
\qquad\delta_{\rm EM}^{\pi^-\pi^0}=-\left(0.186^{+0.114}_{-0.203}\right)\ .
\end{split}
\end{equation}

We see that the model-independent contributions are responsible for 
the relatively large radiative corrections obtained for the $(\bar{K}/\pi)^0\pi^-$ modes. 
For the modes with a $K^-$, 
the dominant (asymmetric) uncertainty comes from the difference between the model 1/2 results, 
which is much larger than the deviation between the model-dependent $2F/3F$ values. 
Instead, for the modes with a $\pi^-$, 
the dominant uncertainty comes from the missing model-dependent virtual-photon corrections. 
Our results for the $\delta_{\rm EM}^{K^-\pi^0/\bar{K}^0\pi^-}$ agree with those in 
Ref.~\cite{Antonelli:2013usa}, and we reduce the uncertainty band by $\sim 45\%$ in the $K^-$ channel. 
We note that the estimate of the errors in this reference yields also an uncertainty band 
in agreement with ours for $\delta_{\rm EM}^{K^-K^0/\pi^-\pi^0}$ 
(our errors are smaller by a factor $\sim 2$ again in the $K^-K^0$ case)~\footnote{A 
former estimation of the $\pi^-\pi^0$ radiative corrections yielded 
$\delta_{\rm EM}^{\pi^-\pi^0}\sim-0.08\%$~\cite{Flores-Baez:2006yiq}, 
where the SD contributions were evaluated using a vector meson dominance model.}.
Although our $\delta_{\rm EM}^{K^-\pi^0}$ and $\delta_{\rm EM}^{\bar{K}^0\pi^-}$ seem to differ 
(the main reason being the scaling of the inner bremsstrahlung contribution with the inverse 
of the charged meson mass), 
the corresponding significance of their non-equality is only $\sim 0.7\sigma$, 
according to our uncertainties. 
Our radiative corrections in Eq.~(\ref{eq:MAINRESULT}) improve over previous analysis
(where, for instance, the structure-dependent corrections were not computed) 
and, as such, should be employed in precision analysis like, 
e.g., CKM unitarity or lepton universality tests \cite{HFLAV:2022esi}
and searches for non-standard interactions.

For completeness, we have also evaluated these corrections for the $K^-\eta^{(\prime)}$ modes. 
In the $G_{\rm EM}^{(0)}$ approximation and using the respective dominance of the 
vector (scalar) form factor \cite{Escribano:2013bca}, we obtain 
\begin{equation}
\delta_{\rm EM}^{K^-\eta}=-\left(0.026^{+0.029}_{-0.163}\right)\ ,
\qquad\delta_{\rm EM}^{K^-\eta^\prime}=-\left(0.304^{+0.422}_{-0.185}\right)\ ,
\end{equation}
where the uncertainty is saturated by the difference between the model 1/2 results in the $\eta$ channel 
and by the non yet computed virtual-photon structure-dependent corrections for the $\eta'$ mode. 
The  $K^-\eta^\prime$ decay mode is the only one (completely) dominated by the scalar form factor, 
which causes the relatively large magnitude of the corresponding radiative correction.

\section{Impact of radiative corrections on NP bounds}
\label{sec:Fits}
In this section, we update the results of Ref.~\cite{Gonzalez-Solis:2020jlh} 
concerning one- and two-meson tau decays including the radiative corrections computed 
in this paper~\footnote{One-meson channels were updated 
in Refs.~\cite{Arroyo-Urena:2021dfe,Arroyo-Urena:2021nil} 
using the improved radiative corrections calculated in those papers. 
It would also be interesting to reanalyse Refs.~\cite{Antonelli:2013usa,Cirigliano:2021yto} 
with our new radiative corrections.}. 
We recall briefly the main aspects here, 
but refer the reader to Ref.~\cite{Gonzalez-Solis:2020jlh} for details.

The low-energy effective Lagrangian describing the $\tau^-\to\bar{u}D\nu_\tau$ decays ($D=d,s$) 
can be written as \cite{Cirigliano:2009wk,Bhattacharya:2011qm}
\begin{equation}
\begin{split}
\mathcal{L}_{\rm eff}&=-\frac{G_FV_{uD}}{\sqrt{2}}
\Big[(1+\epsilon^\tau_L)\bar{\tau}\gamma_\mu(1-\gamma_5)\nu_\tau\cdot\bar{u}\gamma^\mu(1-\gamma_5)D
+\epsilon^\tau_R\bar{\tau}\gamma_\mu(1-\gamma_5)\nu_\tau\cdot\bar{u}\gamma^\mu(1+\gamma_5)D\\[1ex]
&+\bar{\tau}(1-\gamma_5)\nu_\tau\cdot\bar{u}(\epsilon^\tau_S-\epsilon^\tau_P\gamma_5)D
+\epsilon^\tau_T\bar{\tau}\sigma_{\mu\nu}(1-\gamma_5)\nu_\tau\cdot\bar{u}\sigma^{\mu\nu}(1-\gamma_5)D\Big]
+\mathrm{h.c.}\ ,   
\end{split}
\end{equation}
where $G_F$ corresponds to the Standard Model (SM) tree-level definition of the Fermi constant 
and the non-vanishing $\epsilon_i$ $(i=S,P,V,A,T)$ Wilson coefficients 
(assumed to be real for simplicity in what follows) determine the new physics. 
Beyond the SM, super allowed nuclear Fermi $\beta$ decays do not depend on $G_FV_{ud}$ 
but rather on $G_FV_{ud}(1+\epsilon^e_L+\epsilon^e_R)$, as it is accounted for in our analysis.
After using $\mathcal{L}_{\rm eff}$, the relevant (for two-meson decays) scalar, 
vector and tensor hadron matrix elements are computed using dispersion relations, 
nourished with experimental data, keeping track of the associated uncertainties.

We will discuss in the following 
the separate results for the strangeness-conserving and changing channels and, 
finally, those of a joint fit.
\subsection{\texorpdfstring{$\boldsymbol{\Delta S=0}$}{Lg}}
From $\tau^-\to\pi^-\nu_\tau(\gamma)$, we restrict 
\cite{Cirigliano:2018dyk,Gonzalez-Solis:2020jlh,Arroyo-Urena:2021dfe,Arroyo-Urena:2021nil,Cirigliano:2021yto}
\begin{equation}
    \epsilon_L^\tau-\epsilon_L^{e}-\epsilon_R^\tau-\epsilon_R^{e}-
    \frac{m_\pi^2}{m_\tau(m_u+m_d)}\epsilon_P^\tau =-(0.14\pm0.72)\cdot 10^{-2}\ ,
\end{equation}
using $f_\pi=130.2(8)\,\text{MeV}$~\cite{FlavourLatticeAveragingGroupFLAG:2021npn}, 
$\vert V_{ud}\vert=0.97373(31)$~\cite{ParticleDataGroup:2022pth}, $S_{\rm EW}=1.0232$~\cite{Marciano:1993sh}, 
masses and branching ratios from the PDG~\cite{ParticleDataGroup:2022pth}, 
and $\delta_{\text{em}}^{\tau\pi}=-0.24(56)\%$ from Ref.~\cite{Arroyo-Urena:2021nil}.

After performing a fit that includes one ($\pi$) and two meson ($\pi\pi$, $KK$) 
strangeness-conserving exclusive tau decays, the constraints for the non-standard interactions 
(at $\mu=2\text{ GeV}$ in the $\overline{\text{MS}}$ scheme) are 
\begin{equation}
\label{eq:DeltaS_0}
    \left(\begin{array}{c}
        \epsilon_L^\tau -\epsilon_L^e+\epsilon_R^\tau-\epsilon_R^e\\
         \epsilon_R^\tau +\frac{m_\pi^2}{2m_\tau(m_u+m_d)}\epsilon_P^\tau\\
         \epsilon^\tau_S\\
         \epsilon^\tau_T 
    \end{array}\right)=\left(\begin{array}{c}
         \,0.0 \pm 0.6 \,\, ^{+\,\,6.8}_{-\,\,6.4} \pm 0.1 \pm 1.7\,\, ^{+\,\,0.0}_{-\,\,0.2}\\
         \,0.1 \pm 0.5\,\, ^{+\,\,3.4}_{-\,\,3.3}\,\, ^{+\,\,0.0}_{-\,\,0.1} \pm 0.9 \,\, \pm 0.1\\
         10.3 \pm 0.5 \,\, ^{+\,\,\,1.2}_{-\,25.0} \,\, ^{+\,\,0.0}_{-\,\,0.1} \pm 0.9 \,\, 
         ^{+\,\,\,6.2}_{-\,22.4}\\
         0.4 \pm 0.2\,\, ^{+\,\,4.1}_{-\,\,4.4}\,\, ^{+\,\,0.0}_{-\,\,0.1} \pm 1.1\,\, 
         ^{+\,\,0.3}_{-\,\,0.2}
    \end{array}\right)\times 10^{-2}\ ,
\end{equation}
with $\chi^2/{\rm dof}\sim 0.8$, and the associated (statistical) correlation matrix is 
\begin{equation}
    \rho_{ij}=\left(
    \begin{array}{cccc}
        1 & 0.662 & -0.487 & -0.544\\
          & 1 & -0.323 & -0.360\\
          &   & 1 & 0.452\\
          &   &   & 1\\
    \end{array}
    \right)\ .
\end{equation}
The first error in Eq.~(\ref{eq:DeltaS_0}) is the statistical fit uncertainty, 
the second error comes from the theoretical uncertainty associated to the pion form factor, 
the third and fourth ones come from the quark masses and from the uncertainty related to 
the tensor form factor, respectively, 
and the last error is a systematic uncertainty coming from the radiative corrections to two-meson tau decays.

The results obtained are extremely consistent with those in Ref.~\cite{Gonzalez-Solis:2020jlh}. 
In general, the uncertainties induced by the radiative corrections are negligible with respect to others
except for $\epsilon^\tau_S$, where they are leading. 
This is interesting since this is the largest Wilson coefficient, so its compatibility with the SM 
gets even stronger upon including our new radiative corrections computed in this work.

\subsection{\texorpdfstring{$\boldsymbol{\vert\Delta S\vert=1}$}{Lg}}
From $\tau^-\to K^-\nu_\tau$, we bind
\cite{Cirigliano:2018dyk,Gonzalez-Solis:2020jlh,Arroyo-Urena:2021dfe,Arroyo-Urena:2021nil,Cirigliano:2021yto}
\begin{equation}
    \epsilon_L^\tau-\epsilon_L^{e}-\epsilon_R^\tau-\epsilon_R^{e}-\frac{m_K^2}{m_\tau(m_u+m_s)}\epsilon_P^\tau 
    =
    -(1.02\pm0.86)\cdot 10^{-2}\ ,
\end{equation}
using the lattice result $f_K=155.7(3)\,\text{MeV}$~\cite{FlavourLatticeAveragingGroupFLAG:2021npn}, 
$\vert V_{us}\vert=0.2243(8)$~\cite{ParticleDataGroup:2022pth}, 
masses and branching ratios from PDG~\cite{ParticleDataGroup:2022pth}, 
and $\delta_{\text{em}}^{\tau K}=-0.15(57)\%$ from Ref.~\cite{Arroyo-Urena:2021nil}.

The bounds for the non-SM effective couplings resulting from a fit to one ($K$) 
and two meson ($K\pi,K\eta$) strangeness-changing transitions (at $\mu=2\ \text{GeV}$ 
in the $\overline{\text{MS}}$ scheme) 
are 
\begin{equation}
\label{eq:DeltaS_1}
    \left(\begin{array}{c}
        \epsilon_L^\tau -\epsilon_L^e+\epsilon_R^\tau-\epsilon_R^e\\
         \epsilon_R^\tau +\frac{m_K^2}{2m_\tau(m_u+m_s)}\epsilon_P^\tau\\
         \epsilon^\tau_S\\
         \epsilon^\tau_T 
    \end{array}\right)=\left(\begin{array}{c}
         0.4 \pm 1.5 \pm 0.4\,\, ^{+\,\,0.1}_{-\,\,0.0}\\
         0.7 \pm 0.9 \pm 0.2\,\, ^{+\,\,0.1}_{-\,\,0.0}\\
         0.8 \pm 0.9 \pm 0.2\,\, ^{+\,\,0.0}_{-\,\,0.1}\\
         0.5 \pm 0.7 \pm 0.4 \pm 0.0
    \end{array}\right)\times 10^{-2}\ ,
\end{equation}
with $\chi^2/{\rm dof}\sim 0.9$, and
\begin{equation}
    \rho_{ij}=\left(
    \begin{array}{cccc}
        1 & 0.874 & -0.149 & 0.463\\
          & 1 & -0.130 & 0.404\\
          &   & 1 & -0.057\\
          &   &   & 1\\
    \end{array}
    \right)\ .
\end{equation}
In this case, the first error in Eq. (\ref{eq:DeltaS_1}) is the statistical fit uncertainty, 
the second one is due to the tensor form factor, 
and the last uncertainty is associated to the radiative corrections to two-meson tau decays. 
The uncertainties related to the kaon vector form factor and the quark masses are negligible. 
These results comply nicely with those in Ref.~\cite{Gonzalez-Solis:2020jlh}, 
and the uncertainty induced by the radiative corrections is in all couplings negligible.

\subsection{\texorpdfstring{$\boldsymbol{\Delta S=0}$}{Lg} and 
\texorpdfstring{$\boldsymbol{\vert\Delta S\vert=1}$}{Lg} joint fit}
Our previous fits to the strangeness-conserving and changing channels 
were not able to separate $\epsilon^\tau_R$ and $\epsilon^\tau_P$~\footnote{This 
was achieved by combining inclusive and exclusive information 
in Refs.~\cite{Cirigliano:2018dyk, Cirigliano:2021yto}.}. 
In Ref.~\cite{Gonzalez-Solis:2020jlh}, 
we attained this within Minimal Flavour Violation \cite{DAmbrosio:2002vsn} (MFV)~\footnote{MFV 
assumes that flavour mixing within the Standard Model Effective Field Theory (SMEFT) 
is aligned with the SM one (here, in particular, in the quark sector). 
This generally allows for orders of magnitude smaller new physics scales in precision flavour tests.}, 
as we will do here. 
The CKM matrix elements used in this analysis are obtained from the correlation between $\vert V_{ud}\vert$ 
and $\vert V_{us}\vert$, $\vert V_{us}/V_{ud}\vert=0.2313(5)$, and $\vert V_{us}\vert =0.2232(6)$ from 
Ref.~\cite{FlavourLatticeAveragingGroupFLAG:2021npn}, 
which correspond to the region described by the red ellipse in Fig.~10 of 
Ref.~\cite{FlavourLatticeAveragingGroupFLAG:2021npn}.

Performing thus a joint fit that includes both one and two meson strangeness-conserving 
and strangeness-changing tau decays (within MFV), 
the limits for the NP effective couplings (at $\mu=2\ \text{GeV}$ in the $\overline{\text{MS}}$ scheme) are
\begin{equation}\label{eq:Global_fit}
    \left(\begin{array}{c}
        \epsilon_L^\tau -\epsilon_L^e+\epsilon_R^\tau-\epsilon_R^e\\
         \epsilon_R^\tau\\
         \epsilon_P^\tau\\
         \epsilon^\tau_S\\
         \epsilon^\tau_T 
    \end{array}\right)=\left(
    \begin{array}{c}
         \quad2.7 \pm 0.5 \,\,^{+\,\,\,2.3}_{-\,\,\,3.1} \,\,^{+\,\,\,0.4}_{-\,\,\,0.5} \pm 0.0 \pm 0.3
         \,\,^{+\,\,\,0.0}_{-\,\,\,1.3} \pm 0.0\\
         \quad7.1 \pm 4.7 \,\,^{+\,\,\,1.2}_{-\,\,\,1.6} \pm 0.9 \pm 1.8 \pm 
         0.2\,\,^{+\,\,12.3}_{-\,\,\,3.6} \pm 0.0\\
         -7.7 \pm 6.1 \pm 0.0 \,\,^{+\,\,\,1.3}_{-\,\,\,1.2}  \,\,^{+\,\,\,2.4}_{-\,\,\,2.3} \pm 
         0.0\,\,^{+\,\,\,4.1}_{-\,\,17.0} \pm 0.0\\
         \quad5.3 \,\,^{+\,\,\,0.6}_{-\,\,\,0.7} \,\,^{+\,\,\,2.0}_{-\,\,14.9} 
         \,\,^{+\,\,\,0.1}_{-\,\,\,0.0} \pm 0.0 \pm 0.1 
         \,\,^{+\,\,\,0.1}_{-\,\,15.9} \,\,^{+\,\,\,0.1}_{-\,\,\,0.0}\\
         -0.2 \pm 0.2 \,\,^{+\,\,\,3.6}_{-\,\,\,2.9} \,\,^{+\,\,\,0.1}_{-\,\,\,0.0} \pm 0.0 \pm 0.4
         \,\,^{+\,\,\,0.5}_{-\,\,\,0.0} \,\,^{+\,\,\,0.2}_{-\,\,\,0.0}
    \end{array}
    \right)\times 10^{-2}\ ,
\end{equation}
with $\chi^2/{\rm dof}\sim 1.5$, and
\begin{equation}
    \rho_{ij}=\left(
    \begin{array}{ccccc}
        1 & 0.056 & 0.000 & -0.270 & -0.402\\
          & 1 & -0.997 & -0.015 & -0.023\\
          &   & 1 & 0.000 & 0.000\\
          &   &   & 1 & 0.235\\
          &   &   &   & 1 
    \end{array}
    \right)\ .
\end{equation}
Now, the first error in Eq.~(\ref{eq:Global_fit}) corresponds again to the statistical fit uncertainty, 
the second one comes from the uncertainty on the pion form factor, 
the third error is related to the CKM matrix elements $\vert V_{ud}\vert$ and $\vert V_{us}\vert$, 
the fourth one comes from the radiative corrections $\delta^{\tau \pi}_{\text{em}}$ 
and $\delta^{\tau K}_{\text{em}}$, 
the fifth error is associated to the tensor form factor, 
the sixth error comes from the uncertainty of the quark masses, 
and the last one is due to the radiative corrections to two-meson tau decays.

These results accord with Ref.~\cite{Gonzalez-Solis:2020jlh} closely. 
As seen, in this joint fit the uncertainties induced by radiative corrections are always subdominant.

\section{Conclusions}
\label{sec:Concl}
Radiative corrections to the one-meson tau decays have been employed in CKM unitarity,
lepton universality and non-standard interactions tests. 
The corresponding results for the di-pion tau decays allowed tau-based computations 
of the leading-order piece of the hadronic vacuum polarization part of the muon $g-2$. 
Even though the model-independent part of these corrections was available for the $K\pi$ modes, 
the structure-dependent one remained to be calculated. 
We have performed the first step towards bridging this gap, 
by computing the real-photon structure-dependent radiative correction factors with reduced uncertainties, 
and will continue with the required virtual-photon model-dependent corrections in a forthcoming work. 
For completeness, we also quoted our numbers for the $PP$ ($P=\pi,K$) modes 
and estimated them for the $K\eta^{(\prime)}$ cases.

We recapitulate our main results in Eq.~(\ref{eq:MAINRESULT}):
\begin{equation}
\begin{split}
&\delta_{\rm EM}^{K^-\pi^0}=-\left(0.009^{+0.010}_{-0.118}\right)\%\ ,
\qquad\delta_{\rm EM}^{\bar{K}^0\pi^-}=-\left(0.166^{+0.100}_{-0.157}\right)\%\ ,\\[1ex]
&\delta_{\rm EM}^{K^-K^0}=-\left(0.030^{+0.032}_{-0.180}\right)\%\ ,
\qquad\delta_{\rm EM}^{\pi^-\pi^0}=-\left(0.186^{+0.114}_{-0.203}\right)\%\ ,\nonumber
\end{split}
\end{equation}
which reduce previous uncertainties by a factor of $\sim2$ for the $K^-$ modes 
and have similar errors to those quoted for the $\bar{K}^0\pi^-$ channel.

We have put forward the importance of the factorization model for the radiative corrections, 
which saturates the uncertainties in Eq.~(\ref{eq:MAINRESULT}) for the $K^-$ channels.
Analogous relevance shall be found in the radiative corrections for processes including 
diverse final states with hadrons 
(if the resonance regime is allowed by phase space), which calls for further devoted studies. 
While lattice QCD obtains these complicated radiative corrections, 
a deeper understanding of their factorization will probe key in increasing the reach of new physics searches 
through processes including hadrons.

We have finally illustrated the application of our results updating our fits of non-SM interactions 
in di-meson tau decays (also one-meson channels were included in the fits), 
finding results compatible with our earlier work, 
where these corrections and their uncertainties were neglected. 
As interesting outlooks, the reanalysis of Refs.~\cite{Miranda:2018cpf,Rendon:2019awg} 
(looking for NP in $\tau^-\to\pi^-\pi^0\nu_\tau$ and $\tau^-\to(K\pi)^-\nu_\tau$ decays, respectively), 
of inclusive and exclusive semileptonic tau decays\cite{Cirigliano:2018dyk, Cirigliano:2021yto}
and of the impact of kaon data on strangeness-violating tau decays \cite{Antonelli:2013usa}, 
are among the important applications of our results that we can envisage, 
which can be incorporated into the HFLAV analysis \cite{HeavyFlavorAveragingGroup:2022wzx}. 
All these studies will benefit from our future results for the 
virtual-photon structure-dependent radiative corrections.

\begin{acknowledgements}
The work of R.~E.~has been supported by 
the European Union’s Horizon 2020 Research and Innovation Programme under grant 824093 
(H2020-INFRAIA- 2018-1),
the Ministerio de Ciencia e Innovación under grant PID2020-112965GB-I00, and by 
the Departament de Recerca i Universitats from Generalitat de Catalunya 
to the Grup de Recerca `Grup de F\'isica Te\`orica UAB/IFAE' (Codi: 2021 SGR 00649).
IFAE is partially funded by the CERCA program of the Generalitat de Catalunya.
J.~A.~M.~is also supported by MICINN with funding from European Union NextGenerationEU (PRTR-C17.I1) 
and acknowledges Conacyt for his PhD scholarship. 
P.~R.~was partly funded by 
Conacyt’s project within ‘Paradigmas y Controversias de la Ciencia 2022’, number 319395, and by 
Cátedra Marcos Moshinsky (Fundación Marcos Moshinsky) 2020.
\end{acknowledgements}

\appendix
\section{\texorpdfstring{$\boldsymbol{K_{\ell 3}}$}{Lg} decays}
\label{AppKl3}
The most general amplitude for the $K(p_K)\to\pi(p_\pi)\ell(P)\nu_\ell(q')\gamma(k)$ decays 
that complies with Lorentz invariance and the discrete symmetries of QCD can be written as 
\begin{equation}
\begin{split}
    \mathcal{M}&=\frac{e G_F V^{*}_{us}}{\sqrt{2}}\epsilon^{*}_\mu
    \bigg[\frac{H_\nu(-p_K,p_\pi)}{k^2+2k\cdot P}
    \bar{u}(q')\gamma^\nu (1-\gamma^5)(m_\ell -\slashed{P}-\slashed{k})\gamma^\mu v(P)\\[1ex]
    &+(V^{\mu\nu}-A^{\mu\nu})\bar{u}(q')\gamma_\nu (1-\gamma^5) v(P)\bigg]\ ,
\end{split}
\end{equation}
where 
\begin{equation}
\begin{split}
    H^\nu(-p_K,p_\pi)&\equiv
    \left\langle\pi(p_\pi)\left\vert\bar{s}\gamma^\nu u\right\vert K(p_K)\right\rangle\\[1ex]
    &=-C_V F_{+}(t)\left[(p_K+p_\pi)^\nu-\frac{\Delta_{K\pi}}{t}(p_K-p_\pi)^\nu\right]
    -C_S\frac{\Delta_{K\pi}}{t}(p_K-p_\pi)^\nu F_{0}(t)\ ,
\end{split}
\end{equation}
with $t=(p_K-p_\pi)^2$.

The structure-independent term is given by
\begin{equation}
\begin{split}
V^{\mu\nu}_{\rm SI}&=\frac{H^\nu(-p_K+k,p_\pi)(k-2p_K)^\mu}{k^2-2k\cdot p_K}
+\left\{-C_V F_{+}(t^\prime)-\frac{\Delta_{K\pi}}{t^\prime}
\left[C_S F_{0}(t^\prime)-C_V F_{+}(t^\prime)\right]\right\} g^{\mu\nu}\\[1ex]
&+C_V\frac{F_{+}(t^\prime)-F_{+}(t)}{k\cdot (p_K-p_\pi)}
\left[(p_K+p_\pi)^\nu-\frac{\Delta_{K\pi}}{t}(p_K-p_\pi)^\nu\right] (p_K-p_\pi)^\mu\\[1ex]
&+\frac{\Delta_{K\pi}}{tt^\prime}\left\{
2\left[C_S F_{0}(t^\prime)-C_V F_{+}(t^\prime)\right]+\frac{C_S t^\prime}{k\cdot(p_K-p_\pi)}
\left[F_{0}(t^\prime)-F_{0}(t)\right]\right\}\\[1ex]
&\times (p_K-p_\pi)^\mu (p_K-p_\pi)^\nu\ ,
\end{split}
\end{equation}
where $C_V^{K^-\pi^0}=C_S^{K^-\pi^0}=1/\sqrt{2}$ for $K^+\to\pi^0\ell^+\nu_\ell\gamma$, and 
\begin{equation}
\begin{split}
V^{\mu\nu}_{\rm SI}&=\frac{H^\nu(-p_K,p_\pi+k)(k+2p_\pi)^\mu}{k^2+2k\cdot p_\pi}
+\left\{C_V F_{+}(t^\prime)-\frac{\Delta_{K\pi}}{t^\prime}
\left[C_S F_{0}(t^\prime)-C_V F_{+}(t^\prime)\right]\right\} g^{\mu\nu}\\[1ex]
&+C_V\frac{F_{+}(t^\prime)-F_{+}(t)}{k\cdot (p_K-p_\pi)}
\left[(p_K+p_\pi)^\nu-\frac{\Delta_{K\pi}}{t}(p_K-p_\pi)^\nu\right] (p_K-p_\pi)^\mu\\[1ex]
&+\frac{\Delta_{K\pi}}{tt^\prime}\left\{ 
2\left[C_S F_{0}(t^\prime)-C_V F_{+}(t^\prime)\right]+\frac{C_S t^\prime}{k\cdot(p_K-p_\pi)}
\left[F_{0}(t^\prime)-F_{0}(t)\right]\right\}\\[1ex] 
&\times (p_K-p_\pi)^\mu (p_K-p_\pi)^\nu\ ,
\end{split}
\end{equation}
where $C_V^{\bar{K}^0\pi^-}=C_S^{\bar{K}^0\pi^-}=1$ for $K^0\to \pi^-\ell^+\nu_\ell\gamma$,
both with $t^\prime\equiv(p_K-p_\pi-k)^2$. 
All these expressions are obtained from Eqs.~(\ref{eq:tau_decay})--(\ref{eq3:VSI}) by replacing 
$\lbrace^{p_-}_{p_0}\rbrace\to\lbrace^{-p_K}_{\,\,\,\,\, p_\pi}\rbrace$, 
$\Delta_{-0}\to\Delta_{K\pi}$, $P\to -P$ 
and $m_\tau\to m_\ell$ for $K^+\to\pi^0\ell^+\nu_\ell\gamma$, 
and $\lbrace^{p_-}_{p_0}\rbrace\to\lbrace_{-p_K}^{\,\,\,\,\, p_\pi}\rbrace$, $\Delta_{-0}\to -\Delta_{K\pi}$, 
$C_{V,S}\to-C_{V,S}$ $P\to -P$ and $m_\tau\to m_\ell$ for $K^0\to \pi^-\ell^+\nu_\ell\gamma$. 
The structure-dependent terms are analogous to those in Eqs.~(\ref{eq1:VSD}) and (\ref{eq1:ASD1}).

At $\mathcal{O}(p^0)$, we get
\begin{equation}
    V_{\text{SI}}^{\mu\nu}=-C_{K^+}\frac{p_K^\mu}{k\cdot p_K}(p_K+p_\pi)^\nu 
    -C_{K^+}\left(g^{\mu\nu}-\frac{p_K^\mu k^\nu}{k\cdot p_K}\right)\ ,
\end{equation}
for $K^+\to\pi^0$, and 
\begin{equation}
    V_{\text{SI}}^{\mu\nu}=-C_{K^0}\frac{p_\pi^\mu}{k\cdot p_\pi}(p_K+p_\pi)^\nu 
    +C_{K^0}\left(g^{\mu\nu}-\frac{p_\pi^\mu k^\nu}{k\cdot p_\pi}\right)\ ,
\end{equation}
for $K^0\to\pi^-$, where $C_K=C_S=C_V$.
Thus, the overall amplitude at $\mathcal{O}(p^0)$ is given by
\begin{equation}
    \mathcal{M}_\gamma=
    \frac{e G_F }{\sqrt{2}}V_{us}^{*}C_{K^+}\bar{u}(q')(1+\gamma^5)(2\slashed{p}_\pi -m_\ell)
    \left(\frac{\epsilon\cdot P}{k\cdot P}-
    \frac{\epsilon\cdot p_K}{k\cdot p_K}+
    \frac{\slashed{k}\slashed{\epsilon}}{2 k\cdot P}\right) v(P)\ ,
\end{equation}
and
\begin{equation}
    \mathcal{M}_\gamma=
    \frac{e G_F }{\sqrt{2}}V_{us}^{*}C_{K^0}\bar{u}(q')(1+\gamma^5)(2\slashed{p}_K+m_\ell)
    \left(\frac{\epsilon\cdot P}{k\cdot P}-
    \frac{\epsilon\cdot p_\pi}{k\cdot p_\pi}+
    \frac{\slashed{k}\slashed{\epsilon}}{2 k\cdot P}\right) v(P)\ ,
\end{equation}
respectively. These two expressions agree with the Eqs.~(13) and (14) in Ref.~\cite{Cirigliano:2008wn}.
To this order, $V^{\mu\nu}_{\text{SD}}$ and $A^{\mu\nu}_{\text{SD}}$, which are $\mathcal{O}(p^2)$,
can be neglected.

In the Low limit, we obtain
\begin{equation}
    \mathcal{M}_\gamma=\frac{eG_FV_{us}^{*}}{\sqrt{2}}\bar{u}(q')\gamma_\nu (1-\gamma^5) v(P)
    H^\nu(-p_K,p_\pi)\left(\frac{\epsilon\cdot p_{+}}{k\cdot p_{+}}-
    \frac{\epsilon\cdot P}{k\cdot P}\right)\ ,
\end{equation}
where the subscript `$+$' refers to the charged meson.
The spin-averaged squared matrix element is then given by
\begin{equation}
\begin{split}
    \overline{\left\vert\mathcal{M}_\gamma \right\vert^2}&=
    4 C_K^2 e^2 G_F^2 \left\vert V_{us}\right\vert^2 S_{\rm EW}^K
    \left\lbrace\left[\frac{m_\ell^2}{2}(t-m_\ell^2)+2m_{K}^2m_{\pi}^2+
    2u(m_\ell^2-t+m_{K}^2+m_{\pi}^2)-2u^2\right.\right.\\[1ex]
    &-\frac{\Delta_{K\pi}}{t}m_\ell^2(2u+t-m_\ell^2-2m_{\pi}^2)+
    \frac{\Delta_{K\pi}^2}{t^2}\frac{m_\ell^2}{2}(t-m_\ell^2)\left.\right]
    \left\vert F_{+}(t)\right\vert^2\\[1ex]
    &+\frac{\Delta_{K\pi}\,m_\ell^2}{t}\left[2u+t-m_\ell^2-2m_\pi^2+
    \frac{\Delta_{K\pi}}{t}(m_\ell^2-t)\right]\mathrm{Re}\left[F_{+}(t)F_{0}^{*}(t)\right]\left.\right.\\[1ex]
    &+\frac{\Delta_{K\pi}^2\,m_\ell^2}{2t^2}\left(t-m_\ell^2\right)
    \left\vert F_{0}(t)\right\vert^2\left.\right\rbrace \sum_{\gamma\ \rm pols.}
    \left\vert\frac{p_{-}\cdot\epsilon}{p_{-}\cdot k}-\frac{P\cdot \epsilon}{P\cdot k}\right\vert^2 
    +\mathcal{O}(k^0)\ ,
\end{split}
\end{equation}
where $u=(p_K-P)^2$.
The last expression can also be written in terms of $f_{+/-}(t)$,
\begin{equation}
\label{Appx:Kaon_amplitude}
    \overline{\left\vert\mathcal{M}_\gamma \right\vert^2}=
    2 C_K^2 m_K^4 e^2 G_F^2\left\vert V_{us}\right\vert^2 S_{\rm EW}^K\rho^{(0)}(y,z)
    \sum_{\gamma\ \rm pols.}\left\vert\frac{p_{-}\cdot\epsilon}{p_{-}\cdot k}-
    \frac{P\cdot\epsilon}{P\cdot k}\right\vert^2+\mathcal{O}(k^0)\ ,
\end{equation}
where
\begin{equation}
    \rho^{(0)}(y,z)= 
    A_1^{(0)}(y,z)\left\vert f_{+}(t)\right\vert^2+
    A_2^{(0)}(y,z)\mathrm{Re}\left[f_{+}(t)f_{-}^{*}(t)\right]+
    A_3^{(0)}(y,z)\left\vert f_{-}(t)\right\vert^2\ ,
\end{equation}
and the kinematical densities are 
\begin{subequations}
\begin{align}
    A_1^{(0)}&=4(y+z-1)(1-y)+r_\ell (4y+3z-3)-4r_\pi+r_\ell(r_\pi-r_\ell)\ ,\\[1ex]
    A_2^{(0)}&=2r_\ell(r_\ell-r_\pi-2y-z+3)\ ,\\[1ex]
    A_3^{(0)}&=r_\ell(r_\pi -r_\ell+1-z)\ ,
\end{align}
\end{subequations}
with 
\begin{equation}
    z=\frac{2p_\pi\cdot p_K}{m_K^2}=\frac{2E_\pi}{m_k}\ ,\qquad 
    y=\frac{2p_K\cdot p_\ell}{m_K^2}=\frac{2E_\ell}{m_k}\ ,
\end{equation}
$r_\ell=(m_\ell/m_K)^2$, and $r_\pi=(m_\pi/m_K)^2$. 
Here, $E_\pi$ ($E_\ell$) is the energy of the pion (charged lepton) in the kaon rest frame. 
The expression in Eq.~(\ref{Appx:Kaon_amplitude}) can be compared directly with the results in 
Refs.~\cite{Cirigliano:2001mk,Cirigliano:2004pv,Cirigliano:2008wn}.

The $K\to\pi\ell\nu_\ell$ decay width without radiative corrections \cite{Antonelli:2013usa}
is given by 
\begin{equation}
    \Gamma(K\to\pi\ell\nu_\ell)=
    \frac{G_F^2 m_K^5}{192\pi^3}S_{\rm EW}^K\vert V_{us}\vert^2\vert F_{+}(0)\vert I^\ell_K\ ,
\end{equation}
where
\begin{equation}
\begin{split}
    I^\ell_K&=\int_{m_\ell^2}^{t_{\text{max}}}dt \frac{1}{m_K^8}\lambda^{3/2}(t,m_K^2,m_\pi^2)
    \left(1-\frac{m_\ell^2}{t}\right)^2\left(1+\frac{m_\ell^2}{2t}\right)\\[1ex]
    &\times\left[C_V^2\vert \tilde{F}_+(t)\vert^2+
    \frac{3\Delta_{K\pi}^2 m_\ell^2}{(2t+m_\ell^2)\lambda(t,m_K^2,m_\pi^2)}
    C_S\vert \tilde{F}_0(t)\vert^2\right]\ ,
\end{split}
\end{equation}
and $t_{\text{max}}=(m_K-m_\pi)^2$.

\section{Virtual corrections to the hadronic tau decays}
\label{Appx:RC}
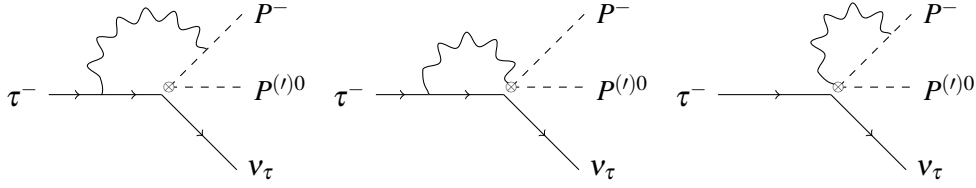
\begin{figure}
\begin{tikzpicture}
	\draw[gauge] (0.6,0.5) arc (30:205:0.781);
	\draw[fill,white] (-0.75,-0.2) circle [radius=0.1];
	\draw[fermion] (-1.5,-0.1) node[left] {\small$\tau^{-}$} --(-0.75,-0.1);
	\draw[fermion] (-0.75,-0.1)--(0,-0.1);
	\draw[scalar] (0.1,0) -- (1.1,1) node[right] {\small$P^-$};
	\draw[scalar] (0.1,0) -- (1.1,0) node[right] {\small$P^{(\prime)0}$};
	\draw[fermion] (0,-0.1) -- (1,-1.1) node[right] {\small$\nu_\tau$};
	\draw[fill,white] (0.1,0) circle [radius=0.06];
	\node at (0.1,0) {\tiny$\otimes$};
\end{tikzpicture}
\begin{tikzpicture}
	\draw[gauge] (0.1,0) arc (0:190:0.55);
	\draw[fermion] (-1.7,-0.1) node[left] {\small$\tau^{-}$} --(-1,-0.1);
	\draw[fermion] (-1,-0.1)--(0,-0.1);
	\draw[scalar] (0.1,0) -- (1.1,1) node[right] {\small$P^-$};
	\draw[scalar] (0.1,0) -- (1.1,0) node[right] {\small$P^{(\prime)0}$};
	\draw[fermion] (0,-0.1) -- (1,-1.1) node[right] {\small$\nu_\tau$};
	\draw[fill,white] (0.1,0) circle [radius=0.06];
	\node at (0.1,0) {\tiny$\otimes$};
\end{tikzpicture}
\begin{tikzpicture}
	\draw[gauge] (0.8,0.7) arc (30:247:0.5);
	\draw[fermion] (-1.5,-0.1) node[left] {\small$\tau^{-}$} --(0,-0.1);
	\draw[scalar] (0.1,0) -- (1.1,1) node[right] {\small$P^-$};
	\draw[scalar] (0.1,0) -- (1.1,0) node[right] {\small$P^{(\prime)0}$};
	\draw[fermion] (0,-0.1) -- (1,-1.1) node[right] {\small$\nu_\tau$};
	\draw[fill,white] (0.1,0) circle [radius=0.06];
	\node at (0.1,0) {\tiny$\otimes$};
\end{tikzpicture}
\centering			
\caption{Photon loop diagrams that contribute to the 
$\tau^-\to (P_1P_2)^-\nu_\tau$ decays.}
\label{RADCOR:Vitualphotons}
\end{figure}

The radiative corrections to the $\tau^-\to (P_1P_2)^{-}\nu_\tau$ decays
at $\mathcal{O}(p^2)$ in ChPT \cite{Weinberg:1978kz,Gasser:1983yg,Gasser:1984gg} 
are depicted in Fig.~\ref{RADCOR:Vitualphotons}.
The overall contribution is given by~\cite{Cirigliano:2001er}
\begin{equation}
\frac{\delta H^\mu(t,u)}{C_V}=
\delta f_{+}(u)(p_{1}-p_{0})^\mu + 
\delta f_{-}(u)(p_{1}+p_{0})^\mu\ ,
\end{equation}
where
\begin{equation}
\begin{split}
\delta f_{+}(u)&=\frac{\alpha}{4\pi}
\left[2+\frac{1}{\epsilon}-\gamma_E+\log4\pi-\log\frac{m_\tau^2}{\mu^2}+
(u-m_{-}^2)\mathcal{A}(u)+(u-m_{-}^2-m_{\tau}^2)\mathcal{B}(u)\right.\\[1ex]
&+2(m_{-}^2+m_{\tau}^2-u)\mathcal{C}(u,M_\gamma)
+2\log\left(\frac{m_{-}m_\tau}{M_\gamma^2}\right)\left.\right] ,
\end{split}
\end{equation}
\begin{equation}
\begin{split}
\delta f_{-}(u)&=\frac{\alpha}{4\pi}
\left[-5-3\left(\frac{1}{\epsilon}-\gamma_E+\log4\pi\right)+\log\frac{m_{-}^2}{\mu^2}+
2\log\frac{m_\tau^2}{\mu^2}+(3u+m_{-}^2-2m_{\tau}^2)\mathcal{A}(u)\right.\\[1ex]
&+(u+m_{-}^2-m_{\tau}^2)\mathcal{B}(u)\left.\right]\ ,
\end{split}
\end{equation}
with
\begin{equation}
\mathcal{A}(u)=
\frac{1}{u}\left(-\frac{1}{2}\log r_\tau + \frac{2-y}{\sqrt{r_\tau}}\frac{x}{1-x^2}\log x\right)\ ,
\end{equation}
\begin{equation}
\mathcal{B}(u)=
\frac{1}{u}\left(\frac{1}{2}\log r_\tau + \frac{2r_\tau - y}{\sqrt{r_\tau}}
\frac{x}{1-x^2}\log x\right)\ ,
\end{equation}
\begin{equation}
\begin{split}
\mathcal{C}(u,M_\gamma)&=
\frac{1}{m_\tau m_{-}}\frac{x}{1-x^2}\bigg[-\frac{1}{2}\log^2 x+
2\log x\log(1-x^2)-\frac{\pi^2}{6}+\frac{1}{8}\log^2 r_\tau\\[1ex]
&+\mathrm{Li}_{2}(x^2)+\mathrm{Li}_{2}\left(1-\frac{x}{\sqrt{r_\tau}}\right)+
\mathrm{Li}_{2}(1-x\sqrt{r_\tau})-\log x\log\left(\frac{M_\gamma^2}{m_\tau m_{-}}\right)\bigg]\ ,
\end{split}
\end{equation}
and $C_V^{\pi\pi, KK, K^{-}\pi^{0}, K^{0}\pi^{-}}=
\left(\sqrt{2},-1,\frac{1}{\sqrt{2}},-1\right)$.
Here, $\mathcal{A}(u)$, $\mathcal{B}(u)$ and $\mathcal{C}(u,M_\gamma)$ 
are written in terms of the variables
\begin{equation}
r_\tau =\frac{m_\tau^2}{m_{-}^2}\ ,\qquad
y=1+r_\tau-\frac{u}{m_{-}^2}\ ,\qquad 
x=\frac{1}{2\sqrt{r_\tau}}\left(y-\sqrt{y^2-4r_\tau}\right)\ ,
\end{equation}
and the dilogarithm
\begin{equation}
\mathrm{Li}_{2}(x)=-\int_{0}^{1}\frac{dt}{t}\log(1-xt)\ .
\end{equation}
 
The radiative corrections to these decays induce a dependence in the $u$ variable.
From a comparison with the results in Ref.~\cite{Antonelli:2013usa}, 
we get the following relation
\begin{equation}
\begin{split}
\delta\bar{f}_{+}(u)&=\frac{\alpha}{4\pi}\frac{1}{f_{+}(0)}
\left[\Gamma_1(u,m_\tau^2,m_{-}^2)+\Gamma_2(u,m_\tau^2,m_{-}^2)\right]+\cdots\\[1ex]
&=\frac{\alpha}{4\pi}\frac{1}{f_{+}(0)}
\left[(u-m_{-}^2)\mathcal{A}(u)+(u-m_{-}^2-m_{\tau}^2)\mathcal{B}(u)\right]+\cdots\ ,
\end{split}
\end{equation}
and 
\begin{equation}
\begin{split}
\delta \bar{f}_{-}(u)&=\frac{\alpha}{4\pi}\frac{1}{f_{+}(0)}
\left[\Gamma_1(u,m_\tau^2,m_{-}^2)-\Gamma_2(u,m_\tau^2,m_{-}^2)\right]+\cdots\\[1ex]
&=\frac{\alpha}{4\pi}\frac{1}{f_{+}(0)}
\left[ (3u+m_{-}^2-2m_{\tau}^2)\mathcal{A}(u)+(u+m_{-}^2-m_{\tau}^2)\mathcal{B}(u)\right]+\cdots\ .
\end{split}
\end{equation}

\section{\texorpdfstring{$\boldsymbol{\tau^{-}\to (P_1P_2)^{-}\nu_\tau}$}{Lg} decays}
\label{AppC}
After the inclusion of the virtual-photon radiative corrections to the form factor 
in Sect.~\ref{Appx:RC},
the amplitude for the $\tau^{-}(P)\to P^{-}_1(p_{-})P^{0}_2(p_{0})\nu_\tau(q')$ decays is given by
\begin{equation}
\mathcal{M}_{0}=\frac{G_{F} V_{uD}\sqrt{S_{\rm EW}}}{\sqrt{2}}H_\nu(p_{-},p_{0})
\bar{u}(q')\gamma^\nu (1-\gamma^5) u(P)\ .
\end{equation}
Thus, the spin-averaged squared amplitude follows 
\begin{equation}
\begin{split}
\overline{\left\vert\mathcal{M}_{0}\right\vert^2}&=
2 G_F^2\left\vert V_{uD}\right\vert^2 S_{\rm EW}
\bigg\lbrace C_{S}^2\left\vert F_{0}(t,u)\right\vert^2 D_{0}^{P^-P^0}(t,u)+C_{S}C_{V}\mathrm{Re}\left[F_{+}(t,u)F_{0}^{*}(t,u)\right] D_{+0}^{P^-P^0}(t,u)\\[1ex]
&+C_{V}^2\left\vert F_{+}(t,u)\right\vert^2 D_{+}^{P^-P^0}(t,u)\bigg\rbrace\ ,
\end{split}
\end{equation}
where we have defined $F_{+/0}(t,u)$ in Eq.~(\ref{F+0tu}), 
and the expressions for $D_{0}^{P^-P^0}(t,u)$, $D_{+0}^{P^-P^0}(t,u)$ and $D_{+}^{P^-P^0}(t,u)$ 
are given in Eqs.~(\ref{SqA:eqDp})--(\ref{SqA:eqDp0}).

The differential decay width in the tau rest frame is
\begin{equation}
\frac{d^2\Gamma}{dt du}=\frac{1}{32(2\pi)^3m_\tau^3}\overline{\left\vert\mathcal{M}_{0}\right\vert^2}\ ,
\end{equation}
where $t=(p_{-}+p_{0})^2$ is the invariant mass and $u=(P-p_{-})^2=(p_0+k+q')^2$. 
The physical region is limited by $(m_{-}+m_0)^2\leq t\leq m_\tau^2$ and $u^{-}(t)\leq u\leq u^{+}(t)$, 
with
\begin{equation}
u^\pm(t)=\frac{1}{2t}\left[2t(m_\tau^2+m_0^2-t)-(m_\tau^2-t)(t+m_{-}^2-m_0^2)\pm
(m_\tau^2-t)\sqrt{\lambda(t,m_{-}^2,m_0^2)}\right]\ ,
\end{equation}
and $\lambda(x,y,z)=x^2+y^2+z^2-2xy-2xz-2yz$.

The invariant mass distribution is obtained integrating upon the $u$ variable
\begin{equation}
\label{eq:AppNR}
\begin{split}
\frac{d\Gamma}{dt}&=\frac{G_F^2 S_{\rm EW}\vert V_{uD}\vert^2 m_\tau^3}{384\pi^3 t}
\bigg\{\frac{1}{2t^2}\left(1-\frac{t}{m_\tau^2}\right)^2\lambda^{1/2}(t,m_{-}^2,m_0^2)\\[1ex]
&\times\left[C_V^2 \left\vert F_{+}(t)\right\vert^2\left(1+\frac{2t}{m_\tau^2}\right)
\lambda(t,m_{-}^2,m_0^2)\left(1+\tilde{\delta}_{+}(t)\right)
+3C_S^2\Delta_{-0}^2\left\vert F_{0}(t)\right\vert^2\left(1+\tilde{\delta}_{0}(t)\right)\right]\\[1ex]
&+C_S C_V \frac{4}{\sqrt{t}}\tilde{\delta}_{+0}(t)\bigg\}\ ,
\end{split}
\end{equation} 
where
\begin{subequations}\label{eq:delta_tilde}\begin{align}
\tilde{\delta}_0(t)&=\frac{\int_{u^{-}(t)}^{u^{+}(t)}D_{0}^{P^{-}P^{0}}(t,u) 
2\mathrm{Re}\left[F_0(t)\delta F_{0}^{*}(t,u)\right]du}
{\int_{u^{-}(t)}^{u^{+}(t)}D_{0}^{P^{-}P^{0}}(t,u)\vert F_{0}(t)\vert^2 du}\ ,\\[1ex]
\tilde{\delta}_+(t)&=\frac{\int_{u^{-}(t)}^{u^{+}(t)}D_{+}^{P^{-}P^{0}}(t,u)
2\mathrm{Re}\left[F_+(t)\delta F_{+}^{*}(u)\right]du}
{\int_{u^{-}(t)}^{u^{+}(t)}D_{+}^{P^{-}P^{0}}(t,u)\vert F_{+}(t)\vert^2du}\ ,\\[1ex]
\tilde{\delta}_{+0}(t)&=\frac{3t\sqrt{t}}{4m_\tau^6}\int_{u^{-}(t)}^{u^{+}(t)}D_{+0}^{P^{-}P^{0}}(t,u)
\left(\mathrm{Re}\left[F_+(t)\delta F_{0}^{*}(t,u)\right]+
\mathrm{Re}\left[F_0(t)\delta F_{+}^{*}(u)\right]\right)du\ .
\end{align}\end{subequations}

\section{Kinematics}
\label{App04:Kine}
As in Refs.~\cite{Cirigliano:2002pv,Antonelli:2013usa,Flores-Tlalpa:2008bws,Miranda:2020wdg},
after an integration over $\mathcal{D}_{\rm IV/III}$ and $\mathcal{D}_{\rm III}$, 
the functions in Eqs.~(\ref{eq:JndK}) are given by (we note that $K_{11}$ is numerically negligible)
\begin{subequations}
\begin{align}
J_{11}(t,u)&=
\log\left(\frac{2x_+(t,u)\bar{\gamma}}{M_\gamma}\right)
\frac{1}{\bar{\beta}}\log\left(\frac{1+\bar{\beta}}{1-\bar{\beta}}\right)\\[1ex]
&+\frac{1}{\bar{\beta}}\left[\mathrm{Li}_2(1/Y_2)-\mathrm{Li}_2(Y_1)
+\log^2(-1/Y_2)/4 -\log^2(-1/Y_1)/4\right]\ ,\\[1ex]
J_{20}\left(t,u\right)&=
\log\left(\frac{M_\gamma(m_\tau^2-t)}{m_\tau\,x_+(t,u)}\right)\ ,\\[1ex]
J_{02}\left(t,u\right)&=
\log\left(\frac{M_\gamma(m_\tau^2+m_{0}^2-t-u)}{m_{-}\,x_+(t,u)}\right)\ ,\\[1ex]
K_{20}\left(t,u\right)&=K_{02}\left(t,u\right)=\log\left(\frac{x_-(t,u)}{x_+(t,u)}\right)\ ,
\end{align}
\end{subequations}
where
\begin{equation}
\label{Appx4:eq18}
\begin{split}
x_\pm\left(t,u\right)=&
\frac{-m_{-}^4+(m_{^0}^2-t)(m_\tau^2-u)+m_{-}^2(m_\tau^2+m_{0}^2+t+u)}{2m_{-}^2}\\[1ex]
&\pm\frac{\lambda^{1/2}\left(u,m_\tau^2,m_{-}^2\right)\lambda^{1/2}\left(t,m_{-}^2,m_{0}^2\right)}
{2m_{-}^2}\ .
\end{split}
\end{equation}
These expressions are written in terms of
\begin{equation}
Y_{1,2}=\frac{1-2\bar{\alpha}\pm\sqrt{(1-2\bar{\alpha})^2-(1-\bar{\beta}^2)}}{1+\bar{\beta}}\ ,
\end{equation}
with 
\begin{subequations}
\begin{align*}
\bar{\alpha}=&
\frac{(m_\tau^2-t)(m_\tau^2+m_{0}^2-t-u)}{(m_{-}^2+m_\tau^2-u)}\cdot
\frac{\lambda(u,m_{-}^2,m_\tau^2)}{2\bar{\delta}}\ ,\\[1ex]
\bar{\beta}=&
-\frac{\sqrt{\lambda(u,m_{-}^2,m_\tau^2)}}{m_{-}^2+m_\tau^2-u}\ ,\\[1ex]
\bar{\gamma}=&
\frac{\sqrt{\lambda(u,m_{-}^2,m_\tau^2)}}{2\sqrt{\bar{\delta}}}\ ,\\[1ex]
\bar{\delta}=&
-m_{0}^4m_\tau^2+m_{-}^2(m_\tau^2-t)(m_{0}^2-u)-tu(-m_\tau^2+t+u)\\[1ex]
&+m_{0}^2(-m_\tau^4+tu+m_\tau^2 t+m_\tau^2 u)\ .
\end{align*}
\end{subequations}

\bibliographystyle{unsrt}
\bibliography{bibl}
\addcontentsline{toc}{section}{References}

\end{document}